\documentclass{aa}

\usepackage{epsfig}
\usepackage{graphicx}
\usepackage{times}
\usepackage{color}
\usepackage{ulem}
\usepackage{hyperref}
\usepackage{url}

\usepackage{xcolor}
\hypersetup{
colorlinks,
linkcolor={red!80!black},
citecolor={blue!80!black},
urlcolor={blue!80!black}
}
\def\cm3{cm$^{-3}$}
\def\kms{km~s$^{-1}$}

\def\lsun{L$_{\odot}$}
\def\rsun{R$_{\odot}$}

\def\msun{M$_{\odot}$}

\def\one{\ts {\,\sc i}}
\def\two{\ts {\,\sc ii}}
\def\three{\ts {\,\sc iii}}

\def\beq{\begin{equation}}
\def\eeq{\end{equation}}

\def\lesssim{\mathrel{\hbox{\rlap{\hbox{\lower4pt\hbox{$\sim$}}}\hbox{$<$}}}}
\def\gtrsim{\mathrel{\hbox{\rlap{\hbox{\lower4pt\hbox{$\sim$}}}\hbox{$>$}}}}

\def\lesssim{\mathrel{\hbox{\rlap{\hbox{\lower4pt\hbox{$\sim$}}}\hbox{$<$}}}}
\def\gtrsim{\mathrel{\hbox{\rlap{\hbox{\lower4pt\hbox{$\sim$}}}\hbox{$>$}}}}

\def\isoni{$^{56}{\rm Ni}$}

\def\one{{\,\sc i}}
\def\two{{\,\sc ii}}
\def\three{{\,\sc iii}}

\def\bec{{\sc bec}}
\def\v1d{{\sc v1d}}

\def\mesa{{\sc mesa}}
\def\cmfgen{{\sc cmfgen}}

\def\ergs{erg\,s$^{-1}$}

\newcommand{\iso}[2]{\ensuremath{^{#1}\rm{#2}}}

\def\aj{AJ}
\def\pasp{PASP}
\def\pasa{PASA}
\def\apj{ApJ}
\def\apjs{ApJS}
\def\apjl{ApJL}
\def\aap{A\&A}
\def\araa{ARA\&A}

\def\mnras{MNRAS}
\def\nat{Nature}

\def\nifs{\iso{56}Ni}

\begin{document}

\title{Supernovae from massive stars with extended tenuous envelopes}

\titlerunning{Explosions in extended stars}

\author{Luc Dessart\inst{\ref{inst1}}
  \and
  Sung-Chul Yoon\inst{\ref{inst2}}
  \and
  Eli Livne\inst{\ref{inst3}}
  \and
  Roni Waldman\inst{\ref{inst3}}
  }

\institute{
    Unidad Mixta Internacional Franco-Chilena de Astronom\'ia (CNRS UMI 3386),
    Departamento de Astronom\'ia, Universidad de Chile,
    Camino El Observatorio 1515, Las Condes, Santiago, Chile\label{inst1}
  \and
    Department of Physics and Astronomy, Seoul National University, Gwanak-ro 1,
    Gwanak-gu, Seoul, 151-742, Republic of Korea\label{inst2}
  \and
    Racah Institute of Physics, The Hebrew University, Jerusalem 91904, Israel\label{inst3}
}

\date{Accepted . Received }

\abstract{
Massive stars with a core-halo structure are interesting objects for stellar physics and hydrodynamics.
Using simulations for stellar evolution, radiation hydrodynamics, and radiative transfer, we study
the explosion of stars with an extended and tenuous envelope (i.e., stars in which 95\% of
the mass is contained within 10\% of the surface radius or less). We consider both H-rich supergiant and He-giant
progenitors resulting from close-binary evolution and dying with a final mass of 2.8-5\,\msun.
An extended envelope causes the supernova (SN) shock to brake and a reverse shock to
form, sweeping core material into a dense shell.
The shock deposited energy, which suffers little degradation from expansion, is trapped in ejecta layers
of moderate optical depth,  thereby enhancing the SN luminosity at early times.
With the delayed \isoni\ heating, we find that the resulting optical and near-IR light curves all exhibit
a double-peak morphology. We show how an extended progenitor can explain the blue and featureless
optical spectra of some Type IIb and Ib SNe. The dense shell formed by the reverse shock leads
to line profiles with a smaller and near-constant width. This ejecta property can explain the statistically narrower profiles
of Type IIb compared to Type Ib SNe, as well as the peculiar H$\alpha$ profile seen in SN\,1993J.
At early times, our He-giant star explosion model shows a high luminosity, a blue colour,
and featureless spectra reminiscent of the Type Ib SN\,2008D, suggesting a low-mass progenitor.
}

\keywords{
radiative transfer --
hydrodynamics --
supernovae: general --
supernovae: individual: SN\,1993J, SN\,2008D.
}

\maketitle
\label{firstpage}

\section{Introduction}

\begin{figure*}
 \includegraphics[width=0.48\textwidth]{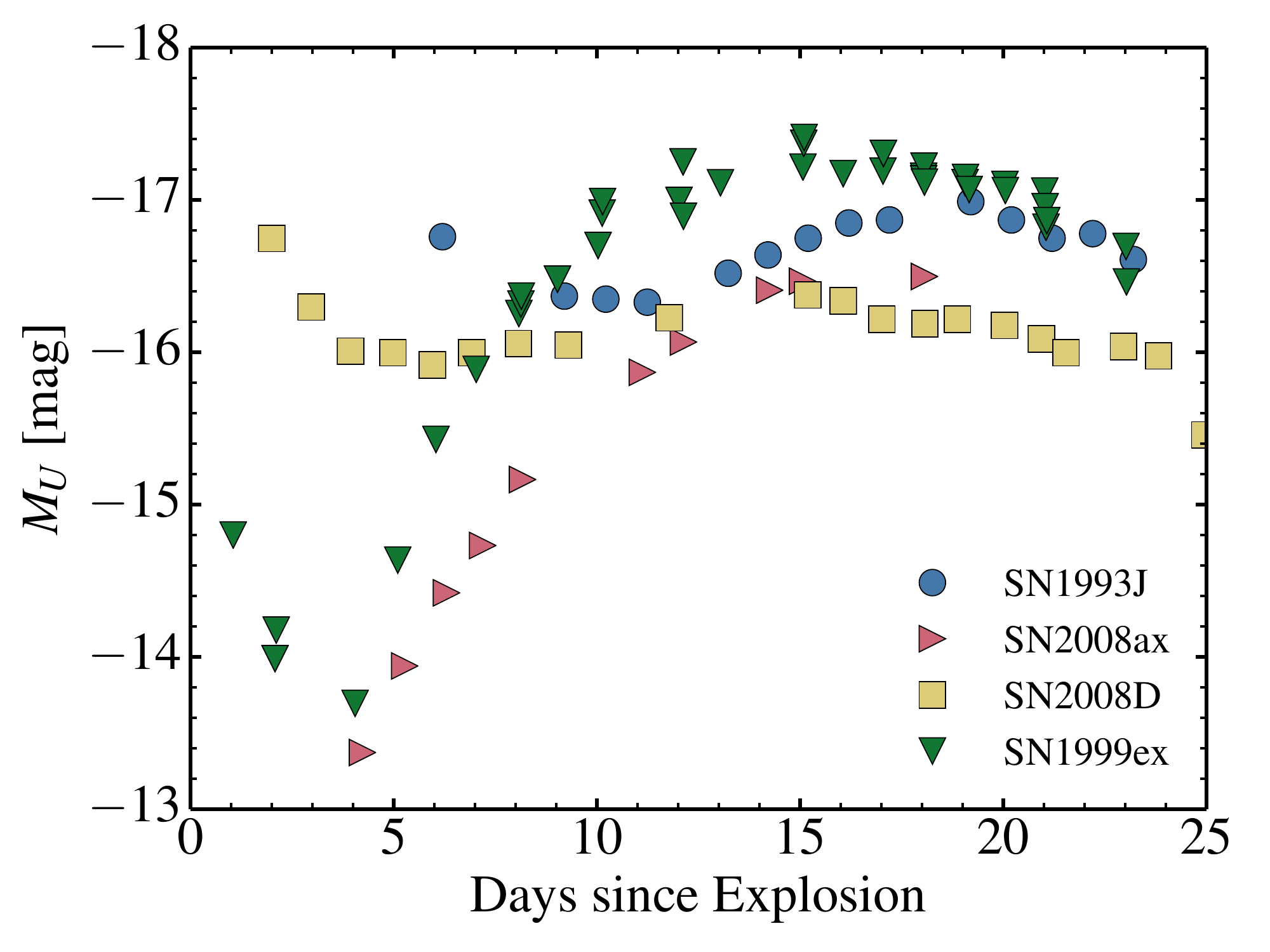}
 \includegraphics[width=0.48\textwidth]{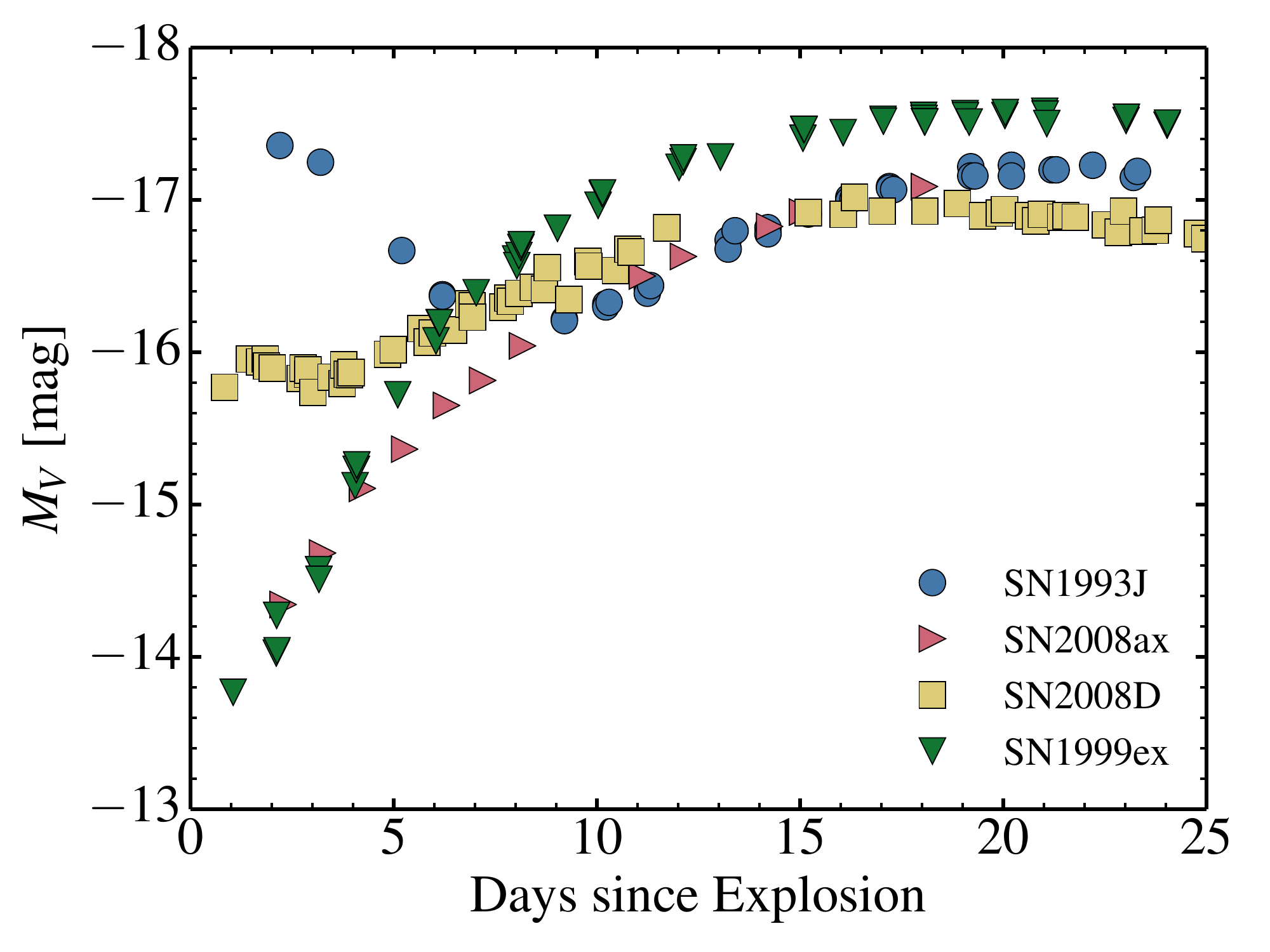}
\caption{Illustration of the evolution of the $U$ and $V$-band absolute magnitudes for two SNe IIb
(1993J and 2008ax) and two SNe Ib (2008D and 1999ex).
For SN\,2008ax, we use the photometry from \citet{taubenberger_08ax}.
We adopt their explosion date MJD\,54528.80. We use a total reddening $E(B-V)$ of 0.302\,mag and
a distance of 8.7\,Mpc \citep{drout_11_ibc}.
For SN\,1999ex, we use the photometric data of \citet{stritzinger_99ex}.
We adopt a reddening $E(B-V)$ of 0.28\,mag, a distance of 46.0\,Mpc \citep{drout_11_ibc}, and
an explosion date of MJD\,51480.0  \citep{stritzinger_99ex}.
See Section~\ref{sect_obs} for details.
\label{fig_obs}
}
\end{figure*}

   The light curves of Type IIb and Ib supernovae (SNe)
   are understood to stem from the decay of  \iso{56}Ni, exhibiting a bolometric maximum
   about 20-30\,d after the  time of explosion
   \citep{ensman_woosley_88,richardson_ibc_06,drout_11_ibc,bianco_etal_14,taddia_ibc_15}.
   Some SNe that exhibit a comparable brightness
   at maximum (compatible with a few 0.01\,\msun\ of \iso{56}Ni) have, in contrast, vastly different
   light curves prior to maximum (Fig.~\ref{fig_obs}). For example, the Type IIb
   SN\,1993J is about 3\,mag brighter than its counterpart SN\,2008ax about 20\,d before
   maximum. This feature is not limited to H-rich progenitors since similar offsets are also seen
   amongst Type Ib SNe (e.g., SNe\,2008D and 1999ex; Fig.~\ref{fig_obs}).\footnote{
   We consider that a model is H-rich if the corresponding progenitor has a large surface H mass fraction.
   Consequently, the observation of the corresponding star would likely reveal
   H\one\ lines in optical spectra. This can occur even if the total H mass is very small.
   In the corresponding Type IIb SN explosion, the H$\alpha$ line may be strong
    even if the total H mass is as low as 0.001\,\msun\ \citep{dessart_11_wr}.}

   Various mechanisms and/or progenitor/ejecta properties may produce this luminosity boost
   at early times, including
   1) an increase in  progenitor radius up to a few 100\,\rsun\
   (e.g., \citealt{nomoto_93j_93}; \citealt{podsiadlowski_93j_93}; \citealt{woosley_94_93j});
   2) an enhanced mixing of \iso{56}Ni into the outer ejecta, as originally inferred for
   SN\,1987A (e.g., \citealt{sn1987A_rev_90});
   3) an interaction with a companion star in close-binary systems \citep{K10,moriya_bin_lc_15};
   4) an interaction with circumstellar material (CSM), in analogy with SNe IIn \citep{schlegel_90}.
   In addition to an impact on the SN luminosity, the ejecta dynamics and shock propagation
   in such progenitor stars are different,
   likely leading to differences in photometric (e.g., color), spectroscopic, and polarisation signatures.
   For example, interaction with CSM may influence line profile formation by producing narrow
   H\one\ and He\one\ lines with extended wings through multiple electron-scattering
   \citep{chugai_98S_01,groh_13cu,grafener_vink_13cu_16,D16_2n}.

    The identification of a few supergiant star progenitors, and sometimes of a companion star,
    on pre/post-explosion  images  (e.g.,  \citealt{aldering_93j_94}; \citealt{vandyk_93j_02};
    \citealt{maund_93j_04,maund_11dh_11};\citealt{vandyk_11dh_13}; \citealt{folatelli_11dh_14})
    supports the notion that, at least in some cases, the diversity in early-time light
    curves may reflect a variation in progenitor radii, perhaps in connection with binarity.

        The supergiant progenitors of Type IIb/Ib SNe have, however,
        a different density structure from the blue-supergiant (BSG)
    and red-surpergiant (RSG) progenitors at the origin of Type II-peculiar SNe (like SN\,1987A) or
    Type II-Plateau SNe (like SN\,1999em). Indeed, SNe IIb/Ib have different light curves. They rise
    to the \iso{56}Ni-powered maximum
    much faster than SN II-peculiar objects and their post-explosion brightness is not sustained for as long
    as SNe II-Plateau. The SN IIb/Ib progenitors may be supergiant stars but their
    mass at core collapse is much smaller than those of Type II-peculiar/II-Plateau SNe,
    and probably in the range $2-5$\,\msun\ (see, e.g., \citealt{ensman_woosley_88}).

    Models of massive stars suggest that large radii are possible. This may happen in single stars
    because of their proximity to the Eddington limit. Because of the iron opacity bump, massive stars
    may expand to lower the envelope opacity, leading to the formation of a tenuous envelope above
    a dense core \citep{ishii_99,petrovic+06,grafener+12}. Tenuous envelopes can also result
    from binary star evolution.
    The envelopes of H-rich supergiants can be drained during mass transfer while retaining a large radius
    \citep{podsiadlowski_92}.  Low-mass massive stars in binary systems may also lead to the formation
    of  He-giant stars, i.e., H-deficient stars with radii of $\sim$\,100\,\rsun, which eventually produce
    a Type Ib SN \citep{yoon_presn_12,eldridge_ib_15,clelland_he_16,yoon_ibc_17}.
    The spatial extent of these tenuous envelopes should also depend
    on metallicity (the abundance of metals influences the gas opacity and impacts
    the energy transport between the core and the stellar surface).

    Radiation-hydrodynamics modelling of SNe originating from extended progenitor stars have
    been done originally in the context of SN\,1993J \citep{woosley_94_93j,blinnikov_94_93j}
    and more recently for SN\,2011dh \citep{bersten_etal_12_11dh,nakar_piro_14}
    or SN\,2008D \citep{bersten_08D_13}, including shock break out \citep{moriya_sbo_15}.
    They have emphasized the impact of the progenitor radius and the envelope structure
    on the early light curve.
    Here, we extend these previous studies to model the bolometric and multi-band light curves
    as well as the spectroscopic properties of explosions arising from H-rich and H-deficient stars
    with an extended envelope. We also devote more time to discuss the importance of the envelope
    structure on the shock dynamics, something that may have relevance in all types of SNe.

   This paper is structured as follows. In Section~\ref{sect_obs}, we briefly review the set of observations
   that we use to confront our model results. In Section~\ref{sect_model}, we review the numerical
   approach used to model the pre-SN evolution, the explosion, and the subsequent ejecta/radiation until
   around the time of bolometric maximum. In Section~\ref{sect_dyn}, we discuss
   the dynamics of the shock until shock breakout as well as the ejecta evolution until the dynamics
   is over (i.e., when homologous expansion is established). In Section~\ref{sect_rad}, we describe
   the photometric and spectroscopic properties of the emergent radiation for our model set, which includes
   both H-rich and H-deficient progenitors. In Section~\ref{sect_comp_obs},  we confront our results to
   the observations of a few SNe. In Section~\ref{sect_conc}, we present our conclusions.


\section{Observational data}
\label{sect_obs}

Our model sample includes both H-rich and H-deficient ejecta, associated with SNe IIb and Ib.
Observed SNe IIb/Ib encompass a range of light
curve properties, in particular in terms of pre-maximum optical brightness.
SN\,1993J (IIb) is bright early-on while SN\,2008ax (IIb) is faint early-on.
Similarly, SN\,2008D (Ib) is bright early on, while SN\,1999ex (Ib) is faint
(Fig.~\ref{fig_obs}).

The focus in the present paper is to see what progenitor/explosion properties
can lead to a bright SN early after explosion, as seen for SNe 1993J and 2008D.
We will thus confront our models to these two SNe.

For SN\,1993J, we use the photometry of \citet{richmond_93j_96} and the spectra from
\citet{matheson_93j_00a}.
We adopt a distance of 3.63\,Mpc and a reddening $E(B-V)=$\,0.08\,mag \citep{richmond_93j_94}.
For the explosion date, we adopt MJD\,49074.0 \citep{lewis_93j_94}.


For SN\,2008D, we use the photometry and optical spectroscopy from \citet{soderberg_08D}
and \citet{modjaz_08D}. For the explosion date, we adopt the time of the X-ray burst, i.e., MJD\,54474.57.
We adopt a reddening $E(B-V)$ of 0.613\,mag and a distance of 30\,Mpc \citep{drout_11_ibc}.
The published optical spectra of SN\,2008D suggest a non-monotonic
colour evolution in the blue because the relative flux of spectra obtained within 0.1\,d of each other
have a distinct colour. These features are suggestive of a problem with the relative flux calibration. For our
comparisons, we only use epochs that yield a smoothly-varying sequence in optical colour.


\begin{figure*}
\begin{center}
 \includegraphics[width=0.45\hsize]{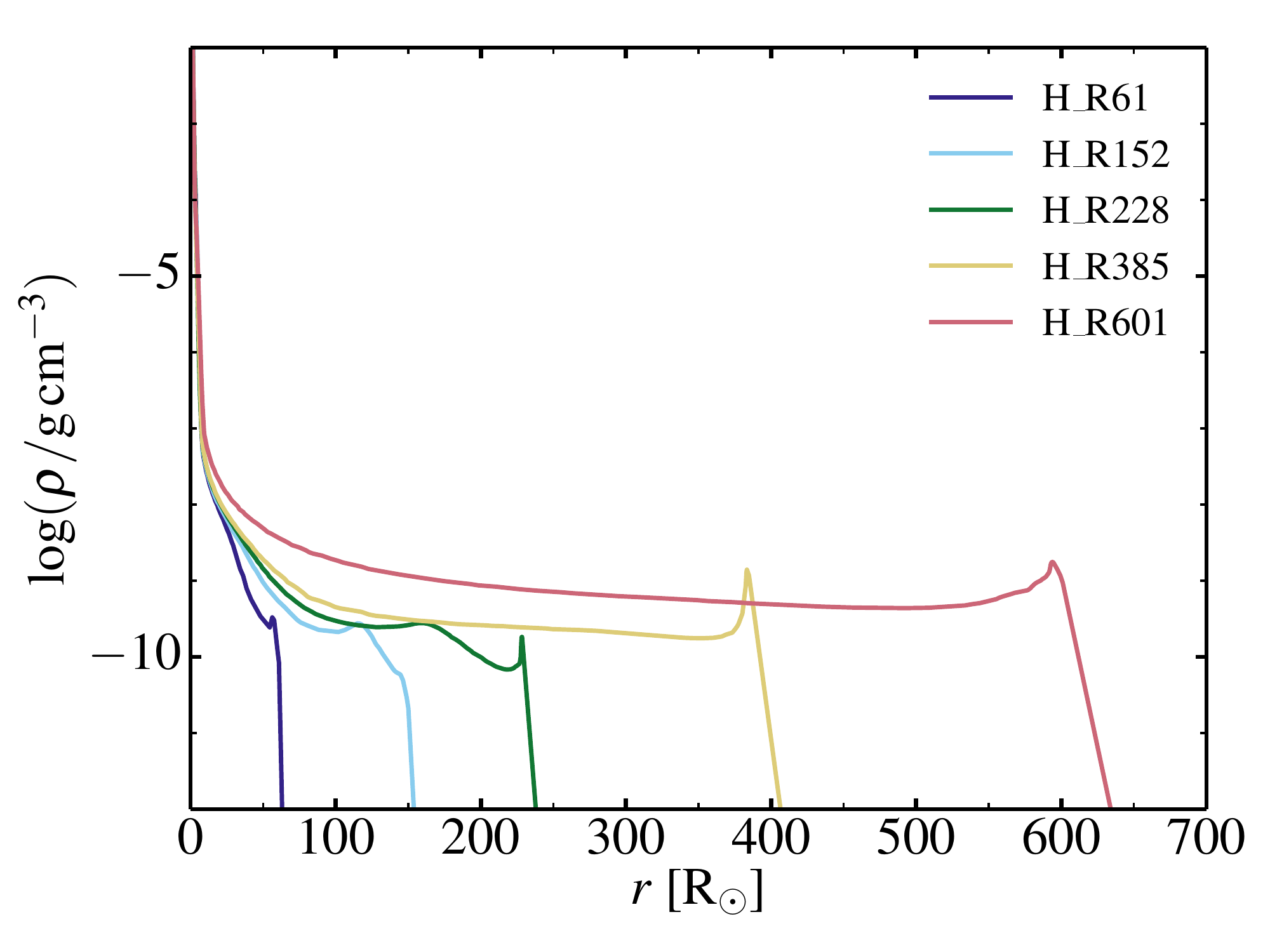}
 \includegraphics[width=0.45\hsize]{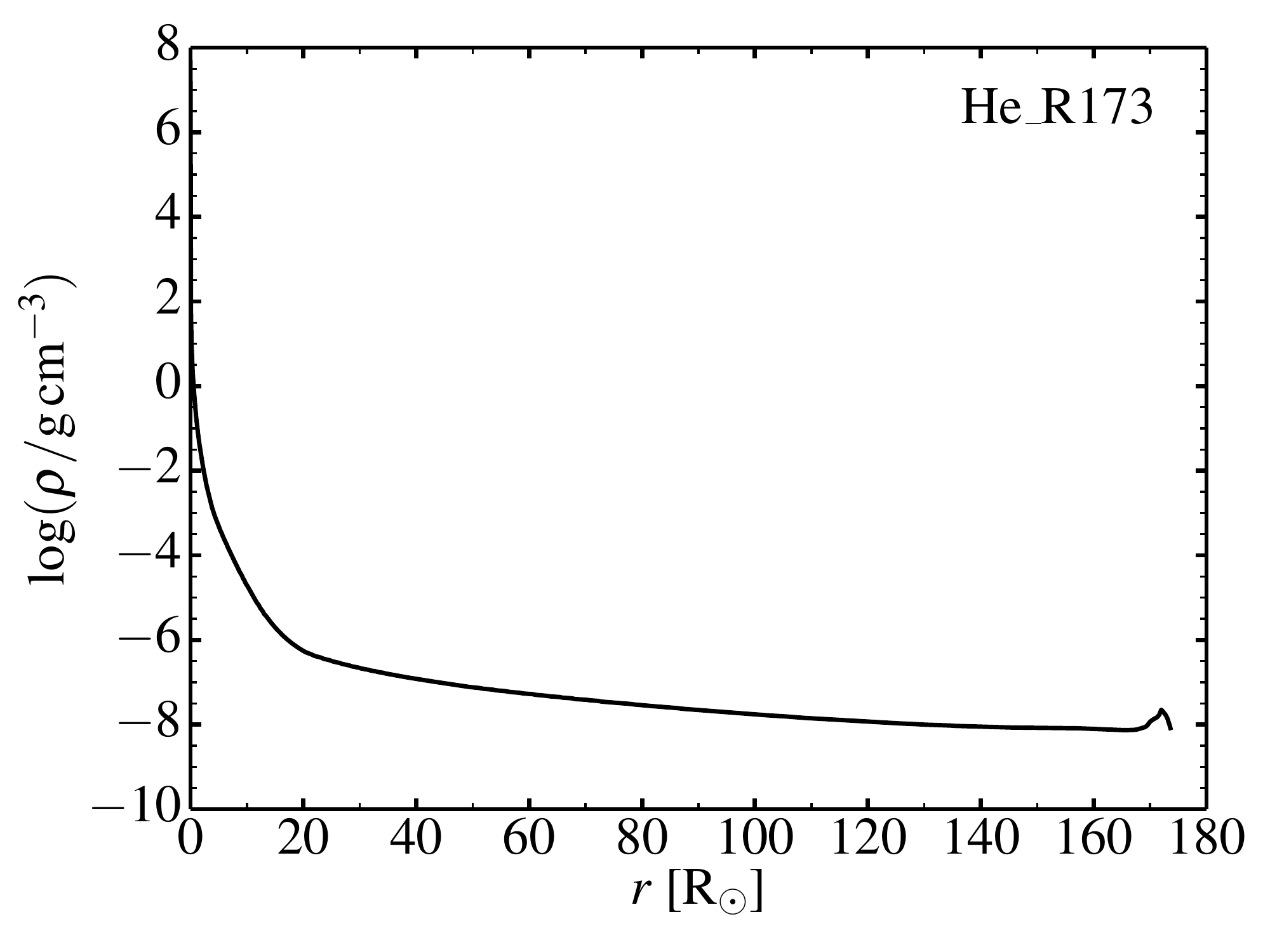}
 \includegraphics[width=0.45\hsize]{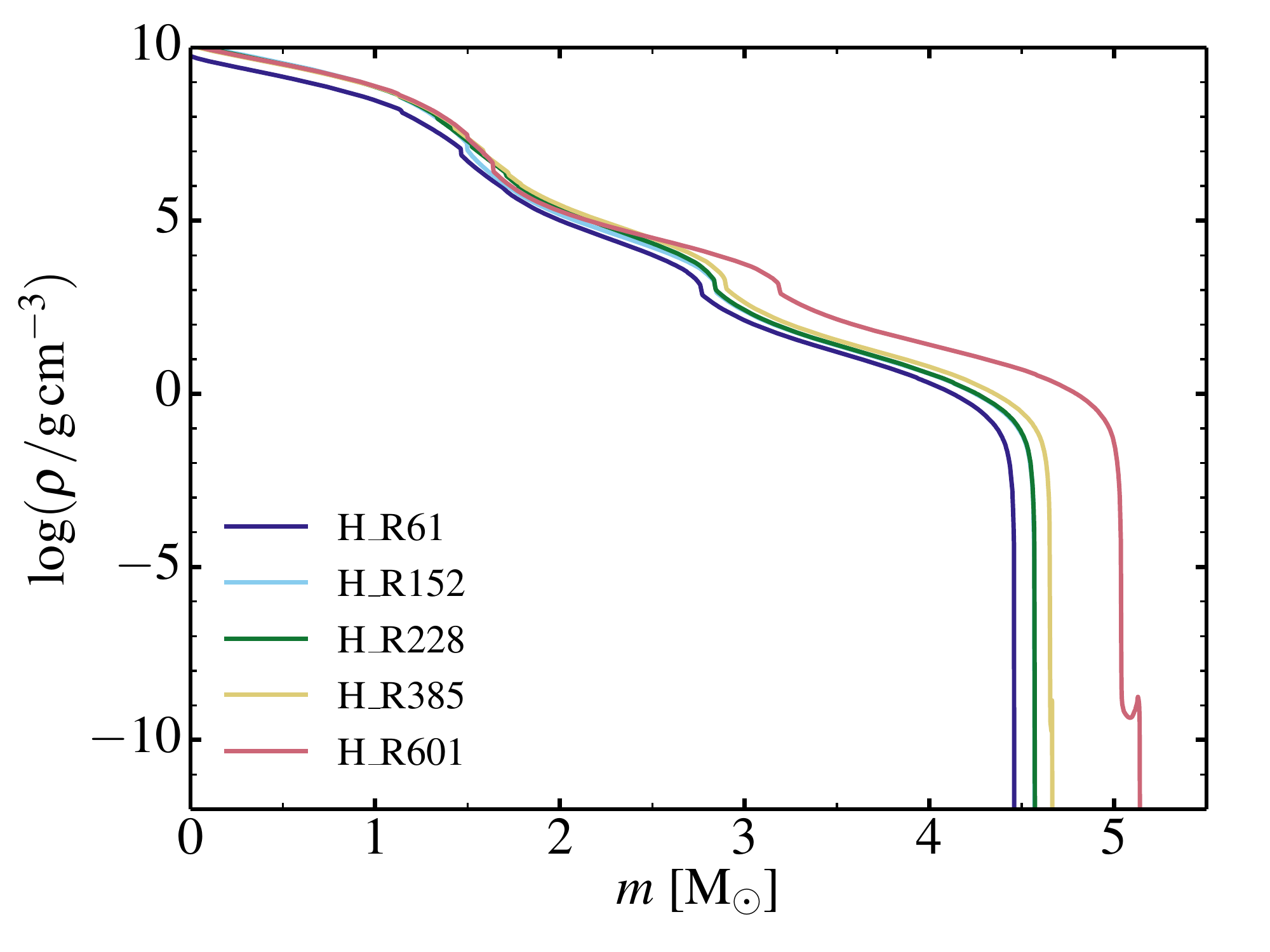}
 \includegraphics[width=0.45\hsize]{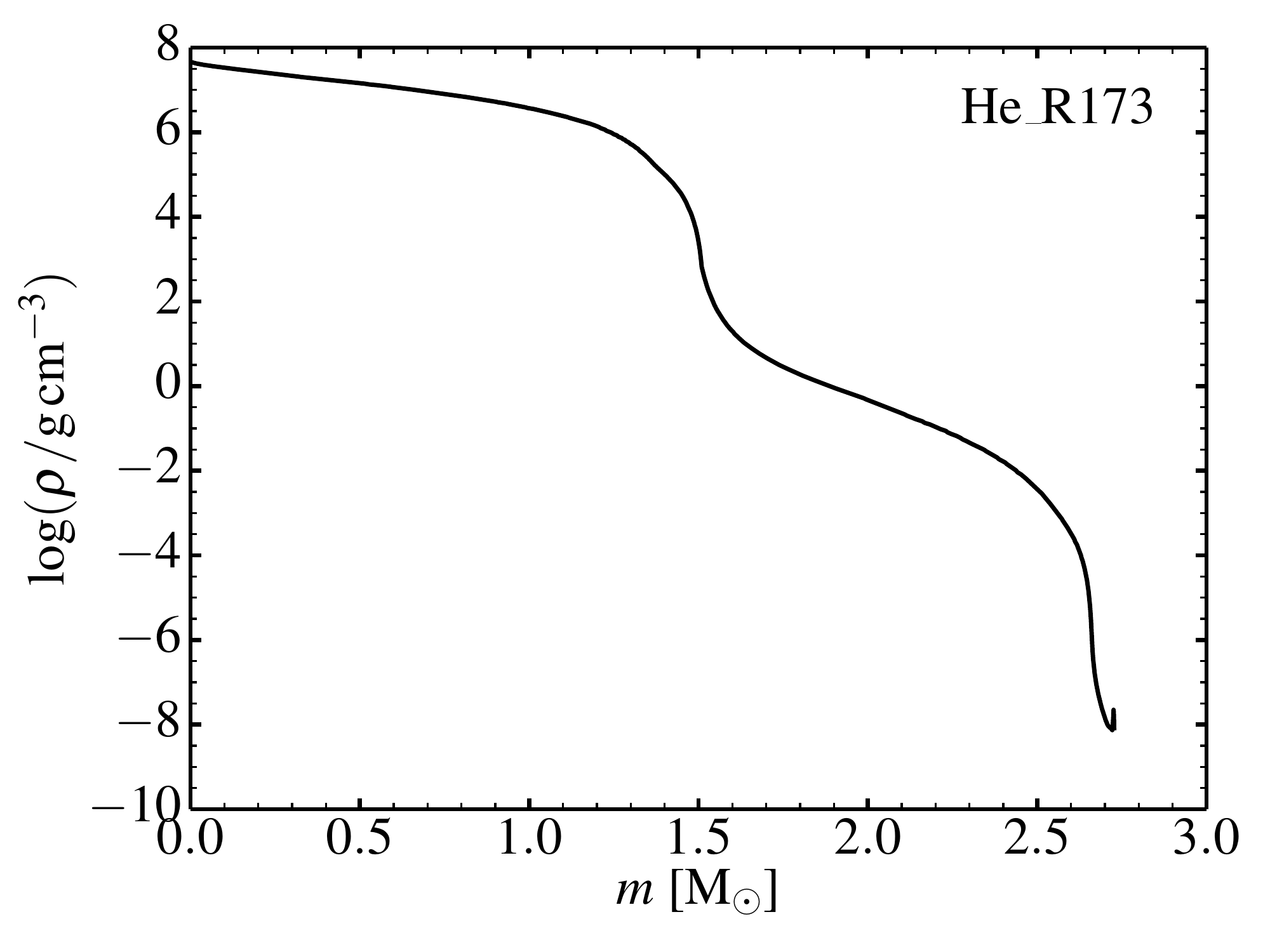}
 \includegraphics[width=0.45\hsize]{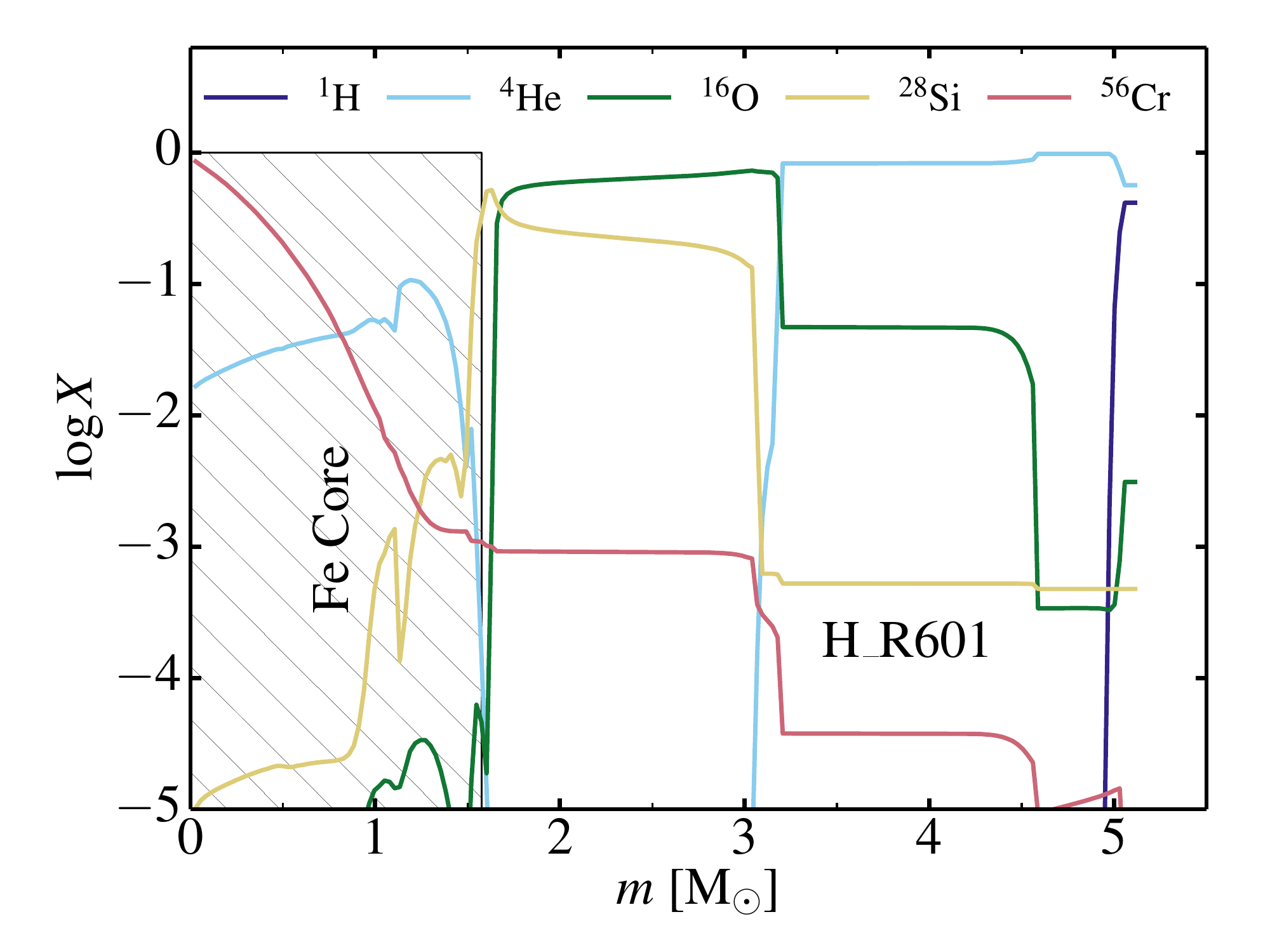}
 \includegraphics[width=0.45\hsize]{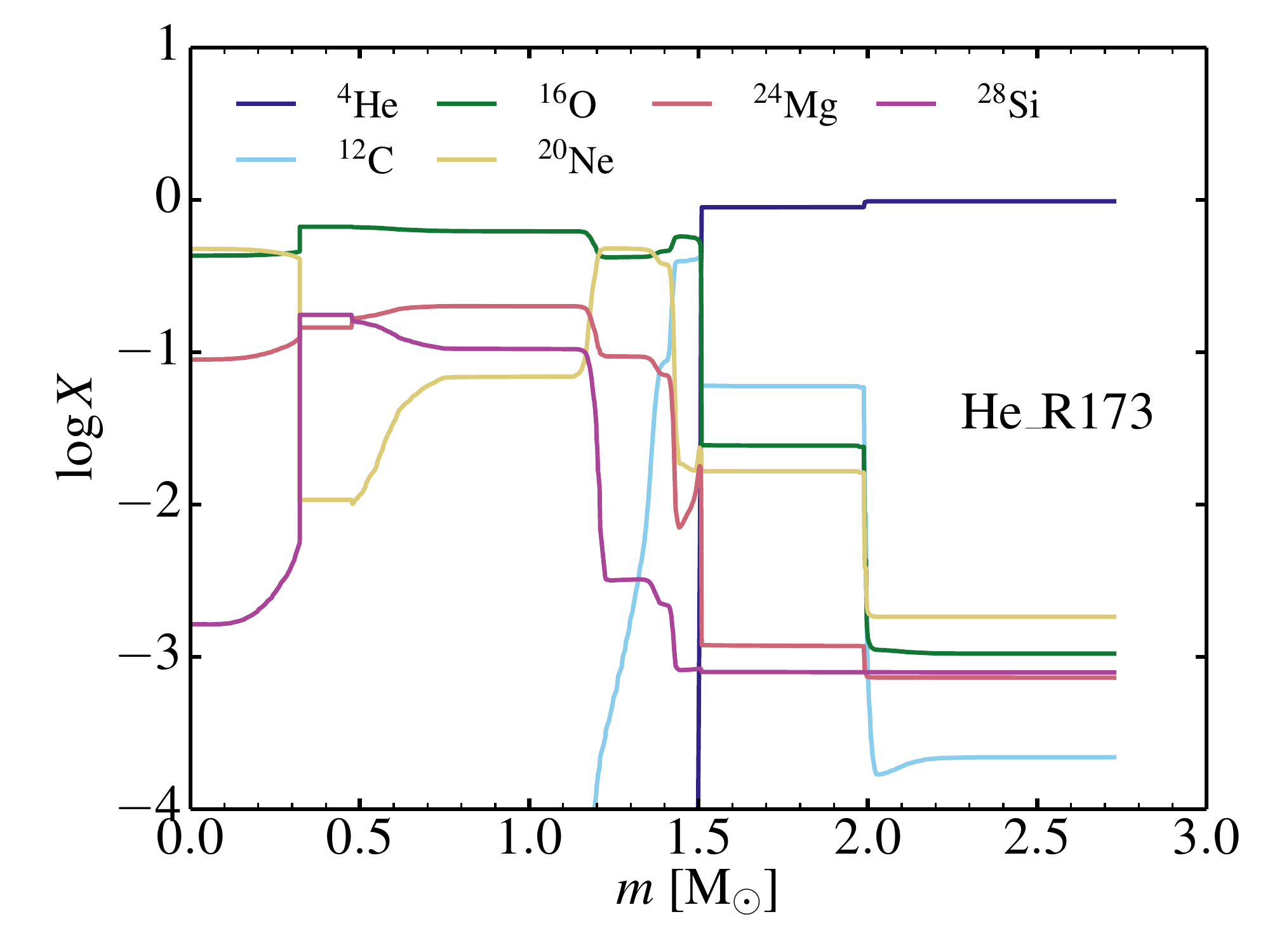}
\caption{Left column: Illustration of the density structure versus radius (top) and mass (middle)
for the model set of H-rich progenitors. We also show the composition for the main species in
the bottom panel.
The composition profiles is shown for all H-rich models in the appendix, in
Fig.~\ref{fig_mesa_comp_h_all}.
All these \mesa\ models are evolved until iron core collapse.
Right column: Same as left column, but now showing the properties of the He-giant progenitor
model He\_R173. This He-giant model is evolved until the onset of neon-core burning only.
The color coding used for the composition (bottom row plots) differs for H-rich (left) and H-deficient
models (right).
\label{fig_summary_prog}
}
\end{center}
\end{figure*}

\section{Numerical setup}
\label{sect_model}

  In this section, we first describe the computation of the evolution of our massive star
  models up to the point when we trigger the explosion (Section~\ref{sect_presn}).
  We then present the approach for simulating the explosion (Section~\ref{sect_v1d})
  and the evolution of the SN radiation and ejecta properties up to a few days/weeks
  (Section~\ref{sect_cmfgen}).

%
%
\begin{table*}
\begin{center}
\caption{
Summary of progenitor and ejecta properties.
$M_{\rm i}$ and $M_{\rm f}$ are the initial and final mass, $R_\star$ is the final surface radius,
$R_{\rm 95}$ corresponds to the radius that contains 95\% of the star mass in the progenitor model.
$M_{\rm env}$ is the mass of the tenuous envelope (we include envelope material with a mass density
lower than 10$^{-5}$\,g\,cm$^{-3}$.
$M_{\rm r}$ is the remnant mass. $M_{\rm e}$ and $E_{\rm kin}$ are the ejecta mass and
kinetic energy.
The following three columns give the ejecta velocity that bounds 99\%
of the corresponding species total mass. The integration is done inwards in velocity space for H and He,
and outwards for \iso{56}Ni. The next three columns give the total mass of H, He, and \iso{56}Ni
initially (\iso{56}Ni$_0$).
The last column give the surface mass fractions of H.
All models were evolved all the way to iron core collapse except models and He\_R11 and He\_R173,
for which the progenitor evolution  was stopped at the onset of neon-core burning.
Model He\_R11 is the same as He\_R173 except that it was trimmed of the extended low-density envelope
prior to explosion (it is used for the comparative study of the shock behaviour presented in Section~\ref{sect_dyn}).
All H-rich models are characterized by a low level of mixing. The H-deficient models He\_R173
and He\_R11 have no \iso{56}Ni.
\label{tab_init}}
\begin{tabular}{
l@{\hspace{-1mm}}c@{\hspace{1mm}}c@{\hspace{1mm}}c@{\hspace{1mm}}
c@{\hspace{1mm}}c@{\hspace{1mm}}c@{\hspace{1mm}}c@{\hspace{1mm}}
c@{\hspace{1mm}}c@{\hspace{1mm}}c@{\hspace{1mm}}c@{\hspace{1mm}}
c@{\hspace{1mm}}c@{\hspace{1mm}}c@{\hspace{1mm}}c@{\hspace{1mm}}
c@{\hspace{1mm}}c@{\hspace{1mm}}c@{\hspace{1mm}}
}
\hline
Model    &    $M_{\rm i}$ &    $M_{\rm f}$ & $L_\star$ & $T_{\rm eff}$ & $R_\star$ & $R_{\rm 95}$ & $M_{\rm env}$&  $M_{\rm r}$ & $M_{\rm e}$ &
$E_{\rm kin}$  &  $V_{\rm 99, H}$&  $V_{\rm 99, He}$&  $V_{\rm 99, Ni}$  & H & He & $^{56}$Ni$_0$ & $X_{\rm H,s}$  \\
               &          [\msun] &        [\msun] & [\lsun] & [K] & [\rsun] &  [\rsun] &     [\msun] &    [\msun] &        [\msun] &      [B] &      [\kms]  &      [\kms]  &      [\kms]   & [\msun]  &  [\msun]  & [\msun]   & \\
\hline
H\_R5        &  16 &   4.13 &  9.8(4) & 4.6(4) &  5.0  &  0.87  & 0.0 & 1.55 &  2.58  &             1.26 &       $\dots$    &       1.75(3) &       6.36(3) &       0.00       &       1.42 &    0.061     &  0.0      \\
H\_R61      &  16 &   4.46 &  \dots  & \dots & 61.1 & 0.80     & 5.0(-4) &  1.60  &  2.86  &             1.23 &       1.19(4) &       3.87(3) &       5.92(3) &       4.10(-3) &       1.51 &  0.059     &  0.062  \\
H\_R152    &  16 &   4.57 &  7.4(4) & 7.7(3) & 152.2 & 0.77 & 1.1(-3) &  1.60  &  2.97  &             1.22 &       1.22(4) &       3.59(3) &       5.74(3) &       3.72(-3) &       1.50 &   0.049   &    0.059 \\
H\_R228    &  16 &   4.57 &  1.0(5) & 6.8(3) & 228.4 & 0.76 &  2.2(-3) &  1.70 &  2.87  &             1.27 &       1.26(4) &       3.64(3) &       5.95(3) &       4.09(-3) &       1.51 &   0.059   &   0.066\\
H\_R385    &  16 &   4.67 &  1.0(5) & 5.3(3) & 385.3 & 0.75 &  1.2(-2)  & 1.70 &  2.97  &             1.21 &       1.23(4) &       3.60(3) &       5.82(3) &       8.38(-3) &       1.53 &    0.049  &   0.12  \\
H\_R601    &  16 &   5.14 &  1.2(5) & 4.3(3) & 601.5 & 0.77 &  0.10 & 1.65 &  3.49  &             1.20 &       8.78(3) &       3.85(3) &       4.53(3) &       5.17(-2) &       1.70 &    0.058  &  0.38  \\
\hline
He\_R11     &  12  & 2.67  & \dots & \dots & 11.0    & 3.4 &    0.0 & $\dots$ & 0.98  &              1.02  &   $\dots$  &       5.00(3)  &  $\dots$   &   0.0   & 0.93  &     0.0   & 0.0    \\
He\_R173   & 12  & 2.73  & 3.0(4) & 5.8(3) & 173.0  & 5.2 &  0.074  & $\dots$ & 1.04  &              0.97  &   $\dots$  &       5.00(3)  &  $\dots$   &   0.0   & 0.99  &     0.0   & 0.0    \\
\hline
\end{tabular}
\flushleft

Notes: a) H-rich models stem from the explosion of the primary in a binary system originally
composed of a 16.0 and a 14.0\,\msun\ component.
The H-deficient models result from the evolution of 2.8\,\msun\ star on the  zero-age He main
sequence --- this star would have had a mass of $\sim$\,12\,\msun\ on the zero-age H main sequence.
b) Model H\_R385 went through Case B mass transfer, and all other models through
Case BB mass transfer.
\end{center}
\end{table*}

\subsection{Progenitor evolution}
\label{sect_presn}

  Two different approaches/codes are used for the evolution of the progenitor star.
The first approach employs the code \mesa\ \citep{mesa1,mesa2,mesa3} and is used to produce H-rich progenitors.
The second approach uses the code \bec\ (see \citealt{yoon_ibc_10} and references therein for
details) and is used to produce He-rich (and H-deficient) stars.
Progenitor properties for our model set at the time of the piston trigger
are illustrated in Fig.~\ref{fig_summary_prog} and, in the appendix,
in Fig.~\ref{fig_mesa_comp_h_all}.

\subsubsection{H-rich binary progenitors}

   We study systems
   consisting on the zero-age-main-sequence (ZAMS) of a 16\msun\ primary and a 14\msun\ secondary star
  at solar metallicity $(Z = 0.02)$, with an initial orbital separation of 50, 100, 200, 300, 500, and 1000\,\rsun.
  We model the evolution of each component from the ZAMS until iron core collapse of the primary star,
  using the binary-evolution implementation in \mesa\ version 7624.
  The `Dutch' recipe is used to estimate the mass loss rate, with a coefficient $\eta=1$.
Convection was followed according to the Ledoux criterion, with a mixing length parameter
$\alpha_{\rm MLT}=3$, a semi-convection efficiency parameter
$\alpha_{\rm SC}=0.1$ \citep[Eq.~12]{mesa2}, and an exponential
overshoot with parameter $f=$\,0.008 \citep[Eq.~2]{mesa1}.
Both components are allowed to rotate, and the resulting effects of
rotation on stellar structure, chemical mixing, angular momentum transport,
and mass loss are accounted for according to the methods available in that \mesa\ version.

Binary evolution was started with both components
at synchronous rotation, and Roche-lobe overflow (RLOF) was followed according
to the Ritter scheme with an implicit method,  as described in \citet{mesa3}.
Similarly, the code follows the effects of tides and angular momentum transfer via RLOF on the rotation of
both components.

The evolution is computed with the {\tt approx21.net} nuclear network, which is composed of the
basic CNO and $\alpha$-chain isotopes up to \iso{56}Ni together with neutrons, protons, \iso{56}Cr,
\iso{54}Fe, and \iso{56}Fe.\footnote{In this nuclear network, \iso{56}Cr serves as a neutron rich end product.
It is produced by a fake electron capture reaction involving \iso{56}Fe and two electrons.}

  In the course of their evolution, such stellar models inevitably become red supergiants (RSGs),
  fill their Roche lobe, and transfer mass to the companion star. This occurs after the He core has
  formed so that the main difference between the models is the structure of the H-rich envelope.

  In the models selected, we obtain primary stars at death with a total mass of 4.46-5.14\,\msun\ ---
  the H-rich envelope was nearly entirely lost during Case B and Case BB mass transfer.
  Their pre-SN radius varies monotonically with the initial orbital
  separation in our set. The values are 61, 152, 228, 385, and 601\,\rsun,
  corresponding in the same order to orbital separations of 50, 100, 200, 300, 500, and 1000\,\rsun.
  The mass of their extended H--rich envelope is $0.01-0.1$\,\msun\ and varies monotonically
  with pre-SN $R_\star$. The envelope density spans a factor of 10 and is higher in more massive
  (or more extended ) progenitors. The same trend applies to the surface H mass fraction,
  which falls in the range $0.062-0.38$. The He core mass increases monotonically with final
  mass and is in the range $4.37-4.95$\,\msun.

  Because one distinguishing property is the final radius, we call these H-rich models
  H\_R61, H\_R152, H\_R228, H\_R385, and H\_R601.
  We also include the model H\_R5, which results from a tighter initial orbital separation
  of 50\,\rsun\ and dies as a compact H-deficient star without a tenuous envelope.
  The model properties are documented in Fig.~\ref{fig_summary_prog} and Fig.~\ref{fig_mesa_comp_h_all},
  as well as in Table~\ref{tab_init}.

\subsubsection{He giant model}
\label{sect_he_giant}

We also consider an H-deficient model in this study in order to gauge the possible influence
on the early-time properties of Type Ib SNe.
Recently, several authors (\citealt{eldridge_ib_15}; \citealt{kim_snibc_prog};
\citealt{clelland_he_16}; \citealt{yoon_ibc_17}; \citealt{hirai_ib_17})
have found that the final stellar properties of some binary low-mass He giants
are compatible with those inferred for the progenitor of Type Ib SN iPTF13bvn (see also
\citealt{bersten_iPTF13bvn_14}).
Here, we expand these studies to discuss the multi-band light curve
and spectral properties of such He-giant star explosions.

We study a He-giant star model that was produced by
the evolution of a pure He star of 2.8~$\mathrm{M_\odot}$ on the zero-age He main sequence
(the initial mass on the zero-age H main sequence should be around 12\,\msun).
The computation was performed with the \bec\ stellar
evolution code.  The adopted mass loss rate is the one given by \citet{yoon_ibc_10}
with a reduction by factor of 5. This mass loss rate prescription is originally
based on \citet{hamann+95} for luminous Wolf-Rayet (WR) stars
($\log L/\mathrm{L_\odot} \ge 4.5$) and \citet{hamann+82}
for less luminous helium stars. The simulations used the OPAL opacity tables of \citet{opal_96}.
Unlike \mesa, the \bec\ code does not go all the way to iron core formation and collapse.
Here, the \bec\ simulation was stopped at the onset of core neon burning, which is merely a few
years prior to core collapse (this model would most likely form an iron core).
Hence the global properties of the star, and in particular its envelope
structure are settled --- only the inner core will evolve in the remaining years of evolution.
At the onset of core neon burning, this H-deficient star has a mass of 2.73\,\msun\ and a surface radius of
173\,\rsun, and helium represents 35\% of its total mass (see  Fig.~\ref{fig_summary_prog}
and Table~\ref{tab_init}).

A He star with such a low mass can only be produced in a close binary system.
By the time of explosion, the He
star would usually have a companion which may be a main sequence star or a
compact object.  The size of the He star would therefore be limited by the
Roche radius, and it would be difficult to evolve into a He giant with
a surface radius of 173\,\rsun\ \citep{yoon_ibc_10}.  However, there exists the
possibility that such a He star can be unbound from the binary system  before
it explodes, if the order in which the two stars explode is reversed
\citep{pols_94}. First, an He star would be produced by mass transfer from the
primary star. Second, the secondary star that gained mass may evolve faster
than the primary star and explode first. The terminal explosion of the secondary star
and the resulting neutron star kick may disrupt the binary system, leaving the
He star isolated. The probability of this SN reversal must be small but
it is not negligible.

In this He-giant star model, the large surface radius is a result of
the very compact (and low mass)
core --- this is the H-deficient counterpart of single stars of $\sim$\,10\,\msun\
stars that die as electron-capture SNe.
Envelope inflation is, however, predicted in higher mass stars like (H-deficient)
WR stars \citep{ishii_99,petrovic+06,grafener+12,clelland_he_16}.
In high mass stars, the concomitance of high luminosity and large envelope opacity places the star close to or above
the Eddington limit at some depths within the envelope. Consequently, to preserve hydrostatic equilibrium,
the stellar envelope may inflate to lower the opacity, producing surface radii of a few \rsun\ \citep{yoon_ibc_15},
and perhaps up to $\sim$\,10\,\rsun\ (see, e.g., \citealt{groh_ib_prog_13}).
These radii are much smaller than for the low-mass He-giant model discussed here.

\subsection{Simulation of the explosion}
\label{sect_v1d}

   For the H-rich models, the \mesa\ simulations were stopped when the maximum infall velocity (somewhere
   within the iron core) reached 1000\,\kms.
   At that time, we remapped the \mesa\ model into \v1d\ \citep{livne_93,DLW10a,DLW10b}.
   The \v1d\ simulations use $1000-1500$ grid points. Near the stellar surface, we use a mass resolution
   of  $10^{-6}-10^{-5}$\,\msun\ to resolve the low-density outer layers --- this is crucial
   in progenitors with extended low-density outer regions.
   Although such massive star progenitors may possess a sizeable wind mass loss rate at the time
   of explosion, we consider only explosions within a vacuum. Interaction with a wind would
   enhance the early time luminosity, produce bluer colors, and might inhibit the presence of
   ejecta material at large velocities. This limitation is to be born in mind when confronting models
   to observations, for example, of SN\,1993J (Section~\ref{sect_comp_obs}).

    The explosion is in all cases triggered by moving a piston at $\sim$\,15000\,\kms\ at the inner boundary,
    which we place at the location where the entropy rises outward from the center to
    4\,k$_{\rm B}$\,baryon$^{-1}$. This location is typically in the outer
    part of the Si-rich shell, just below the O-rich shell, and located around $1.5-1.7$\,\msun\ in these models.

     In these H-rich ejecta (whose progenitor evolution was continued until core collapse), explosive
     nucleosynthesis takes place leading to the production of \iso{56}Ni in the innermost, densest, layers.
     At 1000\,s after the explosion was triggered, we artificially mix the \iso{56}Ni outwards.
     Starting at the base, we step through each ejecta mass shell $m_i$ and mix all mass shells
     within the range [$m_i$, $m_i+\delta m$] with $\delta m=0.3$\,\msun. To preserve
     the normalisation of the sum of all mass fractions to unity in each mass shell, we adjust the H mass fraction.
     In addition, to smear the composition profiles, we repeat the process for all species
     using $\delta m=0.05$\,\msun.
     The reason we use this approach is to preserve the composition of the outer ejecta layers where
     only a small amount of H may remain. In practice, this adopted \iso{56}Ni mixing is quite weak, producing
     a delayed rise of the light curve in compact progenitor models, and lines of moderate width
     at maximum (see below, and also \citealt{d12_snibc}). With the adopted weak mixing, we can
     better gauge the influence of the shock-deposited energy in extended progenitor stars.

   For the He-giant star model, the \bec\ simulation was stopped at the onset of core neon burning. Further
   evolution would impact primarily the core layers. Here, we focus on the early time ejecta and radiation
   evolution (basically the first week after shock breakout) so the exact properties of the inner ejecta
   layers are irrelevant. In \v1d, we set the piston at a Lagrangian mass of 1.4\,\msun\ and produce
   an ejecta with 10$^{51}$\,erg. Because the shock only crosses low-density layers, there is no explosive
   burning and no \iso{56}Ni is produced. We perform no mixing for this model. To test the influence of the
   extended envelope on the SN radiation, we simulate a variant of this model but with the progenitor
   envelope truncated at a radius of 11\,\rsun\ (corresponding to a density of $\sim$\,10$^{-5}$\,g\,cm$^{-3}$).
   The original model is called He\_R173 and its truncated version is called He\_R11.

     A summary of the ejecta properties of our model set is given in Table~\ref{tab_init}.

\begin{figure}
\includegraphics[width=0.48\textwidth]{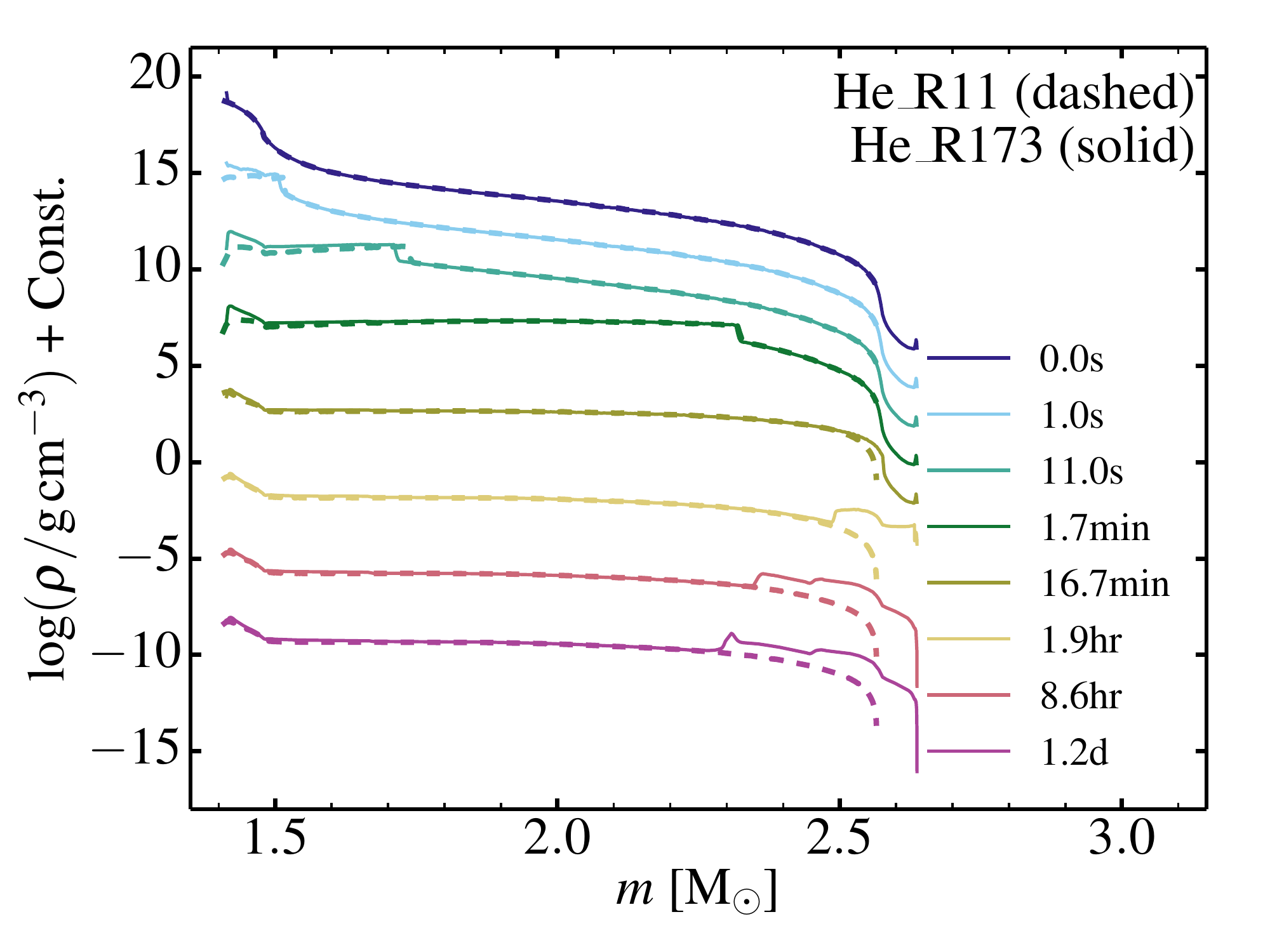}
\includegraphics[width=0.48\textwidth]{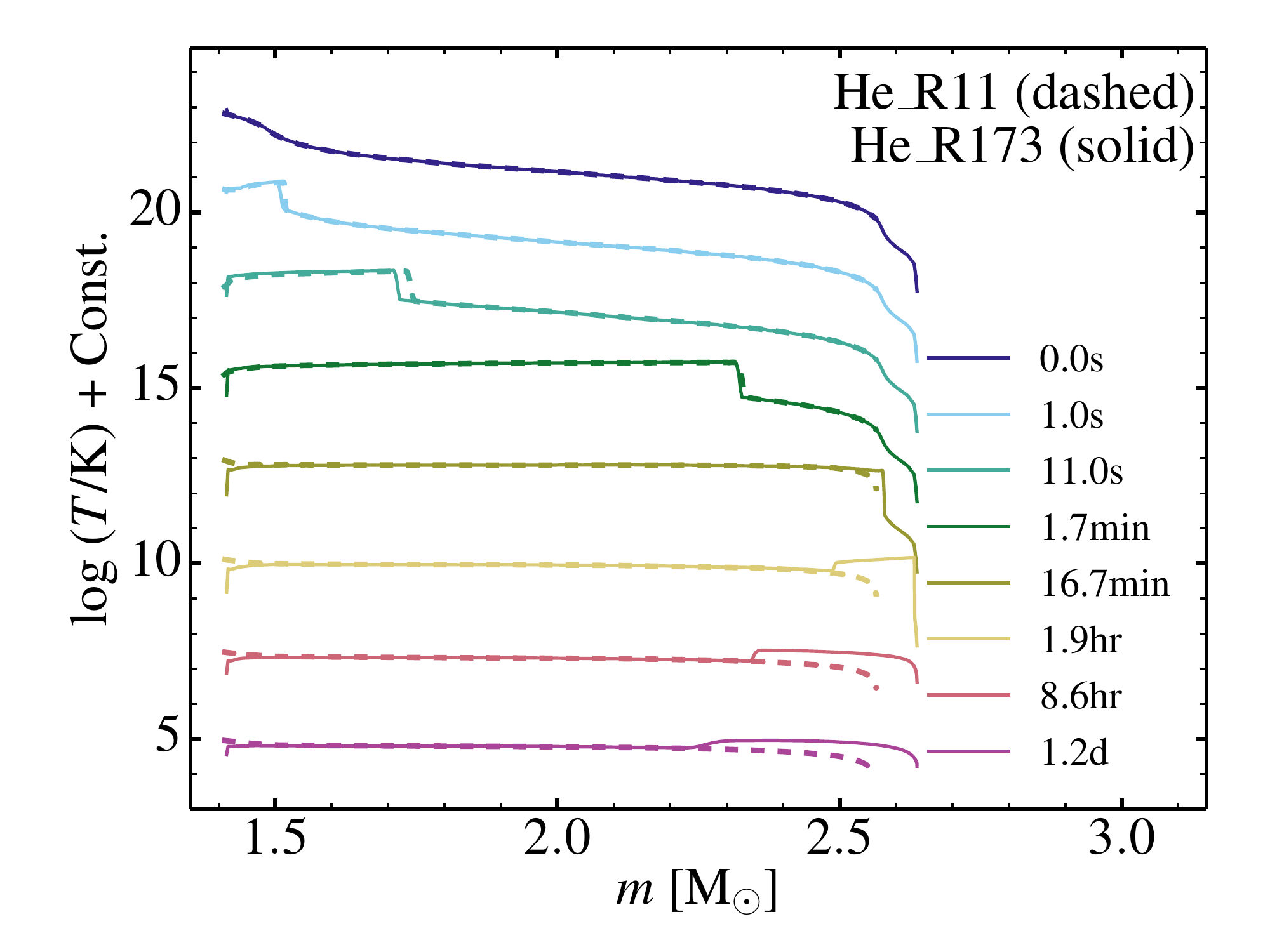}
\includegraphics[width=0.48\textwidth]{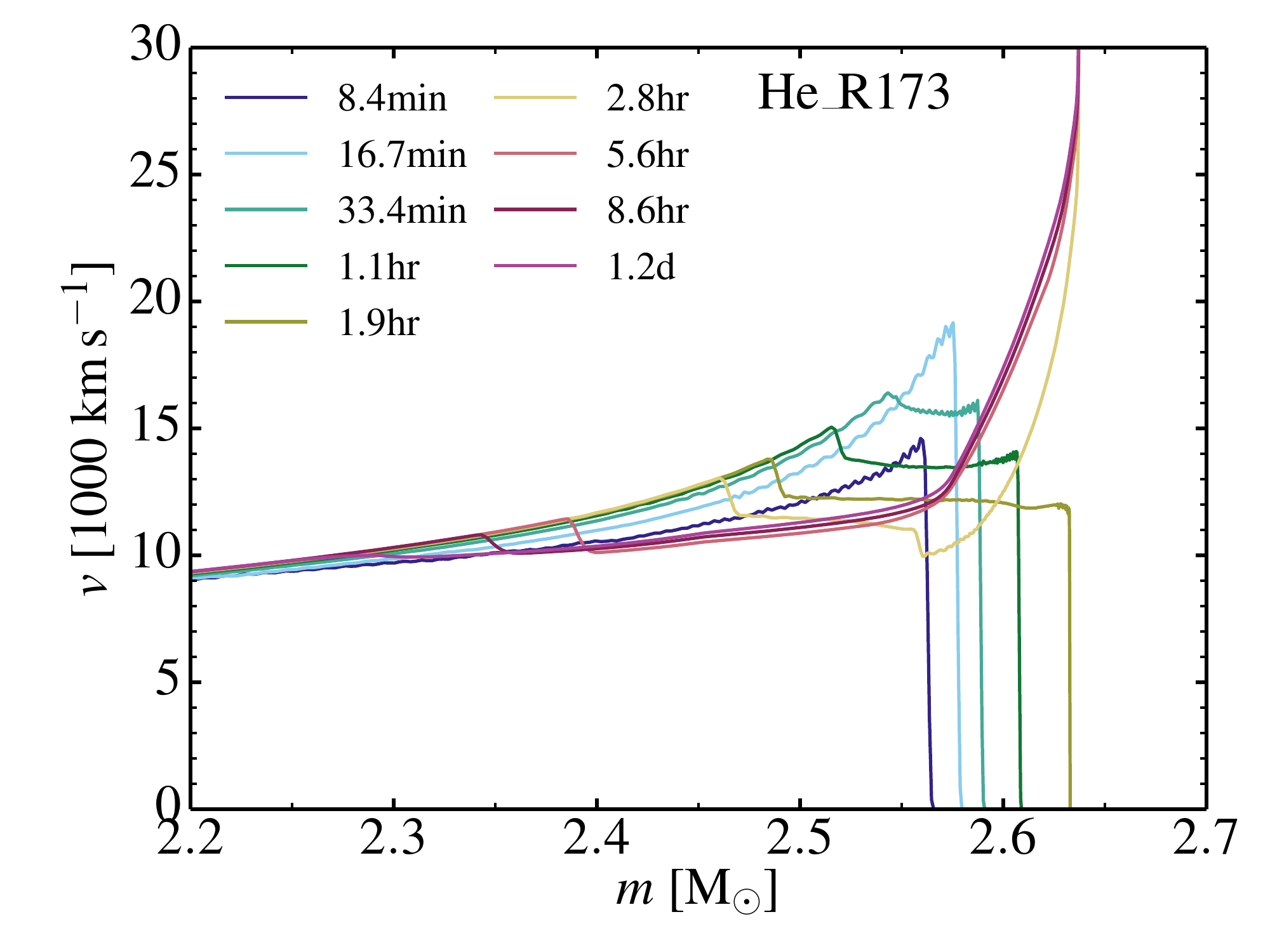}
\caption{Multi-epoch snapshots of the density (top), temperature (middle),
and velocity versus Lagrangian mass for model He\_R173 (solid).
For the top two panels, we include the results for model He\_R11 (dashed).
The bottom panel shows different epochs and only the outer envelope.
The time origin corresponds to the piston trigger (the piston mass cut is located at 1.4\,\msun).
Notice the formation of a reverse shock after $\sim$\,20\,min in the model with the extended
low-density envelope. This reverse shock leads to the formation of a dense shell of 0.2\,\msun\
at $m=$\,2.4\,\msun. For better visibility, all profiles in the top two panels are shifted vertically by a constant.
\label{fig_v1d_seq_he}
}
\end{figure}

\subsection{Radiative transfer simulations}
\label{sect_cmfgen}

      When the \v1d\ simulations have evolved to $\sim$\,1\,d,  we remap them into
   the non-local-thermodynamic-equilibrium (nLTE) time-dependent radiative-transfer code \cmfgen\
   \citep{HM98,DH05a,DH08, HD12,dessart_etal_13}. At this time, the ejecta are in homologous
   expansion. We proceed for these SNe IIb/Ib models as described in \citet{D15_SNIbc_I,D16_SNIbc_II}.
   The notable distinction is that we are only interested in the impact of the progenitor outer envelope
   properties on the SN radiation, which is most visible early after shock breakout.

   For the H-rich models, which synthesize some \iso{56}Ni during the explosion, we discuss
   the evolution up to bolometric
   maximum, which occurs at about 20-30\,d. We had difficulties converging model H\_R601 when
   starting at 1\,d after shock breakout but could proceed when starting
   at 3\,d.\footnote{The problem is with the hot low-density outer ejecta layers
   where the dominance of scattering caused convergence difficulties. This affects only the model
   with the most extended tenuous envelope.} Hence, we present
   results for model H\_R601 starting at 3\,d and for all other models, our \cmfgen\ results start at 1\,d
   after shock breakout.

    To expedite the simulations of the He-giant star model, we trim the {\it inner} ejecta of models He\_R173 and
    He\_R11 to exclude the inner regions below 5000\,\kms. This is justified since there is no \iso{56}Ni in the model,
    the inner regions do not have the right properties since the progenitor star was not evolved until core collapse,
    and finally we focus in that model on the early times when the SN radiation emerges from the outer layers
    (and are not influenced by the properties of the deeper regions). Consequently, these models
    become optically thin within a week of shock breakout and we show their evolution only during that first week.

    Note that similar simulations for smaller progenitor stars ($R_\star \lesssim 10$\,\rsun),
    based on models by \citet{yoon_ibc_10}, have been presented in \citet{dessart_11_wr,D15_SNIbc_I,D16_SNIbc_II}.


\begin{figure*}
\includegraphics[width=\textwidth]{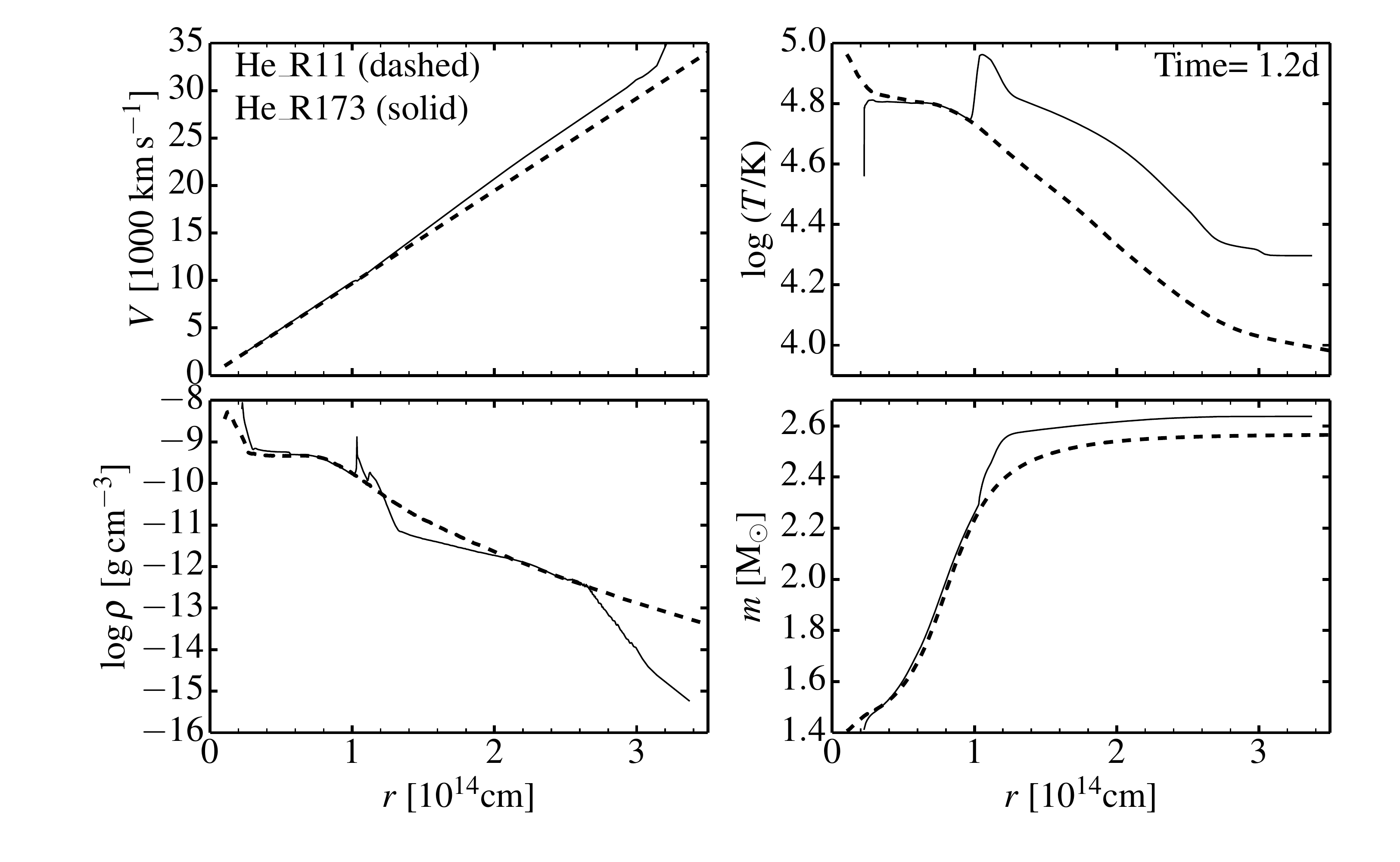}
\caption{Comparison of the velocity, temperature, density, and Lagrangian mass versus radius
for models He\_R173 (solid) and He\_R11 (dashed) at 1.2\,d after the piston trigger.
[See text for discussion.]
\label{fig_v1d_he_1p2d}
}
\end{figure*}

\section{Shock propagation in stars with an extended envelope}
\label{sect_dyn}

\subsection{Preamble}

  The explosion of massive stars  with extended envelopes has been discussed
  in the context of Type IIb SNe.
  \citet{nomoto_93j_93}, \citet{podsiadlowski_93j_93}, \citet{woosley_94_93j},
and \citet{blinnikov_94_93j} presented  radiation-hydrodynamics and/or
radiative-transfer simulations of terminal explosions of supergiant progenitor
models largely stripped of their H-rich envelope.
  The overall consensus for SN\,1993J is on a progenitor star with a 4\,\msun\ He
core surrounded by a $\sim$0.2\,\msun\ H-rich envelope extending out to $\sim$\,500\,\rsun.
More recently, similar supergiant stars with a low total mass of just a few solar masses
have been invoked to explain the  light curve of SN\,2011dh
  \citep{bersten_etal_12_11dh}. Similar results were obtained by \citet{ergon_14_11dh}
using additional constraints from spectra. The impact of an extended envelope on the SN
radiation has been further discussed by \citet{nakar_piro_14}.

  There is thus a consensus concerning the explosions of massive stars with extended
low-density envelopes and their SN light curve. However, little has been shown about
the shock propagation through the envelope and in particular the impact on the temperature
(or internal energy) of the shocked envelope. Secondly, much
  attention has been paid to H-rich progenitors but it is also possible to form He-giant stars
  (\citealt{clelland_he_16}; Section~\ref{sect_he_giant})
  --- the results highlighted for some Type IIb should also apply to some Type Ib SNe.
 In Section~\ref{sect_rad}, we discuss the impact on spectra, which has not been done before.
  The spectral energy distribution, the morphology of line profiles, and the ions influencing
  the spectral formation, all bear an imprint of the shock deposited energy and of the dense shell
  formed by the reverse shock.

\begin{figure*}
\includegraphics[width=0.495\hsize]{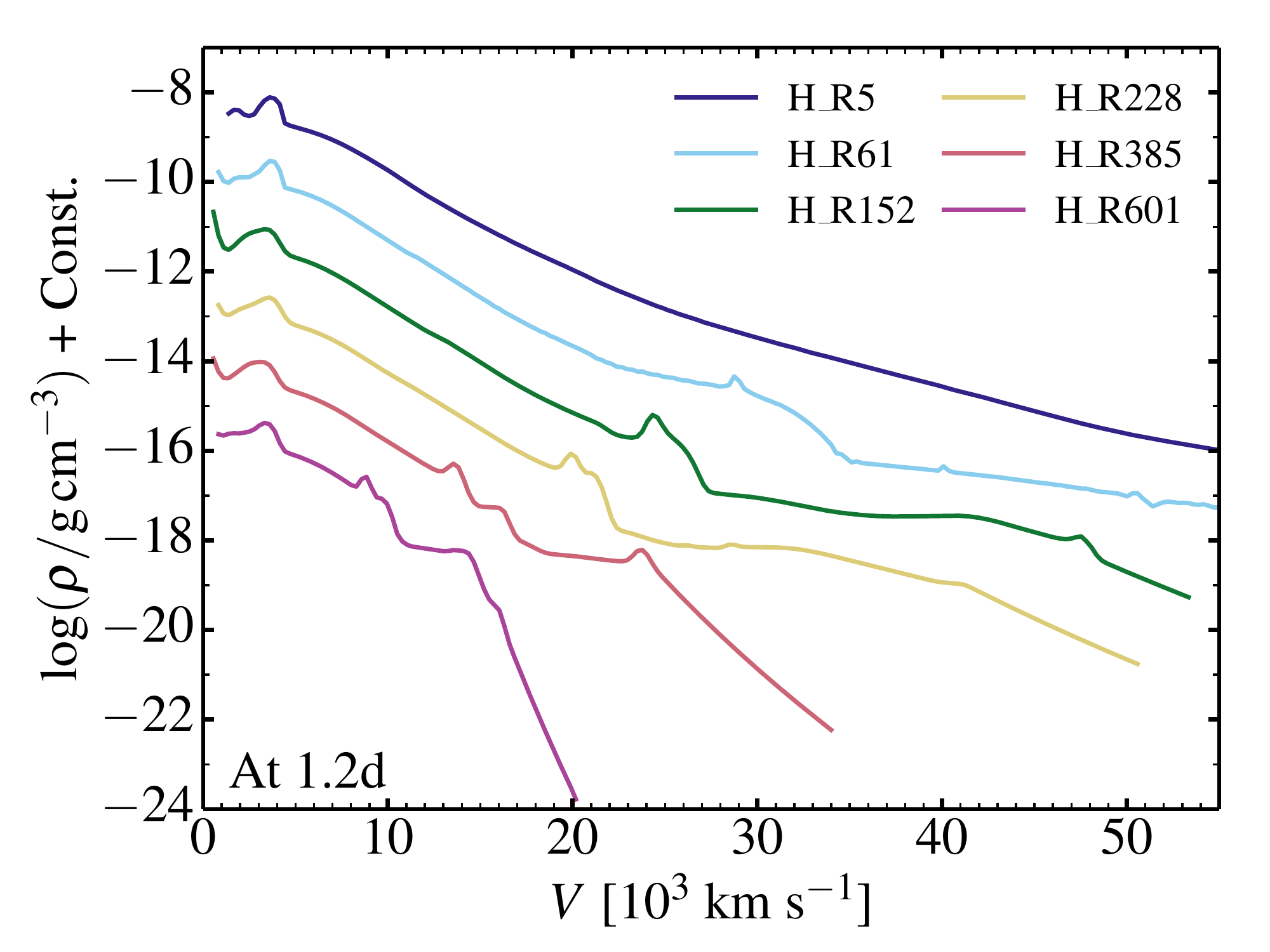}
\includegraphics[width=0.495\hsize]{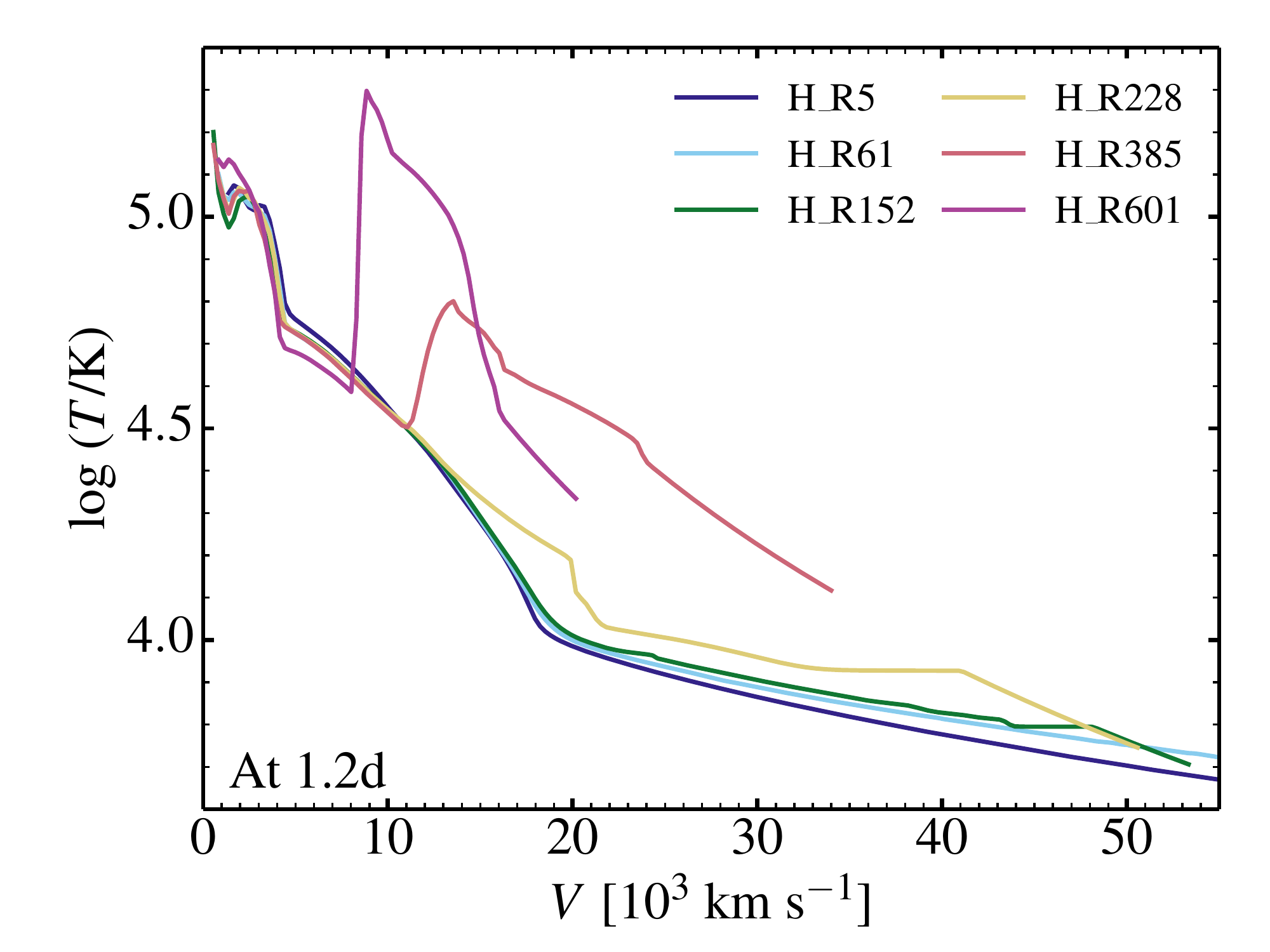}
\includegraphics[width=0.5\hsize]{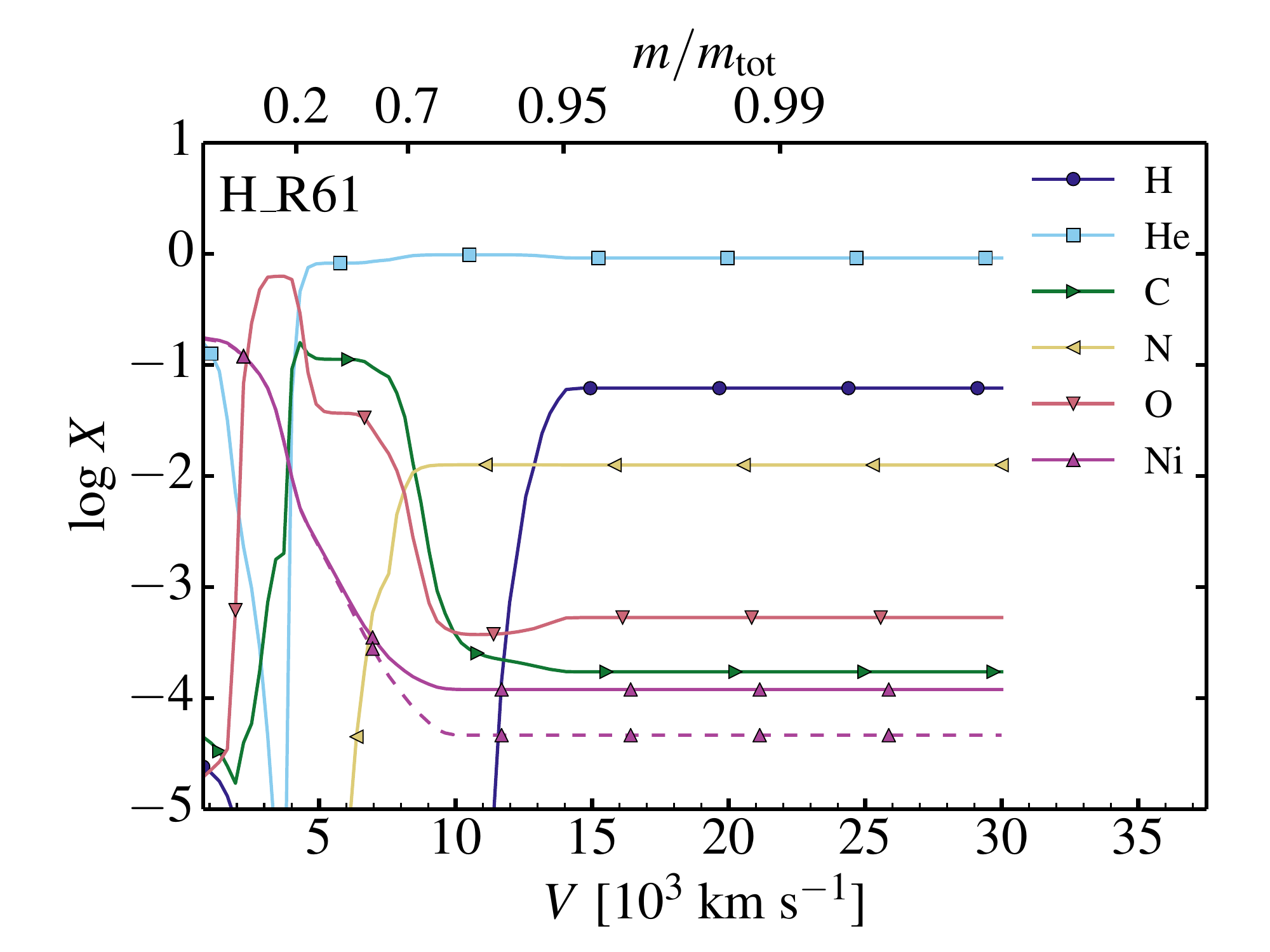}
\includegraphics[width=0.5\hsize]{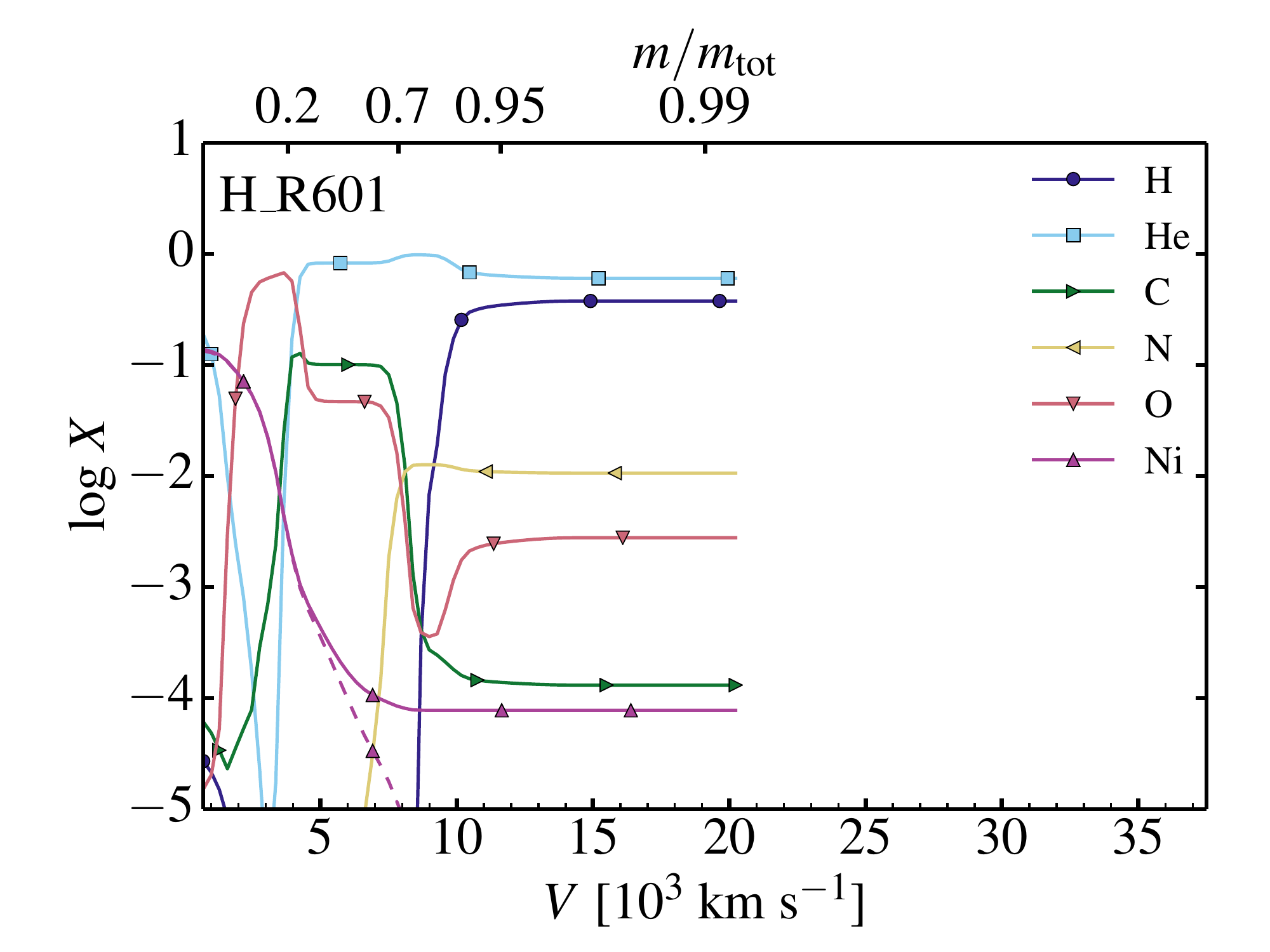}
\caption{Top row: Comparison of the density and temperature profiles at 1.2\,d for the H-rich models.
The more massive and extended is the low-mass H-rich envelope, the greater is the outer ejecta temperature.
Bottom row: Composition profiles for models H\_R61 and H\_R601 at 1.2\.d (see also
Fig.~\ref{fig_v1d_comp_h_all}). The dashed line corresponds to the \iso{56}Ni.
This time corresponds to the time of remapping into \cmfgen.
\label{fig_v1d_1p2d}
}
\end{figure*}

\subsection{Results}

   We first start with the results for the He-giant star model.
   Figure~\ref{fig_v1d_seq_he} shows multi-epoch snapshots of the density (top), temperature (middle),
   and velocity (bottom) versus mass as the shock propagates through the star. For up to about 20\,min,
   the shock propagates within the He core, and steepens as it reaches the steeply declining density
   at its outer edge, around 2.5\,\msun\ (Fig.~\ref{fig_summary_prog}).
   Subsequently, as the shock enters the extended, flat-density He-rich envelope, it decelerates.
   The pressure build-up behind the shock leads to the formation of a reverse shock, which also
   strengthens the forward shock. We thus see the presence of
   both a decelerating forward shock moving outward (in mass space) into the extended envelope and
   a reverse shock moving inwards (in mass space)  into the He-core material.
   In model He\_R173, the forward shock breaks (i.e., emerges) out of the star $\sim$\,2\,hr after the piston
   trigger. At that time, the reverse shock is still present deep inside the progenitor He core, although it is
   by then very weak (see velocity panel).

   The contrast in envelope properties is striking when we compare the density/temperature profiles for model
   He\_R173 (solid line) with the results for the truncated variant (model He\_R11, dashed line).
   This is illustrated at multiple epochs during shock propagation through the envelope (Fig.~\ref{fig_v1d_seq_he}).
   Figure~\ref{fig_v1d_he_1p2d} shows this contrast well after breakout in both models, specifically at 1.2\,d,
   and this time in radial space.
   The model He\_R173 from the extended progenitor is not yet in homologous expansion, primarily because
   of the surviving reverse shock. This reverse shock has led to the formation of a slow dense shell (at
   around 10000\,\kms, and with little mass beyond it), while model He\_R11 shows a smooth density profile
   with more mass at very large velocities (beyond 15000\,\kms).
   The outer ejecta temperature in model He\_R173 is typically $0.2-0.3$ dex greater than in model He\_R11.
   The corresponding energy excess in those outer layers, which suffers little degradation from expansion cooling,
   will have a strong impact on the SN light curve at early times.

   The dynamical properties for the He-giant star explosions also apply to H-rich progenitors
   with an extended envelope. Figure~\ref{fig_v1d_1p2d} shows the density and temperature
   profiles at 1.2\,d after the piston trigger in models H\_R5, H\_R61, H\_R152, H\_R228, H\_R385,
   and H\_R601.
   Progenitors with a greater H-rich envelope mass and extent produce ejecta at 1\,d that have
   a larger temperature (or trapped radiative energy), as well as a more massive and slower dense shell.
   In model H\_R601 at 1.2\,d, the outer ejecta temperature is even greater than in the inner ejecta.
   In that model, the reverse shock went down to
   a velocity of 8000\,\kms, and has therefore decelerated a sizeable fraction of the He rich layers.
   Figure~\ref{fig_v1d_1p2d} shows this and other features for model H\_R601, together with the
   H-rich model with the lowest amount of hydrogen and the smaller progenitor radius
   (model H\_R61 --- model H\_R5 is more compact but it has no residual hydrogen).

    Just like in RSG star explosions leading to Type II-Plateau SNe, the shock deposited energy
    within the envelope is weakly degraded by expansion cooling. The difference with RSGs and SNe II-Plateau
    is the much lower optical depth of these tenuous envelopes, whose mass does not exceed 0.1\,\msun\
    in our model set. By expanding, these layers will radiate their energy on a shorter timescale,
    boosting the SN luminosity for a short while.
    In addition, the reverse shock enhances the internal energy deposited in the He-core layers.
    These layers are located at a greater optical depth. They are also rich in \isoni, and therefore
    influenced by decay heating, so the exact impact of the reverse shock on the light curve
    is harder to discern.

    In a different context, these notions are also relevant to pulsational-delayed detonation models of SNe Ia.
    The progenitor stars are then white dwarfs, which are much more compact, but the occurrence
    of a deflagration ``pulse" followed by a detonation can give rise to an interaction between the
    detonating inner layers and the outer marginally unbound layers \citep{khokhlov_91b}. The interaction
    converts kinetic energy into internal energy, with visible consequences up to bolometric maximum
    \citep{d14_pddel}.

\section{Impact on supernova radiation}
\label{sect_rad}

\subsection{Photometry}

\begin{figure*}
\begin{center}
 \includegraphics[width=0.45\hsize]{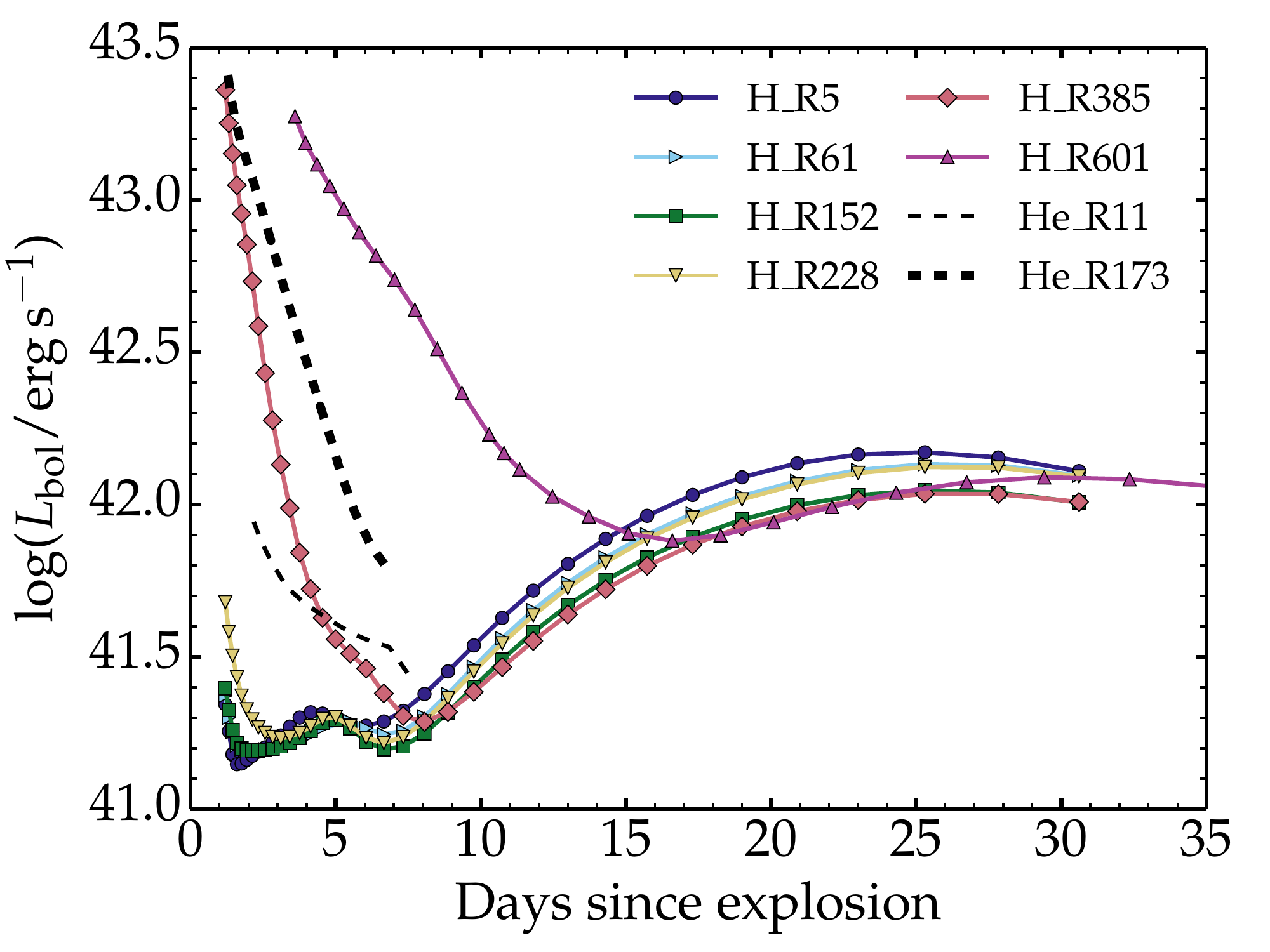}
 \includegraphics[width=0.45\hsize]{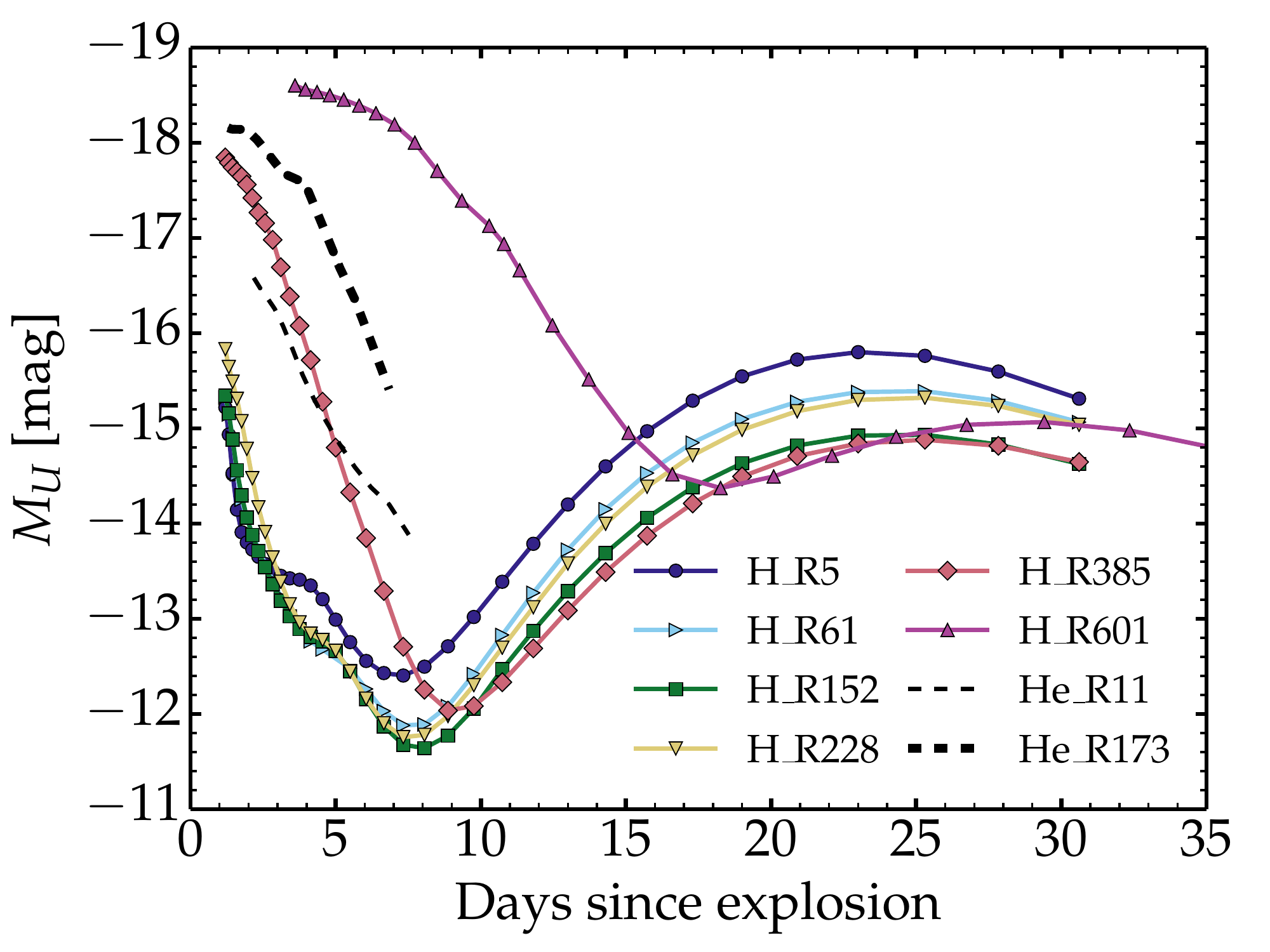}
 \includegraphics[width=0.45\hsize]{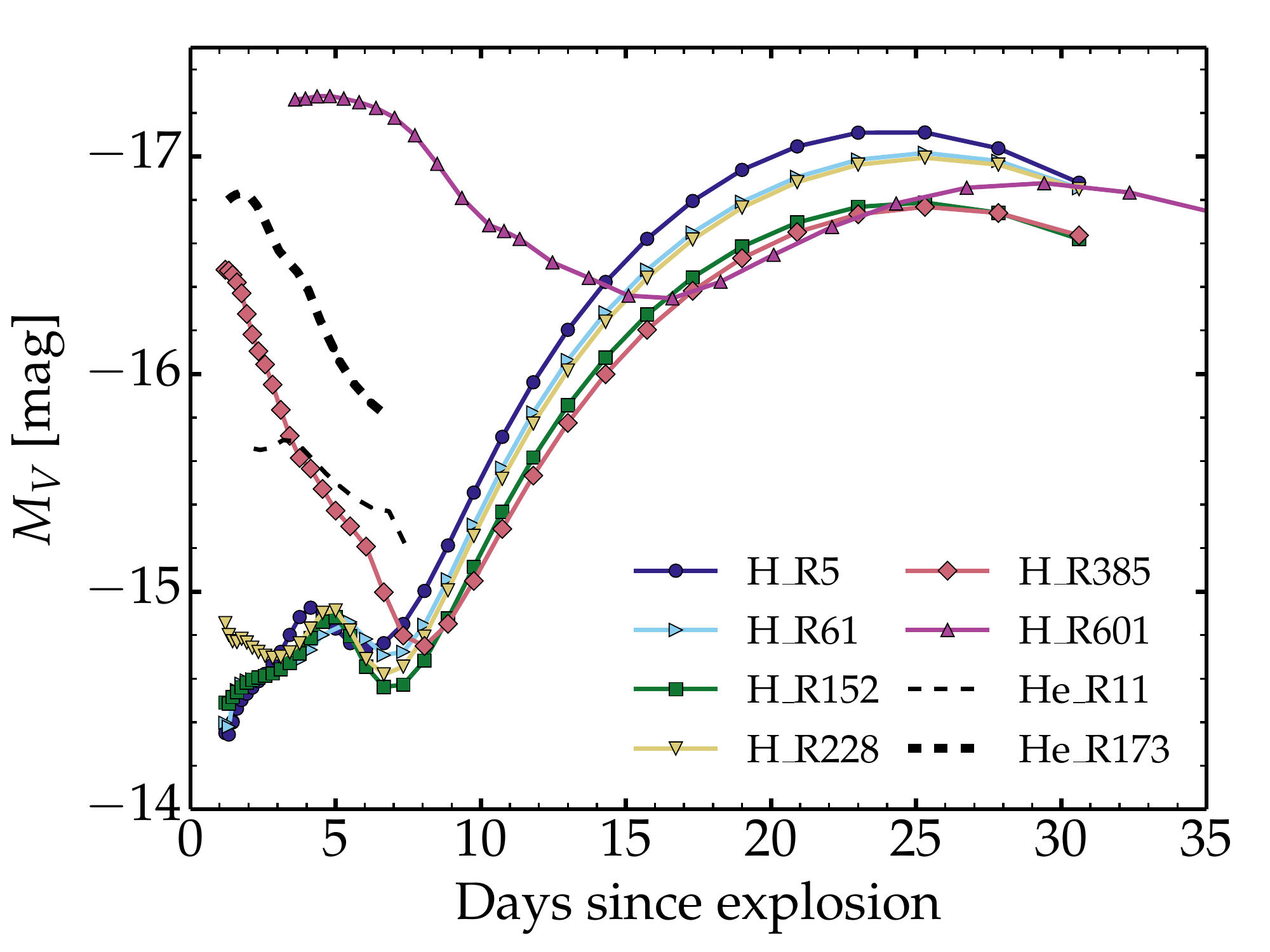}
 \includegraphics[width=0.45\hsize]{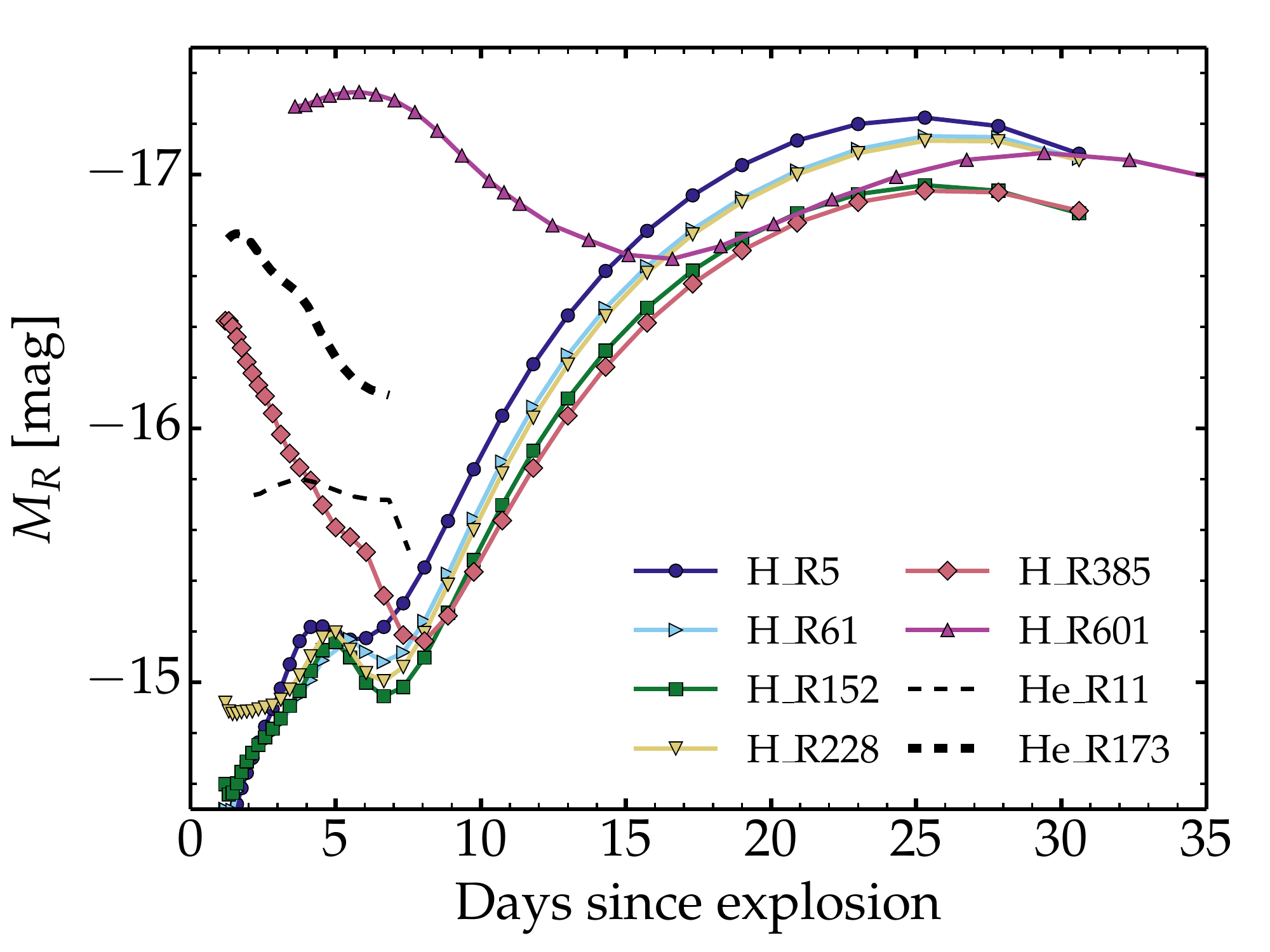}
 \includegraphics[width=0.45\hsize]{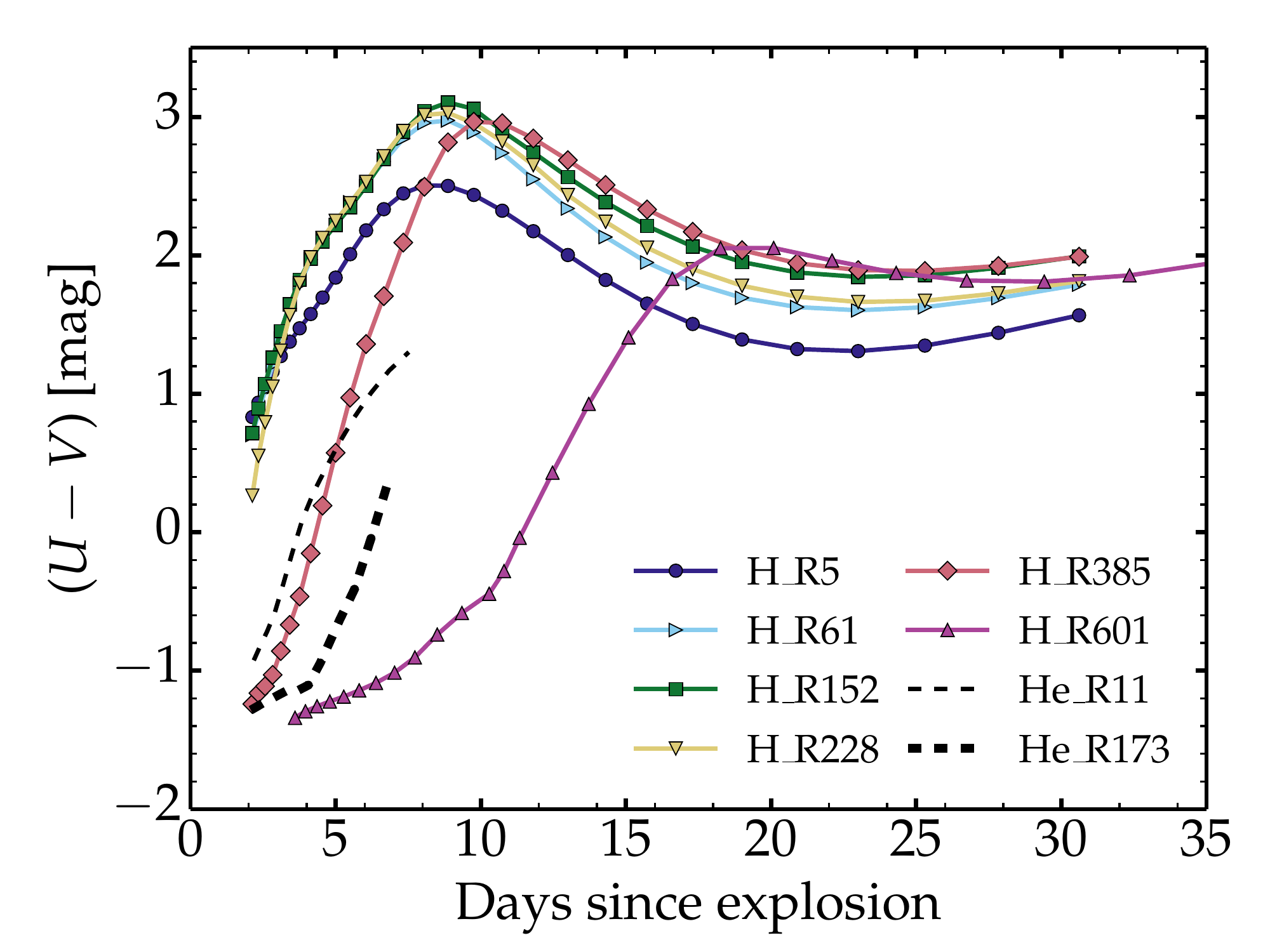}
 \includegraphics[width=0.45\hsize]{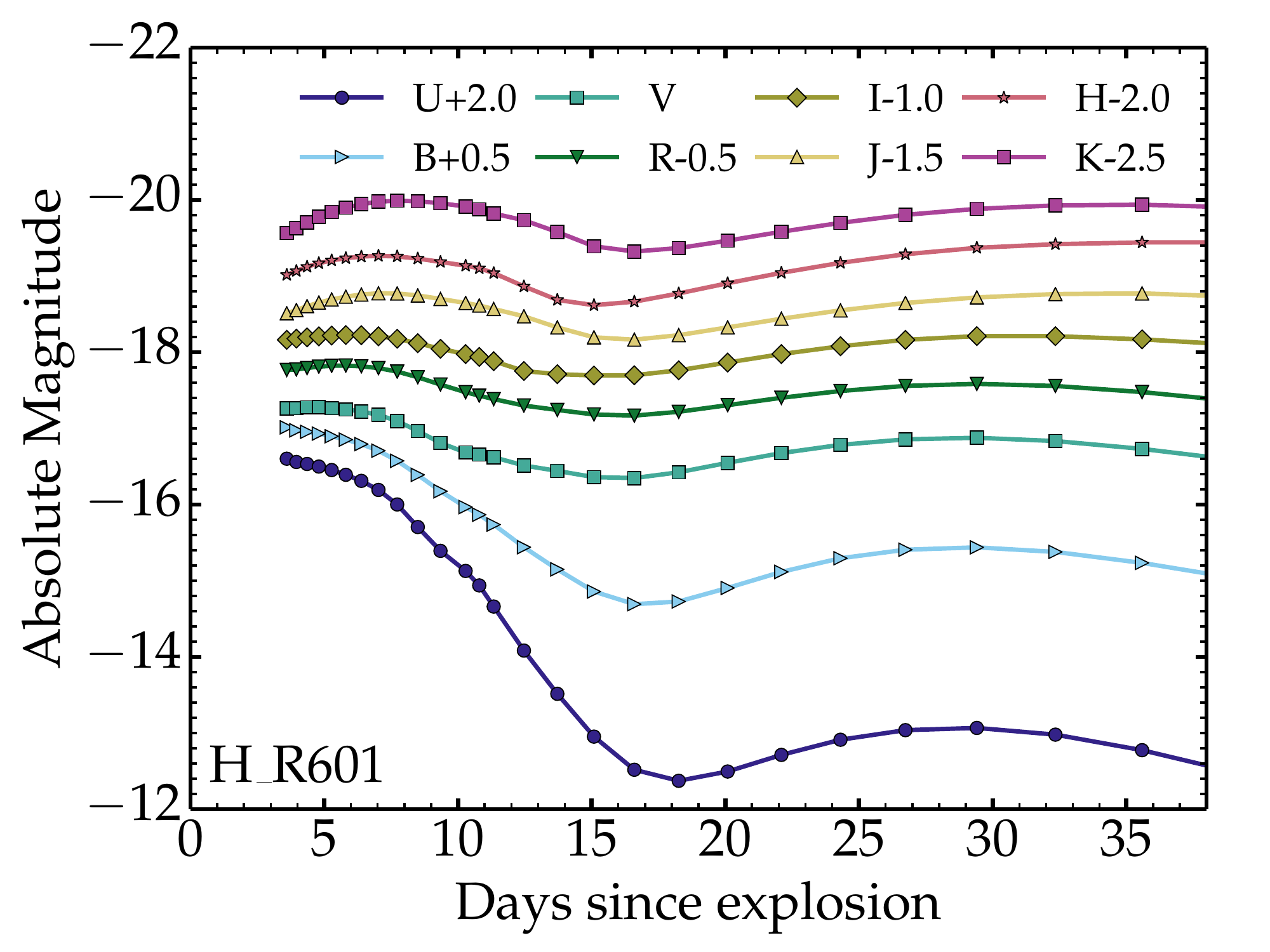}
\caption{Evolution of the bolometric luminosity (top left),
the absolute magnitude in the $U$ (top right), $V$ (middle left), $R$ (middle right) bands,
and the ($U-V$) color for our set of models.
We also add the $UBVRIJHK$ light curves for model H\_R601 (bottom right).
\label{fig_h_phot}
}
\end{center}
\end{figure*}

Figure~\ref{fig_h_phot} shows the bolometric light curve for our set of H-rich and H-deficient
models computed with \cmfgen.
The rise time of 20-30\,d to a peak luminosity of $\sim$\,10$^{42}$\,\ergs\ is comparable for all models
because the ejecta mass, the ejecta kinetic energy, and the \iso{56}Ni mass are similar (Table~\ref{tab_init}).

In our set, progenitors with surface radii less than 300\,\rsun\ have comparable light curves
beyond a day after shock breakout. Indeed, the \v1d\ bolometric light curves (not shown), which are calculated
at all times through shock breakout, show variations between these models only up to a day, with the
luminosity being greater for more extended progenitors.
This impact is limited to times less than a day because the extended H-rich envelope
is small, both in mass and extent.

For model H\_R385, the luminosity boost is clearly visible at 1\,d and persists for about a week. In
model H\_R601, the luminosity is $\gtrsim$\,10$^{42}$\,\ergs\ for about two weeks after shock breakout,
and is in excess of the subsequent \iso{56}Ni powered peak that occurs at $\sim$30\,d.
The amount of stored radiation energy
in the ejecta layers  between 8000 and 16000\,\kms\ at 1.2\,d after explosion is 5.8$\times$\,10$^{48}$\,erg.
This value is about one hundred times greater than for model H\_R61.
Such an energy radiated over a time scale of 20\,d corresponds to a mean luminosity of
$\sim$\,3.3$\times$\,10$^{42}$\,\ergs\ or $\lesssim$10$^9$\,\lsun, which corresponds roughly to the
value seen in the light curve.

A luminosity boost is also obtained for the explosion of the He-giant star.
Model He\_R173 shows a luminosity $\gtrsim$\,10$^{42}$\,\ergs\ for about a week. This value is
intermediate between that of models H\_R385 and H\_R601. In all three cases, the luminosity boost stems from
the excess internal energy available in the outer ejecta layers at 1\,d, which itself scales with the extent and mass
of the tenuous progenitor envelope (Figs~\ref{fig_v1d_he_1p2d}--\ref{fig_v1d_1p2d}).

 The multi-band light curves of model H\_R601 show a large brightness in all bands at early times (top-right panel
 of Fig.~\ref{fig_h_phot}).
 From $U$ to $K$, the rise time to the first maximum increases (the blue filters have their maximum
 at the first epoch computed, at 3.6\,d, and redder filters peak later, with the latest being 8\,d for the $K$ band).
 The SN is initially very blue and reddens
 with time as the ejecta/photosphere cools and expands after breakout. It fades in the blue (for example, the
 $U$-band magnitude fades by 4\,mag during the first two weeks) but its magnitude
 is nearly constant in the near-IR (for example, the $K$-band magnitude is between $-19.5$ and $-20.0$\,mag
 for the first 30 days).
 After about two weeks, the energy/power from \iso{56}Ni decay causes the rebrightening of the SN in all bands,
 but with a negligible associated change in color.

 All optical and near-IR light curves have therefore two peaks during the first 30 days.
 The first peak is a consequence of cooling and expansion after shock breakout. The second peak is caused
 by reheating from \iso{56}Ni decay and expansion. The two peaks are more obvious in redder filters
and for progenitors with more extended/massive envelopes.
  This double-peak morphology is a natural consequence of the core-halo structure of the progenitor
 star and has been clearly observed in SN\,1993J
 \citep{nomoto_93j_93,podsiadlowski_93j_93,woosley_94_93j,nakar_piro_14}.

 The middle and bottom rows of Fig.~\ref{fig_h_phot} compare the photometric properties of all models,
 illustrating further what has been presented above. Relatively compact progenitors follow a similar photometric
 path. In contrast, explosions of extended stars stand apart and show a strong sensitivity to the initial conditions.
 This is seen in brightness, but perhaps even more vividly in terms of color. Model H\_R601 stays 2-3\,mag
 bluer than more compact progenitor models.

\begin{figure*}
\begin{center}
 \includegraphics[width=0.45\hsize]{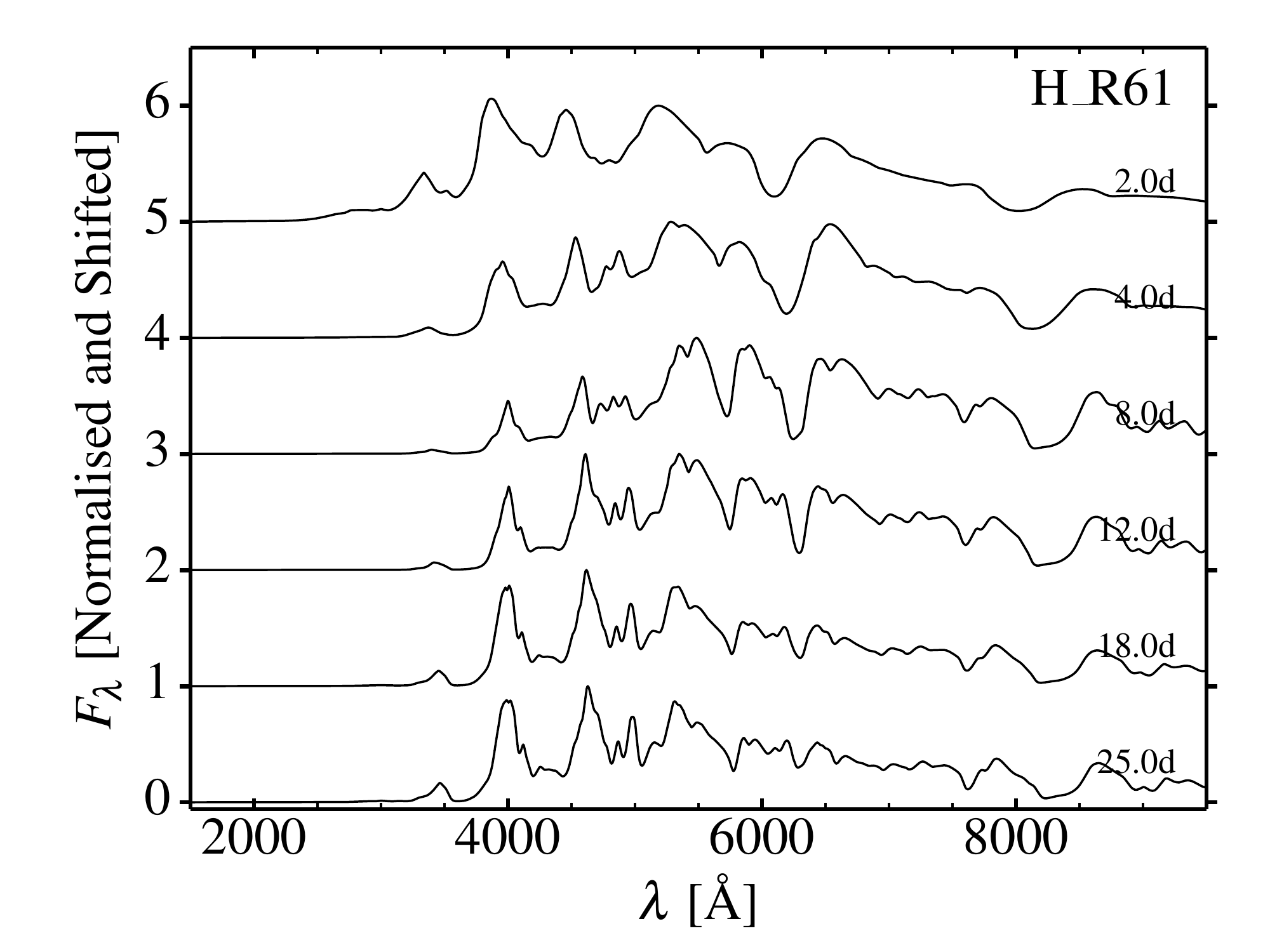}
 \includegraphics[width=0.45\hsize]{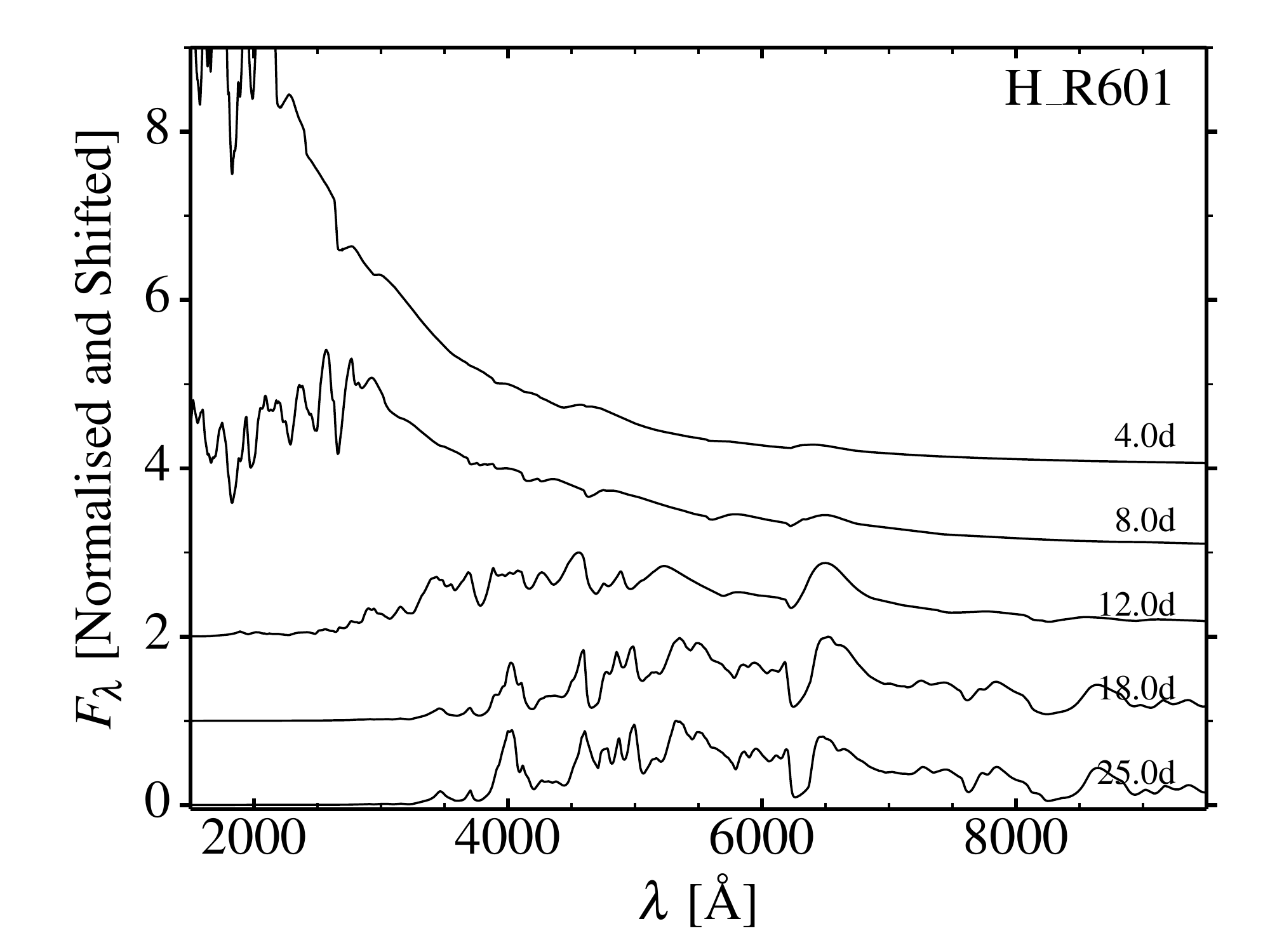}
 \includegraphics[width=0.45\hsize]{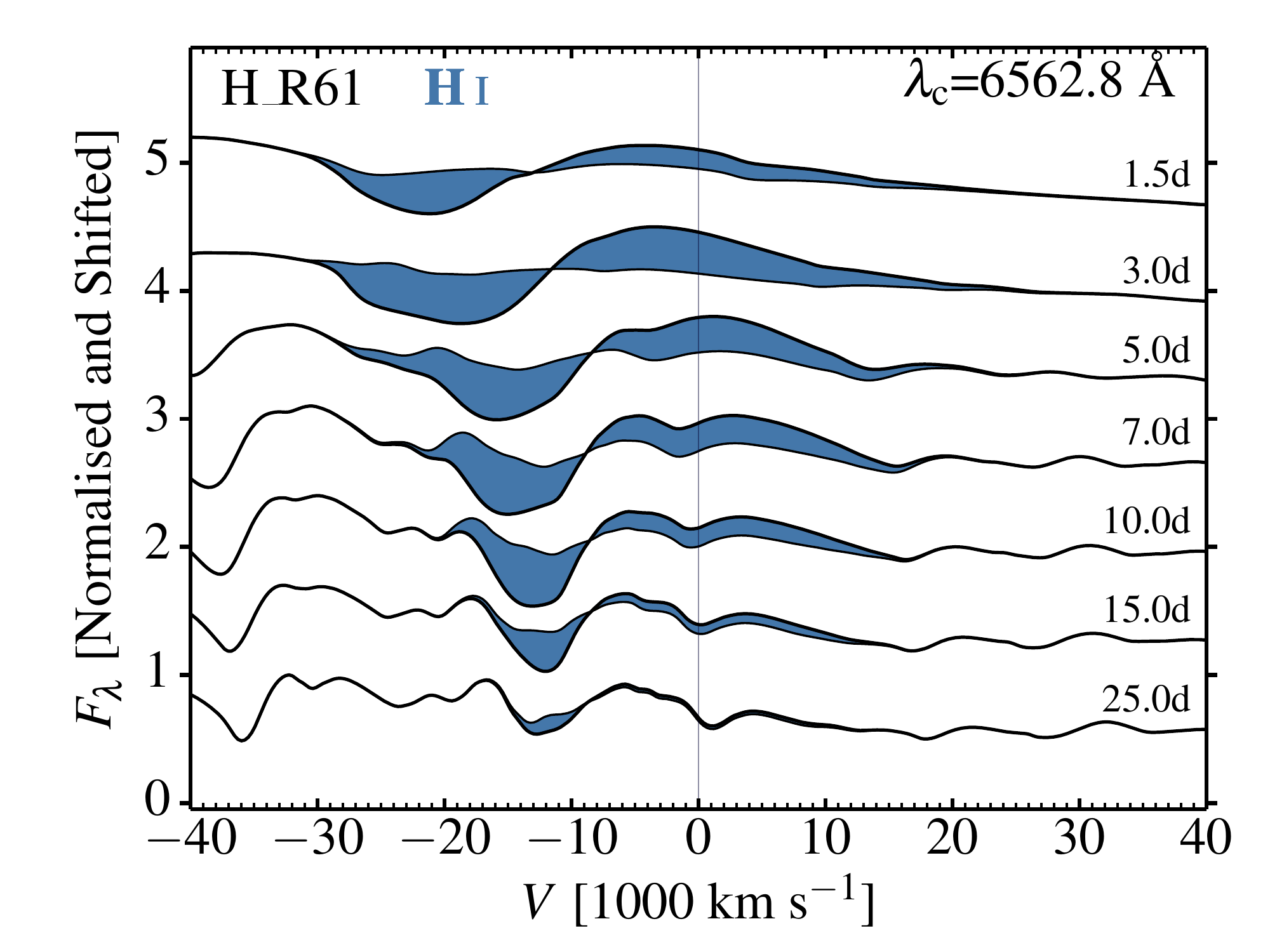}
 \includegraphics[width=0.45\hsize]{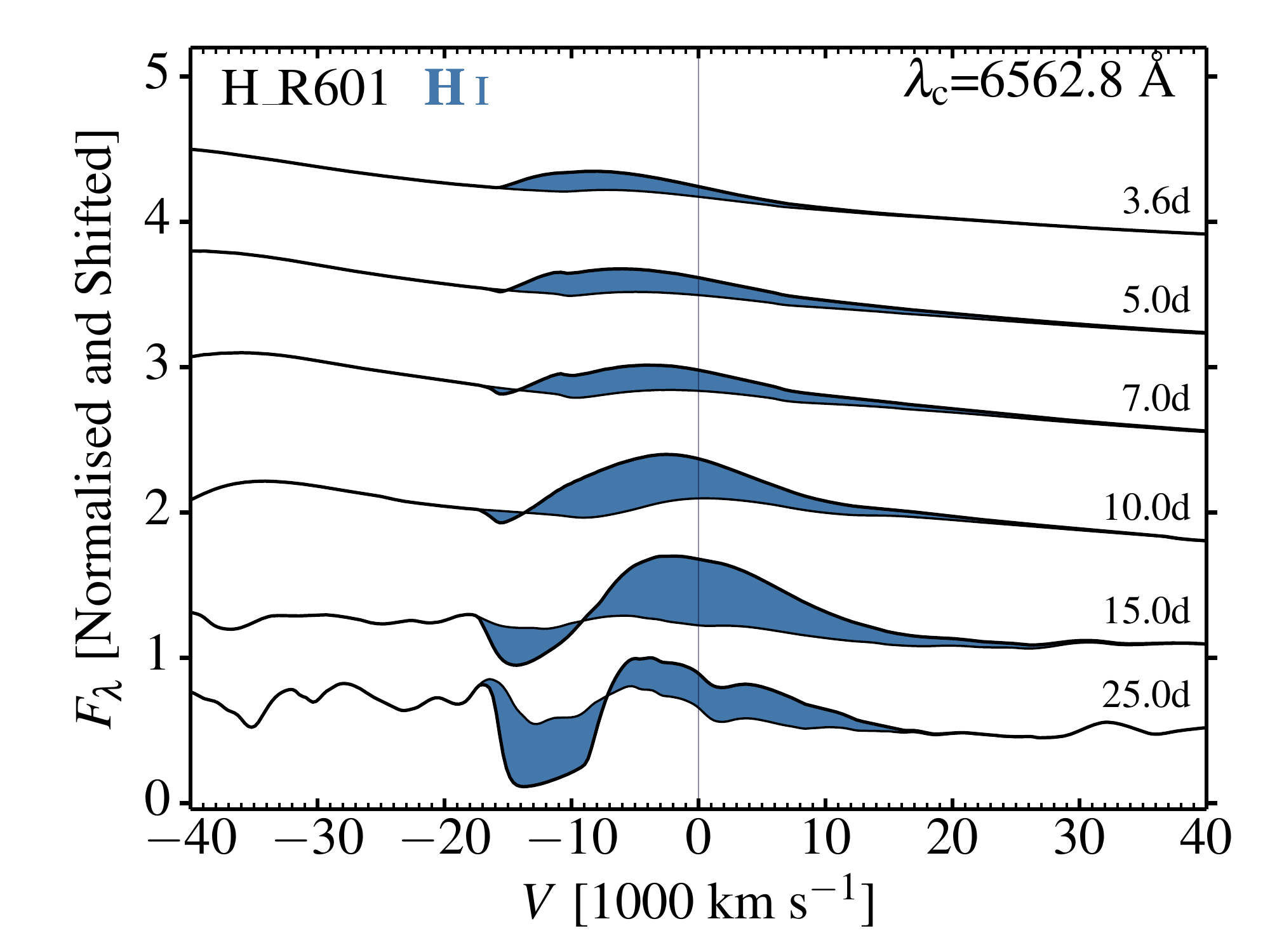}
\caption{Top row: Spectral evolution of models H\_R61 (left) and H\_R601 (right) in the ultraviolet
and optical ranges from 2-4\,d until 25\,d after explosion.
Bottom row: Evolution of the H$\alpha$ region in velocity space up to 25\,d after explosion
for model H\_R61 (left) and H\_R601 (right).
The shaded area corresponds to the contribution from H$\alpha$. In practice, we compute
the formal solution of the radiative transfer equation for each model  but without the bound-bound
transitions from H\one. The shaded area corresponds to the flux offset between that solution, and
the solution for the model that accounts for all bound-bound transitions.
\label{fig_spec}
}
\end{center}
\end{figure*}

\begin{figure*}
\begin{center}
\includegraphics[width=0.45\hsize]{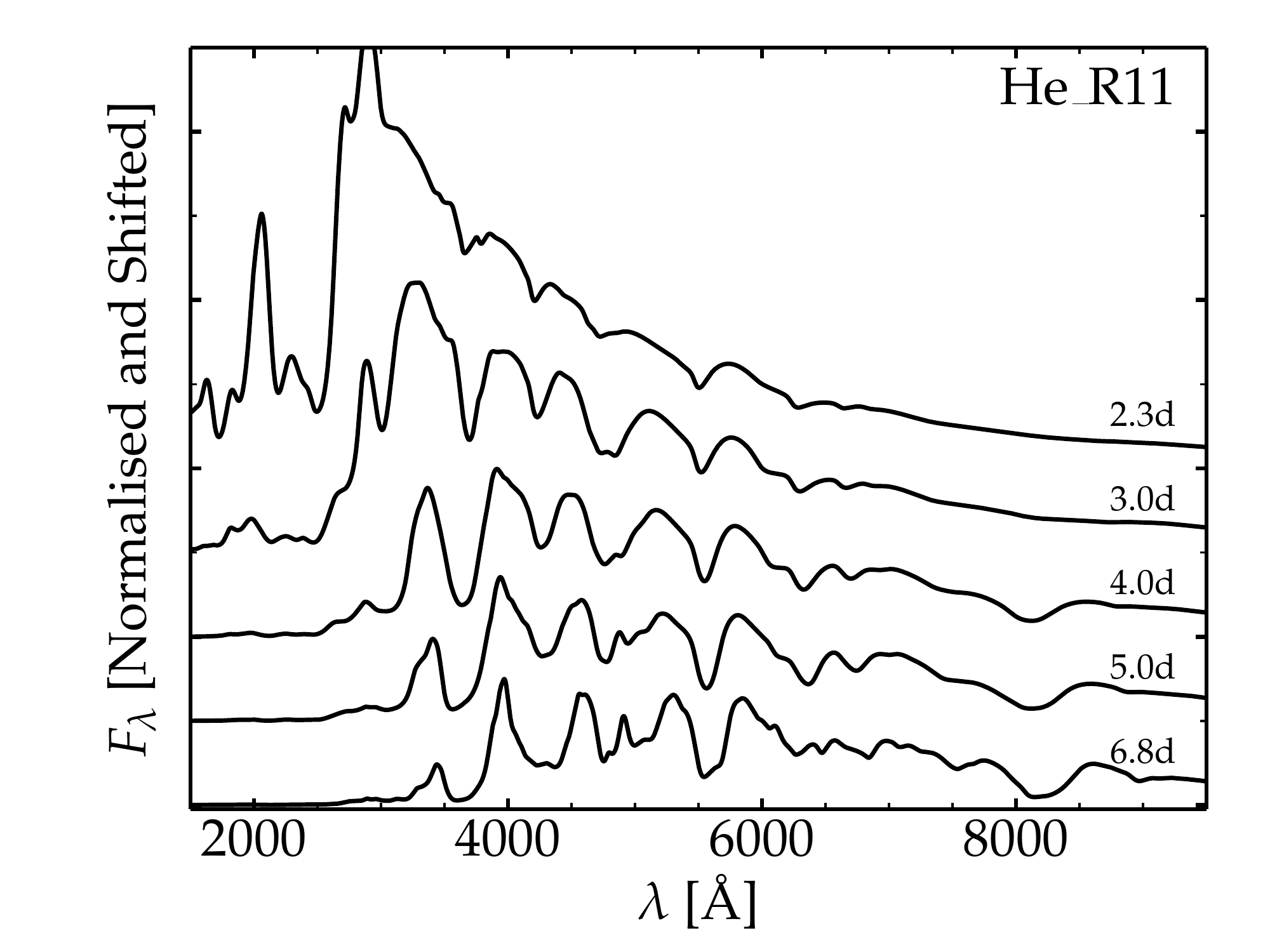}
\includegraphics[width=0.45\hsize]{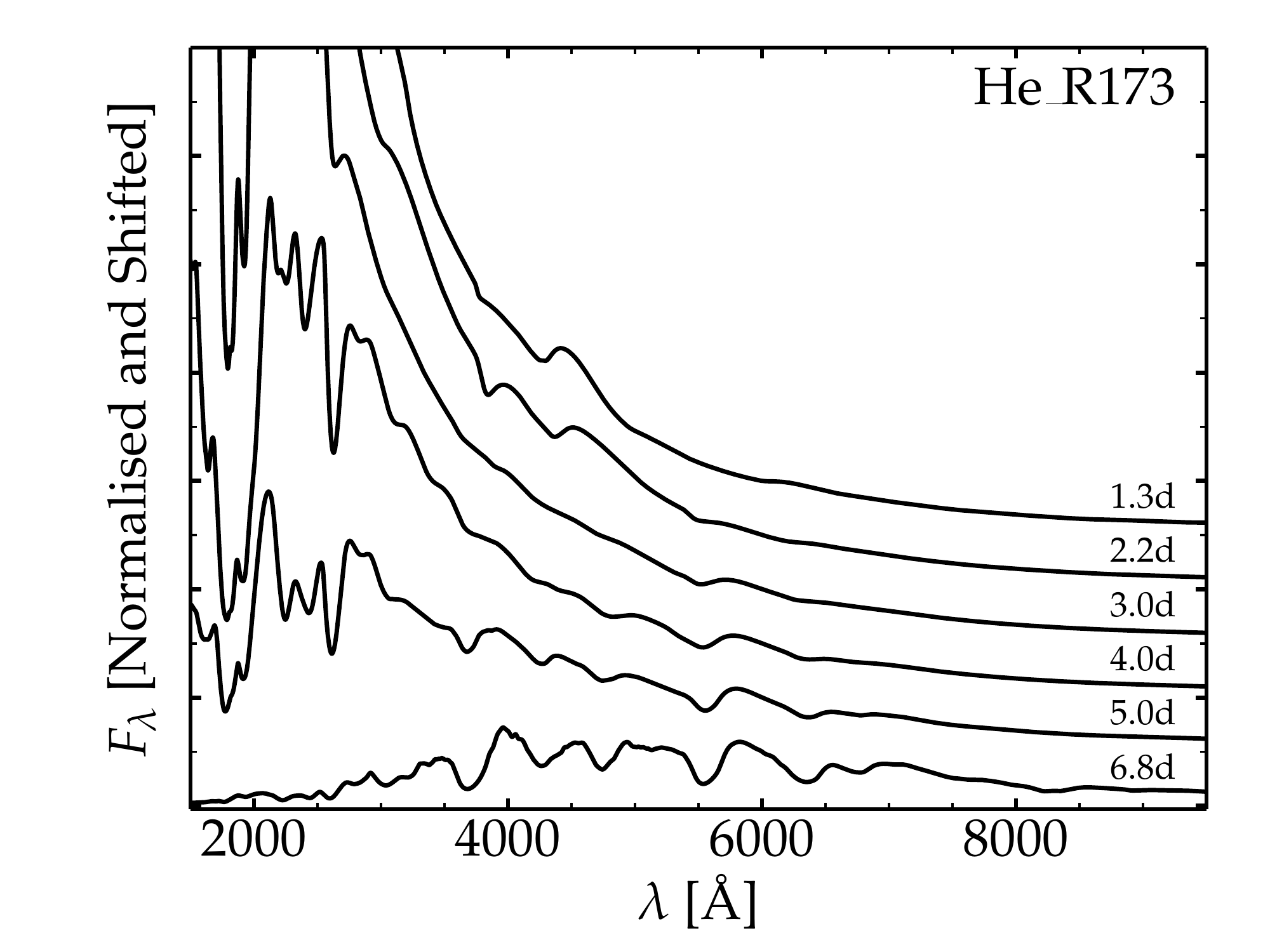}
\includegraphics[width=0.45\hsize]{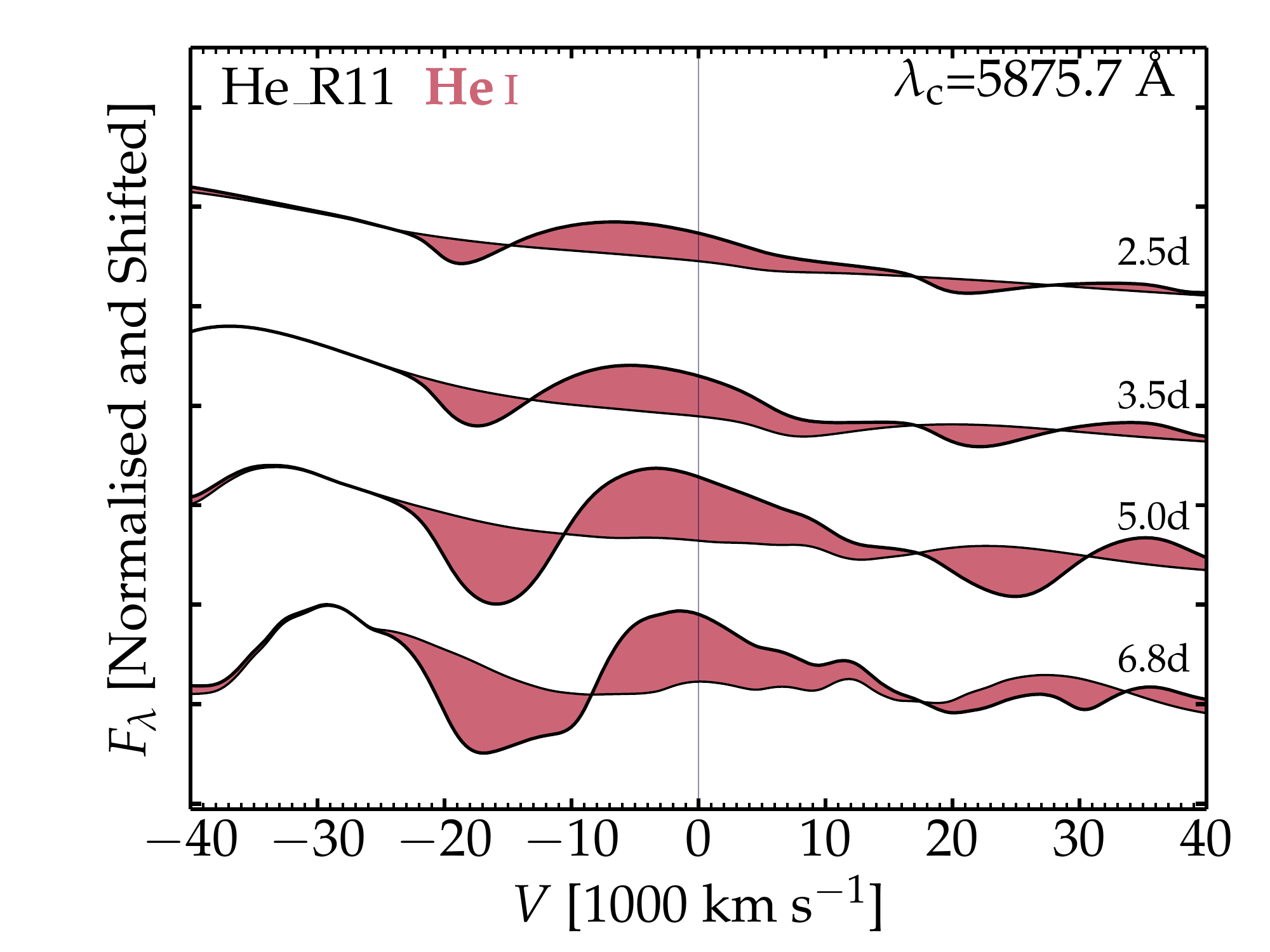}
\includegraphics[width=0.45\hsize]{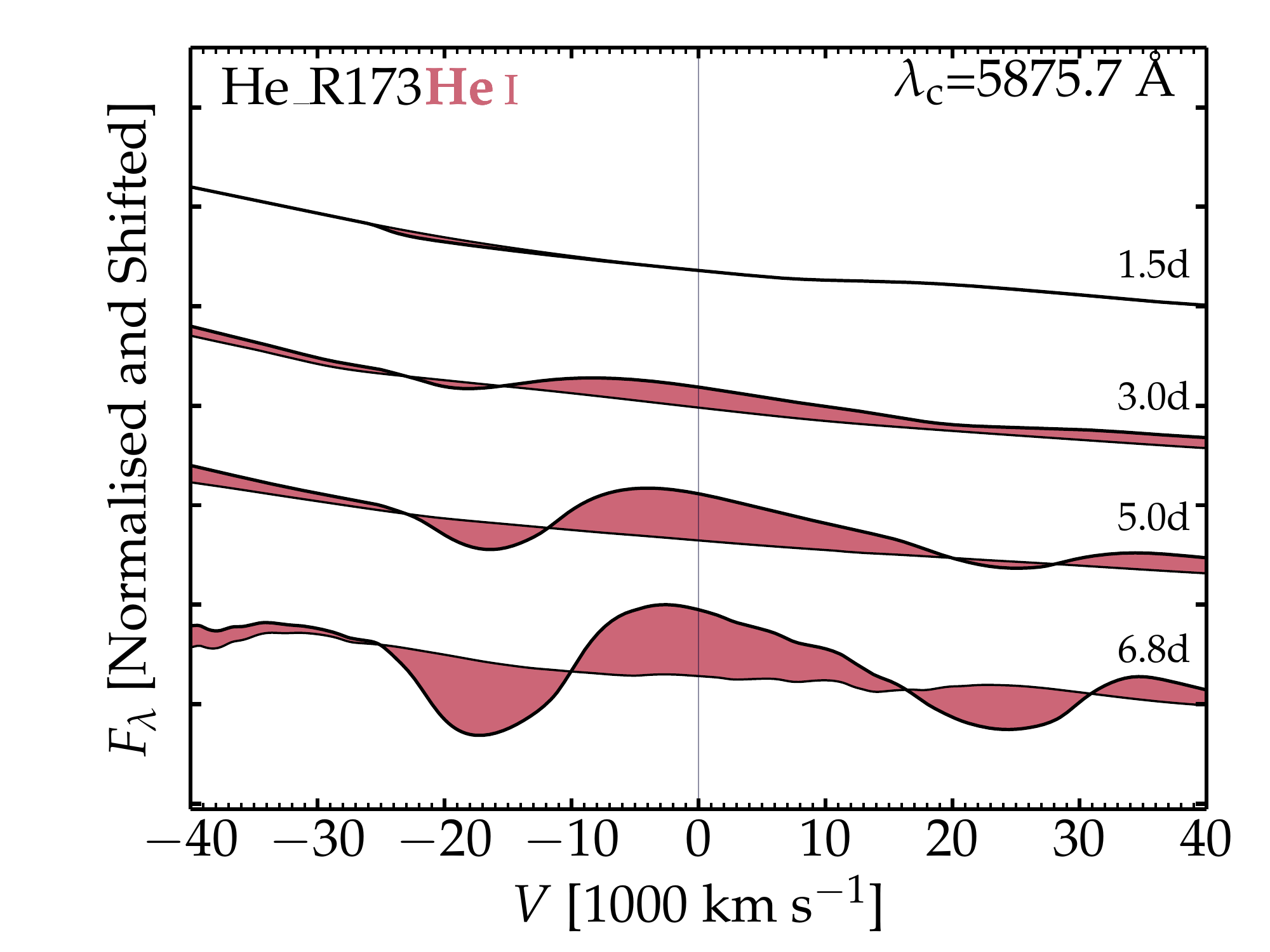}
\caption{
Spectral evolution for models He\_R11 (left) and He\_R173 (right)
covering the first week after explosion.
The top-row panels show the ultraviolet and optical ranges and the
bottom-row panels show the region around 5875\,\AA\ --- the P-Cygni
profile to the red of He\one\,5875\,\AA\ is due to He\one\,6678\,\AA.
Since these ejecta models do not contain any \nifs, the presence of He\one\
lines is not a result of non-thermal effects (see also \citealt{dessart_11_wr}).
Blue-featureless spectra in Type Ib SNe are compatible with a giant
progenitor star, like our model He\_R173.
\label{fig_spec_he}
}
\end{center}
\end{figure*}

   \subsection{Spectroscopy}
\label{sect_spec}

So far, time-dependent radiative transfer simulations of explosions from extended stars have been
limited to light curves. In this section, we discuss the associated spectral evolution, and start with
a description of the results for models H\_R61 and H\_R601 (Fig.~\ref{fig_spec}).

Model H\_R61 shows properties that are analogous to the Type IIb SNe from compact WR progenitor
stars (see, e.g., \citealt{D16_SNIbc_II}). The small amount of hydrogen (about 0.004\,\msun)
present in the outermost ejecta layers (beyond $\sim$\,14000\,\kms; Fig.~\ref{fig_v1d_comp_h_all})
gives rise to a strong H$\alpha$ line for $\sim$\,10\,d, which survives as a small absorption dip at 25\,d
(bottom row of Fig.~\ref{fig_spec}). This dip recedes in velocity space down to about $-14000$\,\kms,
which is the velocity at the base of the H-rich ejecta. As expected from the discussion above, the model color
is quite red at all times, with little flux blueward of the $U$ band.

In contrast, model H\_R601 shows completely different spectral properties from model H\_R61
(compare top right and top left panels of Fig.~\ref{fig_spec}). These are also different from the results based
on the compact WR progenitors of \citet{yoon_ibc_10}.

Model H\_R601 is extremely blue early on, with the peak of the flux
shortward of the $U$ band, well into the ultraviolet at $\sim$\,1200\,\AA\ at 4\,d. At the photosphere, the velocity
is 16200\,\kms, the temperature is 17500\,K, and the density profile is very steep (Fig.~\ref{fig_v1d_1p2d}).
Such high temperatures are typical of RSG star explosions at $1-2$\,d after shock breakout.
Consequently, the spectral features are very weak (little absorption and emission), little line blanketing
occurs in the optical, and the spectrum appears blue and featureless. This is reminiscent of the earliest spectrum
of SN\,1993J (\citealt{matheson_93j_00a}; see Section~\ref{sect_93j}).
At 3-4\,d in model H\_R601, we see spectral lines similar to those seen in the earliest spectra of
Type II-Plateau SNe \citep{quimby_06bp_07,dessart_05cs_06bp,galyam_early_2p_11},
with He\two\,4686\,\AA, He\one\,5875\,\AA\ and H$\alpha$.
It takes model H\_R601 two weeks to redden to the same color as model H\_R61 (see also Fig.~\ref{fig_h_phot}).
H$\alpha$ reveals other differences between these two models (bottom row panels of Fig.~\ref{fig_spec}).
In model H\_R601, H$\alpha$ is visible even at 25\,d, with an absorption that covers between $-16000$\,\kms\
and $-8000$\,\kms. This corresponds to the velocity span of the H-rich dense shell (Fig.~\ref{fig_v1d_comp_h_all}).

 The reverse shock in model H\_R601 slows down the He-core material so that the He-rich
shell stretches out to about 8000\,\kms. In the models with lower envelope radius/mass,
the outer edge of the He core is located at greater velocities since the reverse shock is weaker
or even absent (model H\_R5). Although the formation of He\one\ lines is complicated (sensitive
to non-thermal excitation and \isoni\ mixing), this different stratification in velocity space
provides a possible explanation for the smaller absorption velocities of He\one\,5875\,\AA\ in
Type IIb compared to Type Ib SNe \citep{liu_snibc_15}.

Also shown in Fig.~\ref{fig_spec} is the sizeable peak blueshift of the H$\alpha$ line. This originates
from an optical depth effect \citep{DH05a} and is a generic property of SN spectra \citep{anderson_etal_14},
even present in Type IIn SNe \citep{zhang_10jl_12,D15_2n}. This effect is predicted in
spherically symmetric ejecta --- asphericity is  not necessary.

For He-giant star explosion model He\_R173, the spectral evolution is comparable
to that of model H\_R601, the obvious difference being the absence of H\one\ Balmer lines
(Fig.~\ref{fig_spec_he}).
The spectrum is blue and featureless at early times. At 2\,d, the spectral energy
distribution peaks in the ultraviolet, over the range $1000-2000$\,\AA.
At $1-2$\,d, we see He\two\,4686\,\AA\ (the peak emission in that line is blueshifted
by about $-15000$\,\kms) as well as multiplets of N\three\ at 3940\,\AA\ and 4640\,\AA.
Later on, as the photosphere cools, the spectrum shows lines of He\,\one, as typical of Type Ib SNe
(Fig.~\ref{fig_spec_he}).
In model He\_R11 (i.e., the trimmed version of the He-giant model He\_R173),
the spectra are redder at all times and exhibit the presence of numerous lines,
including He\one\ but also from metals --- the spectra are never as blue and featureless
as for model He\_R173.

Contrary to standard SNe Ib around maximum light, these He\one\ lines
are produced without any contribution from non-thermal processes --- there is no \isoni\ in
the ejecta model He\_R173.

For both H-rich and H-deficient progenitors, we therefore find that the presence
of an extended and tenuous envelope has a strong influence not only on the light curve
(modulation of the early-time luminosity and color), but also on the spectra (modulation in
chemical stratification in velocity space and ionisation that impact the spectral
energy distribution and line profiles).

\begin{figure*}
\includegraphics[width=0.49\textwidth]{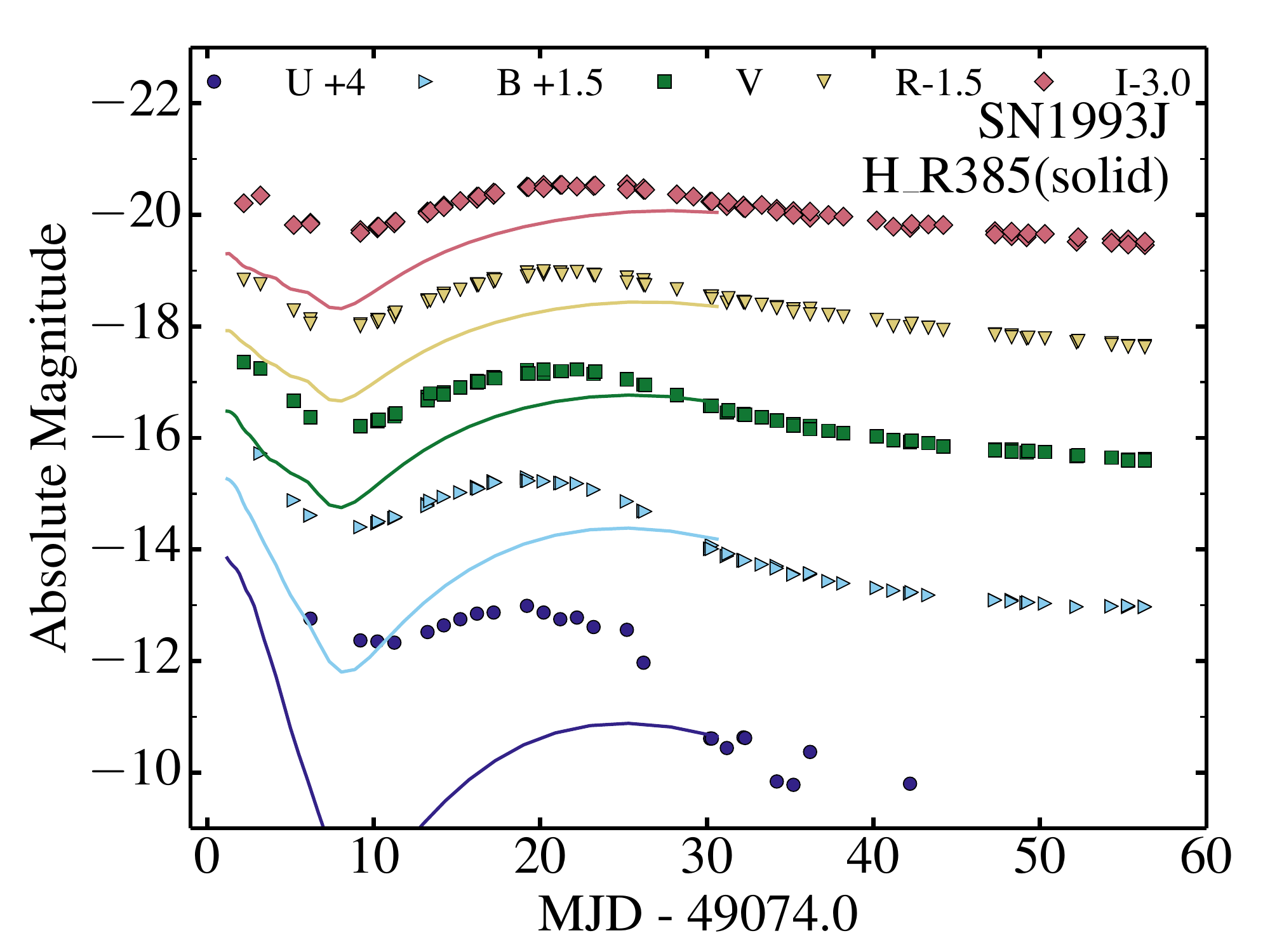}
\includegraphics[width=0.49\textwidth]{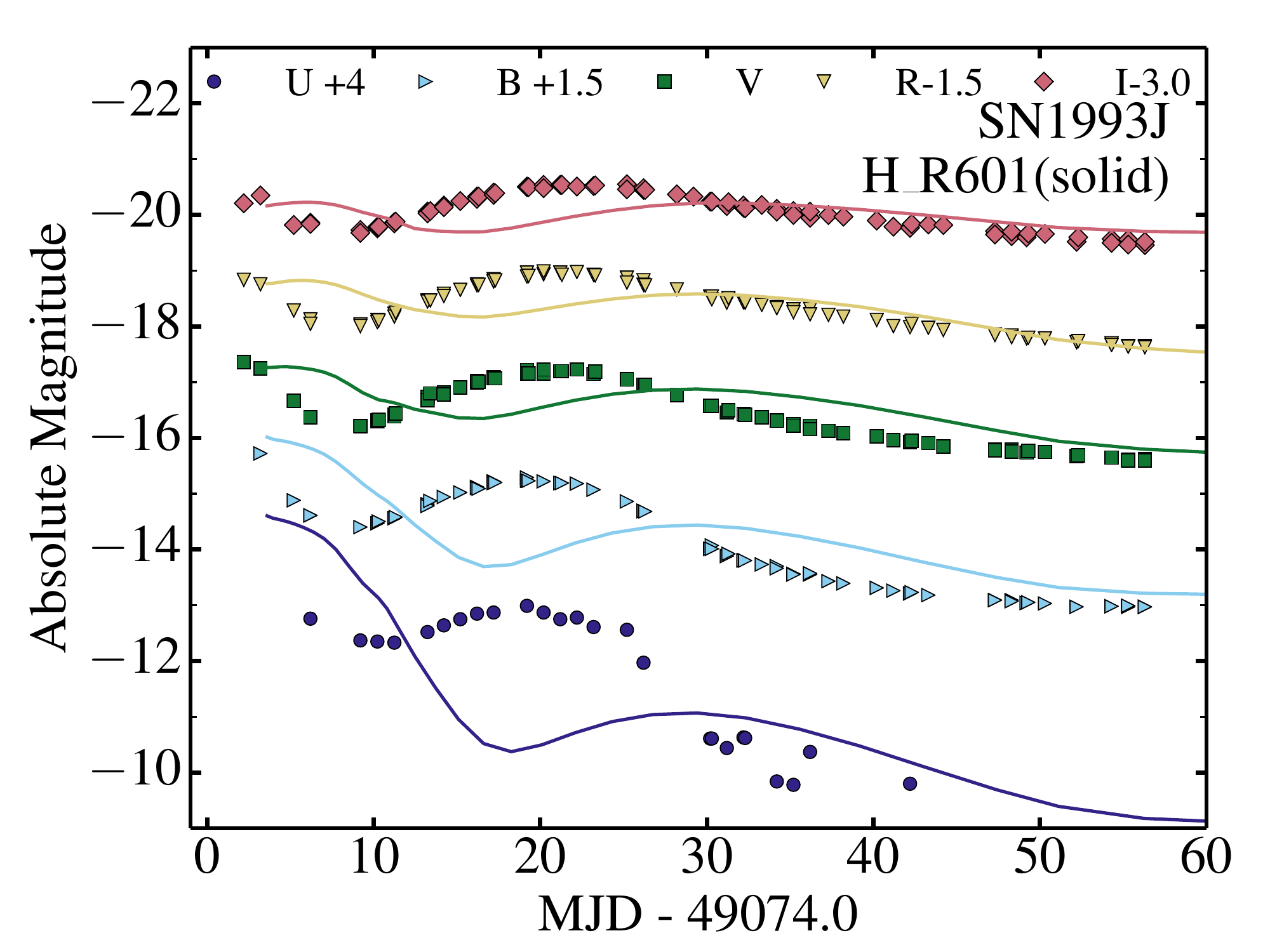}
\includegraphics[width=\textwidth]{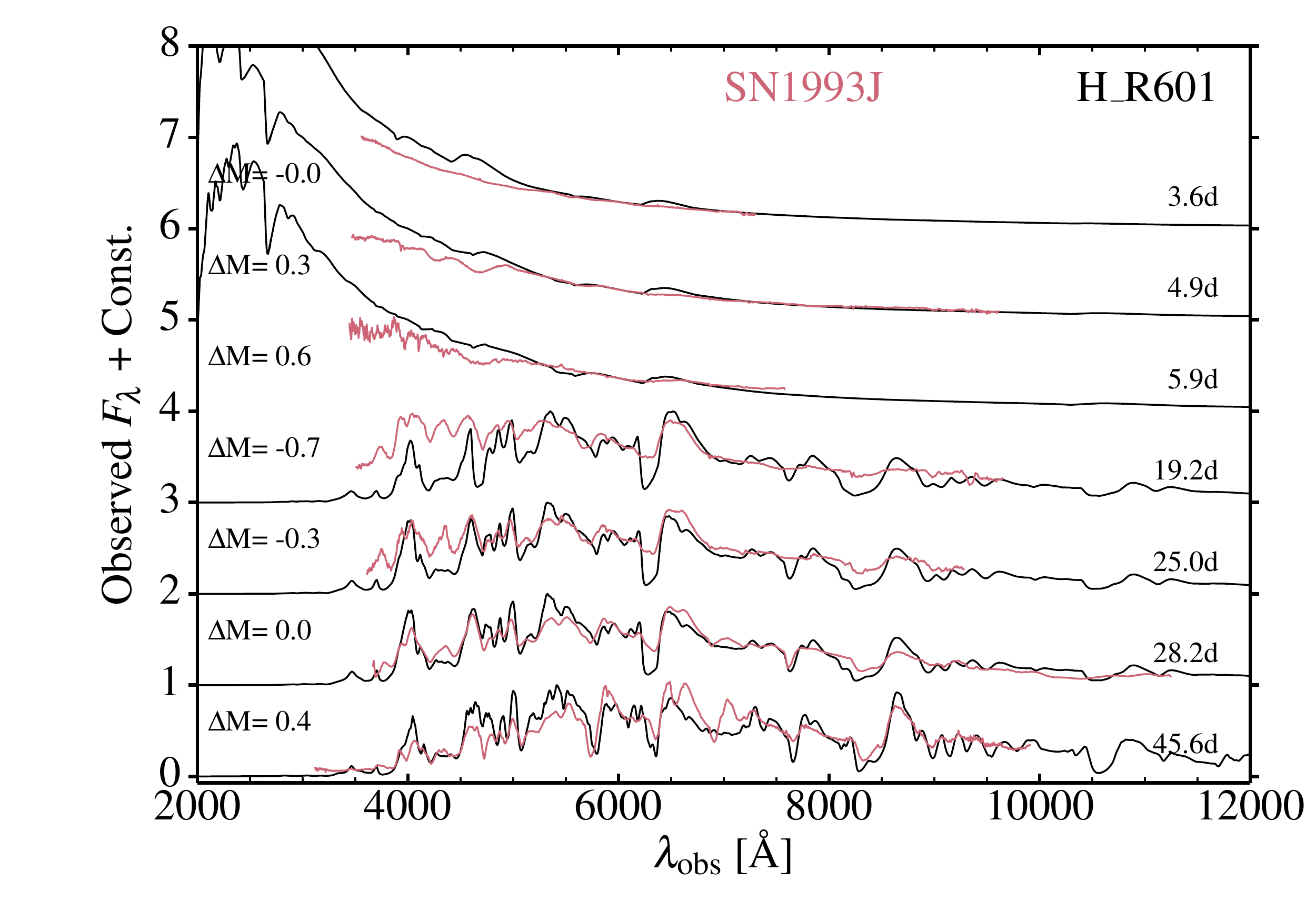}
\caption{Comparison of the multi-band light curves of models H\_R385 and H\_R601
and of multi-epoch optical spectra of model H\_R601 with the observations of SN\,1993J
(the photometry is from \citet{richmond_93j_96} and the spectra from \citet{matheson_93j_00a}).
In this figure, we either correct the model or the observations for distance, reddening, and redshift
of SN\,1993J (see Section~\ref{sect_obs} for details).
\label{fig_comp_93J}
}
\end{figure*}

\section{Comparison to observations}
\label{sect_comp_obs}

   Explosions from progenitors with a core-halo structure can produce SNe with much diversity
   in light curve and spectral properties.
   Progenitors may vary in their global properties (e.g., mass, radius, composition), but also in their
   internal structure (e.g., density profile within the low-density extended H-rich or He-rich shell).
   Once exploded, these stars may produce ejecta with a different kinetic energy at infinity, \iso{56}Ni mass,
   and different composition mixing. Finally, the explosion may depart significantly
   from spherical symmetry (see, e.g., evidence for this from light echoes; \citealt{rest_casA_echo_11}).

   Because of this expected diversity and our limited set of simulations, we use our results to study
   qualitative aspects of observations.
   Quantitative offsets that can be understood by invoking slight adjustments to the
   above parameters are not a major issue. Slight variations in explosion energy
   and ejecta mass can impact the rise time to maximum, the spectral line widths or the light curve width.
   Variations in \iso{56}Ni mass and mixing  may impact the photometric brightness, colors, and spectral line
   widths. The impact of such variations has been explored in \citet{D15_SNIbc_I,D16_SNIbc_II}.

\subsection{SN\,1993J}
\label{sect_93j}

  Figure~\ref{fig_comp_93J} shows the comparison of our models with the observations of SN\,1993J.
  This SN shows two phases of high brightness. The first one extends over the first 10 days and is
  powered by shock deposited energy in the outer ejecta. The second is powered by \iso{56}Ni decay
  and causes the bolometric maximum at about 20\,d after explosion. During the fading from the
  first peak, SN\,1993J reddens with time. During the second maximum, its colour changes little.

  In our model set, only the models from extended progenitors (H\_R385 and H\_R601) yield properties
  qualitatively similar to SN\,1993J.
  They show a relatively high brightness in all bands during the early post-breakout
  phase, for one week in H\_R385 and for about two weeks in H\_R601. This excess is lower (higher) than observed
  for model H\_R385 (H\_R601). The \iso{56}Ni powered peak is of comparable brightness for both
  models and observations. It occurs $10-15$\,d later in the models (the models also overestimate
  the light curve width since it correlates with the rise time).
  In both model H\_R601 and observations, the $R$-band magnitude is comparable during the first and the second
  maximum.
  However, SN\,1993J is bluer in the optical during the \iso{56}Ni powered peak.
  Overall, our models show a greater variation in color as they evolve through the first and the second
  peaks. While the color offset during the first peak can be resolved by invoking different progenitor
  envelope properties, the color offset during the \iso{56}Ni powered peak is harder to explain. The mixing
  adopted here affects only \iso{56}Ni and therefore should not alter much the magnitude of line blanketing
  (i.e., we do not mix other metals into the outer ejecta). One possibility is the extra contribution of radiation
  from the SN shock going through the progenitor wind \citep{fransson_93j_96,matheson_93j_00a} --- this
  interaction is ignored in our work (we assume that the ejecta expands  in a vacuum).

    \begin{figure*}
\includegraphics[width=0.48\textwidth]{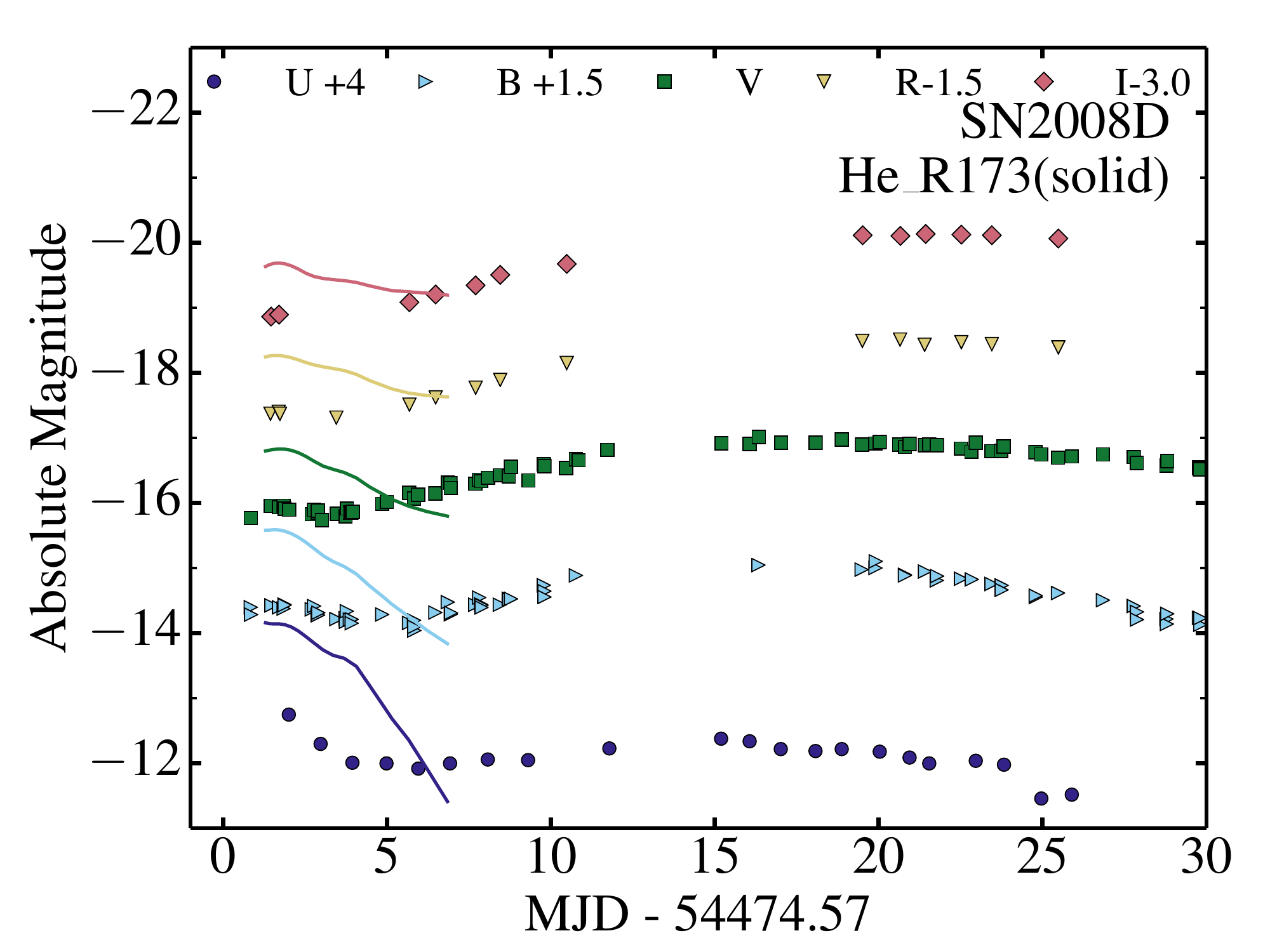}
\includegraphics[width=0.48\textwidth]{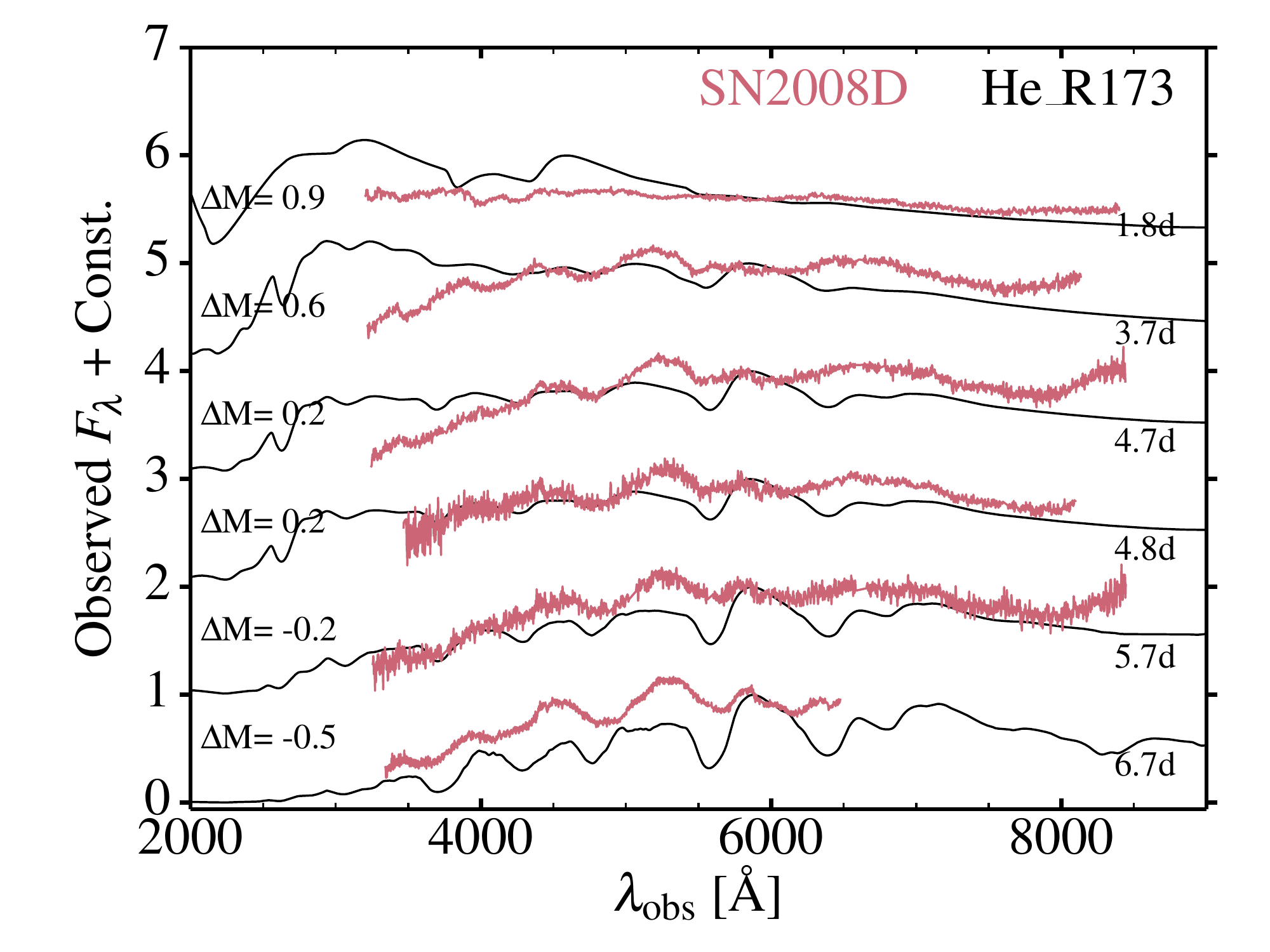}
\caption{Multi-band light curves (left) and multi-epoch spectra (right)
for the He-giant star explosion model He\_R173 (solid) and the observations (symbols) of SN\,2008D.
The time sequence is computed between 1 and 7\,d after explosion.
In this figure, we either correct the model or the observations for distance, reddening, and redshift
of SN\,2008D (see Section~\ref{sect_obs} for details).
\label{fig_comp_08D}
}
\end{figure*}

  Spectroscopically, model H\_R601 shares a number of properties with the observations of SN\,1993J.
  The model reproduces the blue featureless spectra at early times although the model is too blue during the first week.
  The weak absorption/emission of spectral lines is a consequence of forming in an atmosphere with a steep
  density profile. Also, just like in type II-P SNe detected very early, we see the presence of
  He\two\,4686\,\AA\ (see, e.g., the early-time observations of SN\,2006bp or SN\,1999gi and associated models
  in \citealt{dessart_05cs_06bp}).
  Unlike Type II-P SNe, the colour evolution is rapid, although the model reddens more than observed during the first
  two weeks (this may in part be due to the weak \isoni\ mixing in our model, which causes a delayed rebrightening).
  The reddening is associated with a visible strengthening of line blanketing in the optical range.
  At the last epoch shown (i.e., 45.6\,d), the correspondence between the observations and the model H\_R601 is
  quite good.

  We can also identify more subtle properties.
  The absorption trough of H$\alpha$ shows the same width at all times.
  That width is a good tracer of the velocity extent of the dense shell formed by the reverse shock.
  A progenitor without an extended tenuous
  envelope shows instead a broad H$\alpha$ absorption that continuously narrows until 20-30\,d
  (see model H\_R61 in Fig.~\ref{fig_spec}).
  Furthermore, the strength of the H$\alpha$ absorption is overestimated by the model, and the same holds for other lines
  (e.g., the Ca\two\, 8500\,\AA\ triplet). This might be additional evidence that some excess radiation, perhaps
  from an interaction with the progenitor wind, is contaminating the SN radiation.
  Also, our model underestimates the strength of He\one\,6678\,\AA\  and 7068\,\AA, perhaps
  because \iso{56}Ni is insufficiently mixed or the progenitor He-shell mass is too small.
  Interestingly, model H\_R601 overestimates the
  strength of O\one\,7774\,\AA\ and of C\one\ lines located red-ward of Ca\two\, 8500\,\AA.
  Our progenitor model may have a too large CO core.

  Despite these moderate deficiencies, model H\_R601 is unlike any other SN IIb
  resulting from compact WR progenitors.
  The greatest difference is not seen during the \iso{56}Ni
  powered peak (whose luminosity is not affected by the presence of a low-density extended progenitor envelope)
  but during the early post-breakout phase.
  All the models we presented in \citet{dessart_11_wr, d12_snibc, D15_SNIbc_I,D16_SNIbc_II} have
  red colors and show strong signs of line blanketing at early times (i.e., during the first two weeks after explosion).
  Model H\_R601 shows that an extended
  progenitor can produce blue and featureless spectra (as well as blue optical colors) at early times, as seen most vividly in
  SN\,1993J.

   Finally, the progenitor model of H\_R601 corresponds to a star with an effective temperature
  of  4300\,K. This value is compatible with the inferences of \citet{aldering_93j_94} based on pre-explosion
  photometry of the SN site.

\subsection{SN\,2008D}
\label{sect_08D}

In this section, we compare model He\_R173 with the observations of SN\,2008D.
The optical maximum at 17\,d after explosion \citep{soderberg_08D,modjaz_08D} is probably
powered by $\sim$\,0.07\,\msun\ of \iso{56}Ni \citep{mazzali_08D_08}. If this is the case,
the contribution
from decay power is already strong at 5\,d  since it causes the SN to be only 1\,mag fainter then
than at maximum. This implies that \iso{56}Ni is efficiently mixed into the outer ejecta layers.

Our simulation is limited to the first week after explosion and does not include \isoni\ so
the comparison is pertinent prior to the rebrightening of SN\,2008D.
Figure~\ref{fig_comp_08D} shows the multi-band light curves and multi-epoch spectra
for the model He\_R173 and the observations of SN\,2008D.
With our adopted distance and reddening, we see that model He\_R173
is too luminous and also too blue at the earliest times.
An He-giant star with an envelope of lower mass/extent could remedy this problem.

Model He\_R173 shows a very blue and quasi featureless spectrum early on, which is
also a striking property of SN\,2008D.
The model shows He\two\,4686\,\AA\ and N\three\ lines at 3940\,\AA\, and 4640\,\AA.
It is not clear whether these are observed because of the offset in wavelength and the mismatch
in line strength.  Our approach assumes spherical symmetry, while light echoes of the Type IIb
SN associated with Cas A clearly shows that the SN spectral appearance can vary with the
viewing angle \citep{rest_casA_echo_11}. Furthermore, the detected polarisation around light
curve maximum and at nebular times suggests the ejecta of SN\,2008D is
asymmetric \citep{maund_08D_09,tanaka_08D_09b}.
\citet{modjaz_08D} suggests the presence of C\three\ and O\three\ lines. Their {\sc synow} spectrum
matches well the observations. It is however surprising that a Type Ib SN progenitor, rich in He and Nitrogen (and
thus depleted in C and O at the surface) could show C\three\ and O\three\ lines at the earliest times, and
at the same time not show He\two\,4686\,\AA.
Spectral lines have their peak emission blueshifted when the ejecta is optically thick, especially early-on
(see Section~\ref{sect_spec}).
This feature, not predicted by {\sc synow}, may impact line identifications done with this code
when the photospheric velocity is large and the ejecta optically thick.
Through this first week, the He\,\one\ lines strengthen in the model and remain very broad.
These He\one\ lines are present despite the absence of \iso{56}Ni in the model
(see also \citealt{dessart_11_wr}).

\citet{bersten_08D_13} propose the mixing of 0.01\,\msun\ of \iso{56}Ni
out to 20000\,\kms\ to reproduce the SN\,2008D light curve. They also consider explosions from extended stars,
but their models are distinct from the low-mass He-giant star that we consider here.
The weak and narrow He\one\ lines observed in SN\,2008D may be in tension with the strongly mixed
model of \citet{bersten_08D_13} since the \iso{56}Ni that causes the early luminosity boost in their
model would also enhance non-thermal effects in the outer  layers. Consequently, very broad He\one\ line
absorption troughs could be expected at one week in this model --- such broad lines are not seen.

 Here, we have performed a prospective study that shows encouraging similarities with
the observations of SN\,2008D, both photometrically and spectroscopically.
In particular, a high luminosity, blue optical colors, and featureless spectra at early-times
can be explained with a He-giant star progenitor.
Such a progenitor structure may arise from a low mass primary in a binary
system \citep{pols_94,yoon_ibc_10,clelland_he_16,yoon_ibc_17}.
For SN\,2008D, this represents an alternative scenario to a high mass
progenitor \citep{mazzali_08D_08,tanaka_08D_09a}.


\section{Conclusions}
\label{sect_conc}

  We have presented models for the explosion of massive stars characterized by extended
  and tenuous envelopes. Our models are computed with the stellar evolution codes \mesa\ and \bec,
  exploded with the radiation-hydrodynamics code \v1d, and followed from $\sim$\,1\,d after shock breakout
  until 20-30\,d with the radiative transfer code \cmfgen. We focus here on a limited set.
  Our H-rich progenitors result from the explosion of the primary star, evolved from a
  16\,\msun\ $\oplus$ 14\,\msun\ system with an initial orbital separation between 100 and 1000\,\rsun.
  We also present simulations for a 2.8\,\msun\ He-giant star that may result from binary evolution.

  These pre-SN models have a core-halo density structure. Outside of a dense core
  ($\lesssim$\,1\,\rsun) containing the bulk of the mass, a tenuous envelope ($\lesssim$\,0.1\,\msun) with a flat density
  profile extends out to a radius as large as a few hundred \rsun.  The larger the mass/extent of the tenuous
  envelope, the stronger is the reverse shock, the more massive is the dense shell, and the greater is
  the trapped radiation energy stored in the envelope at 1\,d.
  While this energy is only a tiny fraction of the total ejecta kinetic energy at
  infinity, it is stored in an envelope of low/moderate optically depth. This energy is radiated
  prior to the \iso{56}Ni powered peak at 20-30\,d and can therefore dramatically alter the SN luminosity
  at early times.

  Our model H\_R601 with a progenitor radius of 601\,\rsun\ and an ejecta mass of 3.5\,\msun\ share many
  similarities with the observations of SN\,1993J, including the luminosity boost for 1-2 weeks after shock
  breakout, the existence of two light-curve maxima in optical and near-IR bands,
  the blue and featureless spectra at early times, and the rather
  narrow (and constant width) of the H$\alpha$ profile for the first 20-30\,d. Our model shows some discrepancies
  which might be cured by invoking modest changes in ejecta mass, progenitor radius, chemical mixing,
  as well as invoking the contribution of radiation from a persistent ejecta/wind interaction.

   We also use our He-giant explosion model to compare to the early time observations of SN\,2008D.
   Our model He\_R173, with its progenitor radius of 173\,\rsun, produces a much larger luminosity
   than compact WR star explosions at early times. Our model overestimates the early-time luminosity of SN\,2008D --
   our He-giant model is somewhat too big.

   We also present the first spectral calculations for such He-giant star explosions.
   The model predicts a blue spectrum at early
   times, as observed, but not as featureless. In particular, the model predicts the presence at $1-2$\,d of lines
   from He\two\ and N\three. This is more compatible with the expectations for a nitrogen-rich
   Wolf-Rayet star progenitor than the proposed identification of C\three\ and O\three\ \citep{modjaz_08D}
   with no He\two. Our model does not include any \iso{56}Ni but predicts the presence of strong He\one\ lines as
   soon as He\two\,4686\,\AA\ vanishes.
    As long as the photospheric conditions remain hot and ionized, He\one\ lines can be thermally excited
   and may therefore be present without the influence of non-thermal processes \citep{lucy_91}.

   He\one\ lines are seen in SN\,2008D, but they appear quite narrow at one week.
   This seems in tension with the model of \citet{bersten_08D_13}, which invokes
   the presence of 0.01\,\msun\ of \iso{56}Ni at 20000\,\kms\ in order to explain
   the SN\,2008D bolometric light curve.  This should
   enhance the width of He\one\ lines around bolometric maximum \citep{d12_snibc}.

   Overall, a low-mass He-giant star offers attractive properties for Type Ib SNe that
   show a large luminosity during the first week after explosion, as well as a very rapid photometric evolution and
   a short rise time to bolometric maximum. Indeed, such
   He giants are fundamentally low-mass stars, that should have a very small iron core, and should
   synthesize modest amounts of \iso{56}Ni. They are also compatible
   with the inferences for Type Ib iPTF13bvn
   \citep{bersten_iPTF13bvn_14,clelland_he_16}.
    If the extent of the low-density envelope is also a function
    of the opacity \citep{clelland_he_16},
    He-giant stars may statistically have a smaller radius at lower metallicities.
    Consequently, Type Ib SNe from He-giant stars could have early-time colors
    that are redder at lower metallicity.

 \section*{Acknowledgements}

LD thanks ESO-Vitacura for their hospitality.
This work utilized computing resources of the mesocentre SIGAMM,
hosted by the Observatoire de la C\^ote d'Azur, Nice, France.
SCY acknowledges support from the Korea Astronomy and Space Science Institute
under the R\&D program (Project No. 3348-20160002) supervised by the Ministry of Science,
ICT and Future Planning.


\appendix

\section{Composition profiles for all models}

\begin{figure*}
\begin{center}
 \includegraphics[width=0.45\hsize]{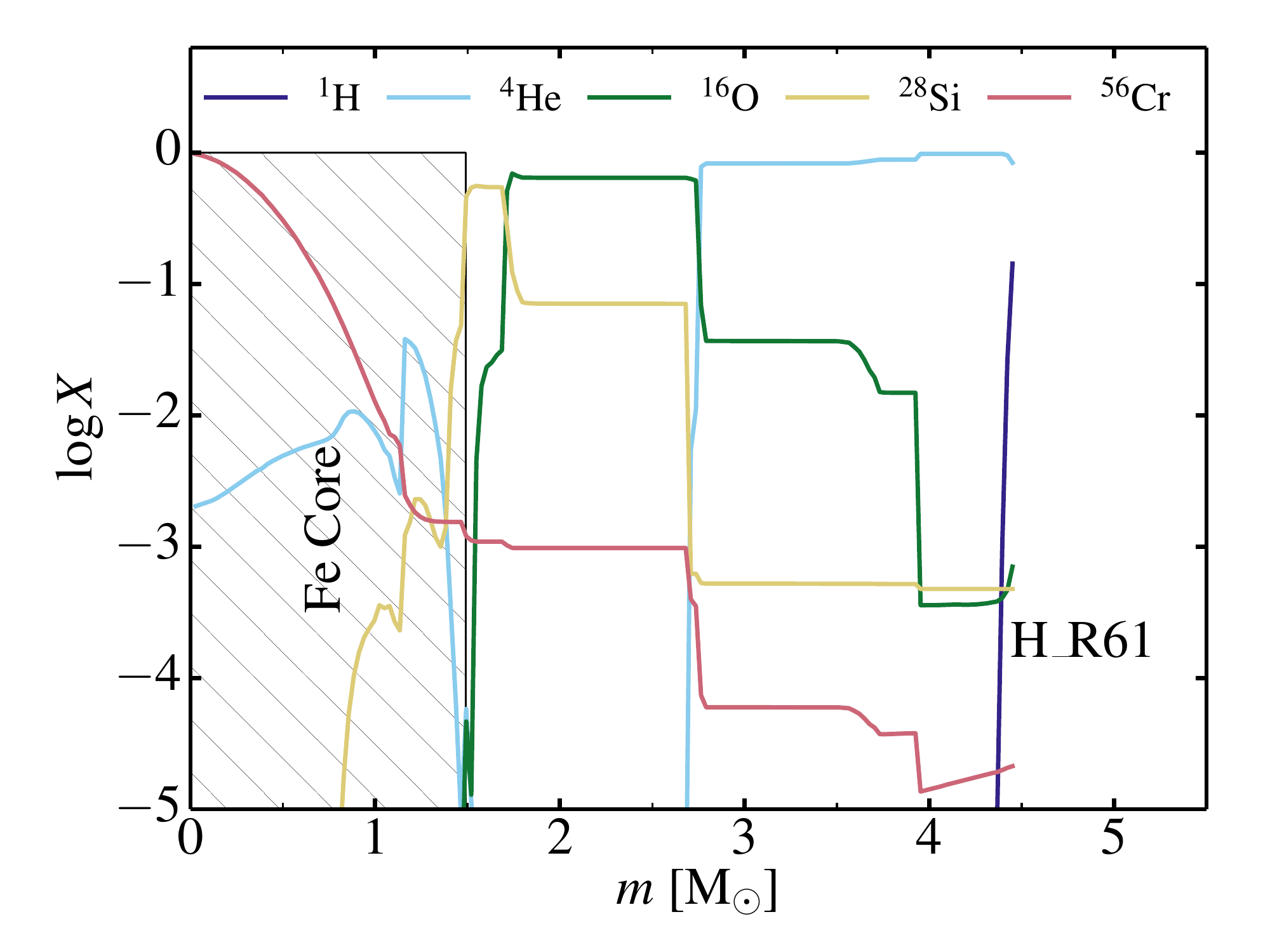}
 \includegraphics[width=0.45\hsize]{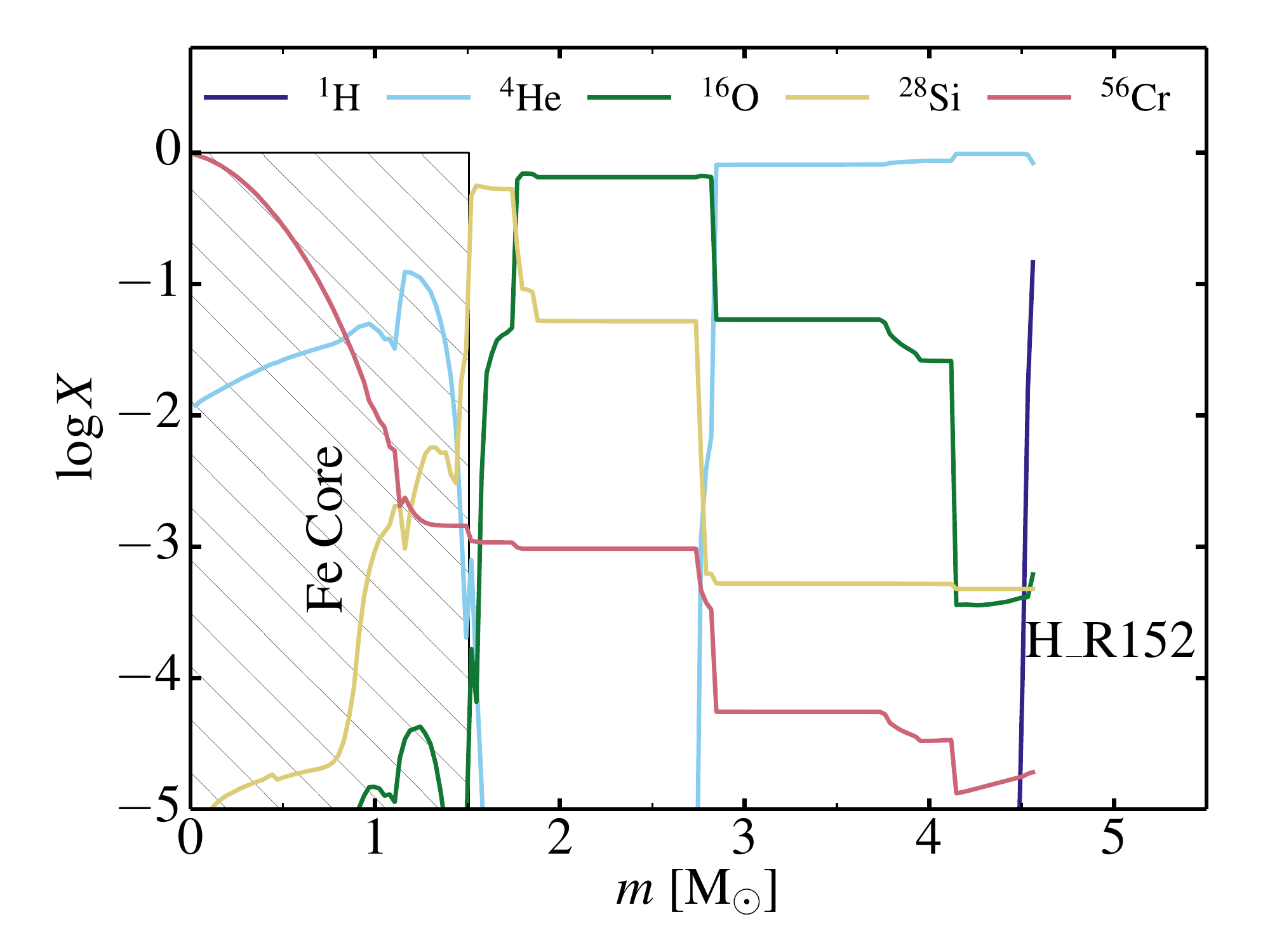}
 \includegraphics[width=0.45\hsize]{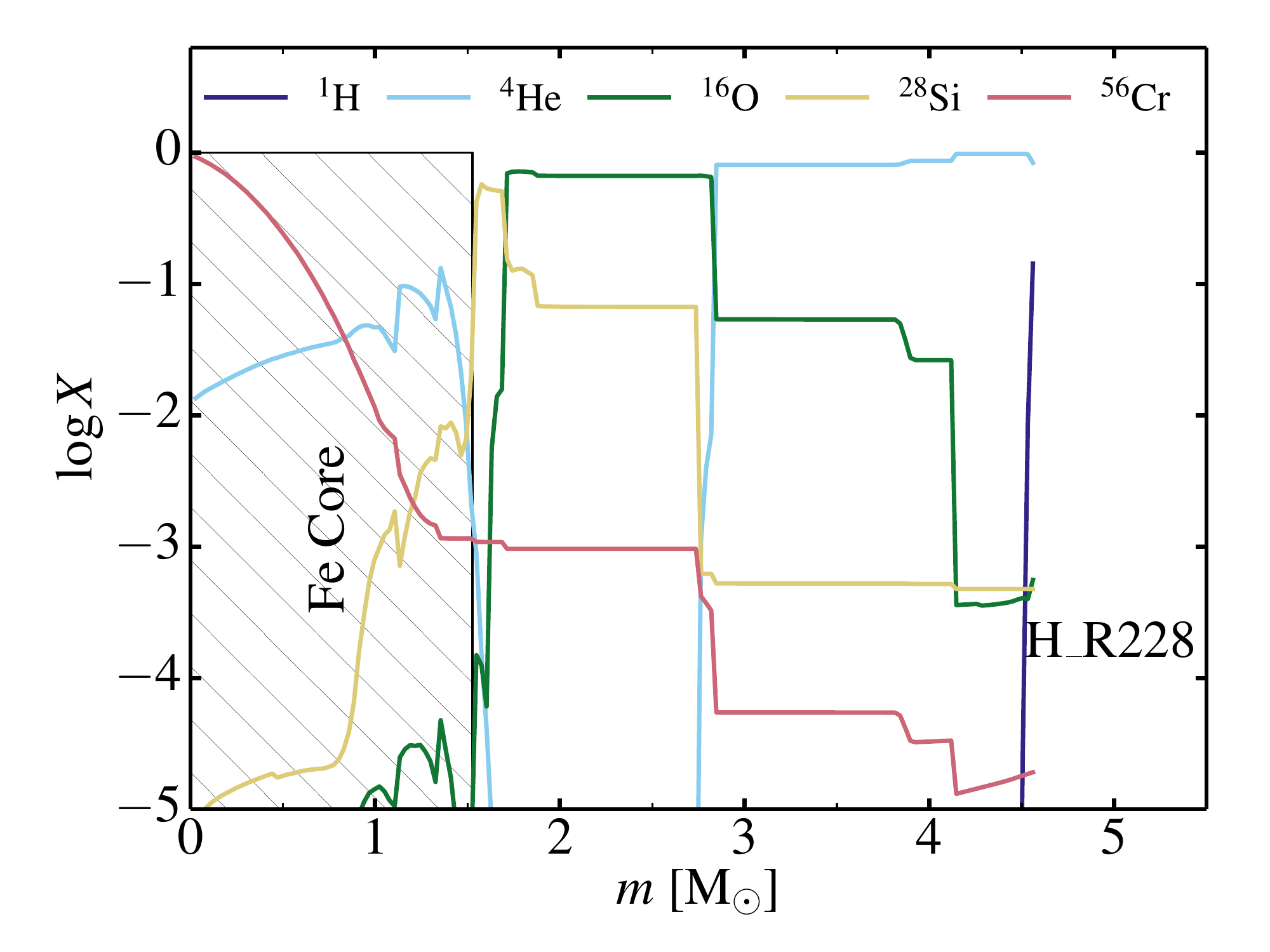}
 \includegraphics[width=0.45\hsize]{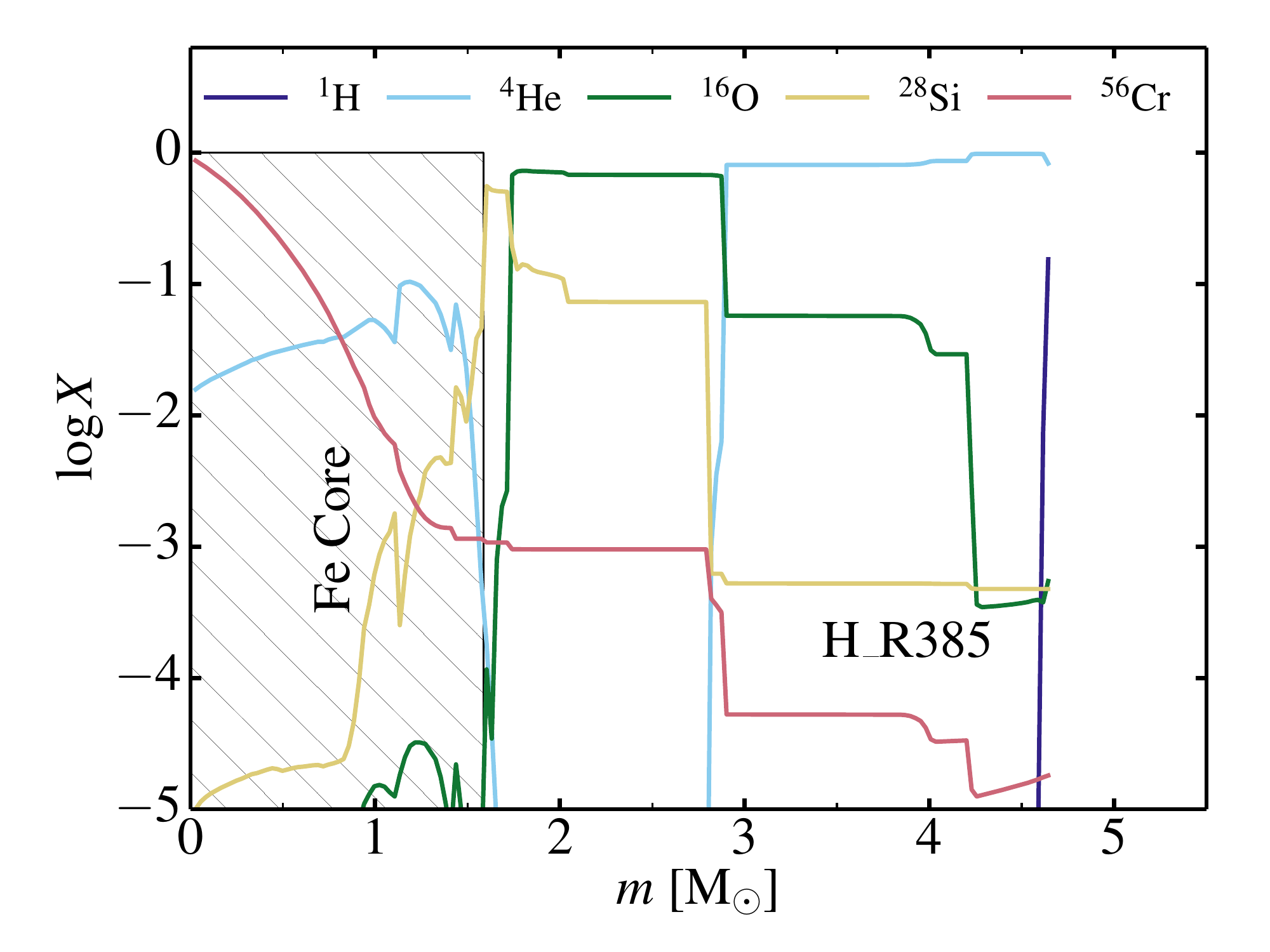}
 \includegraphics[width=0.45\hsize]{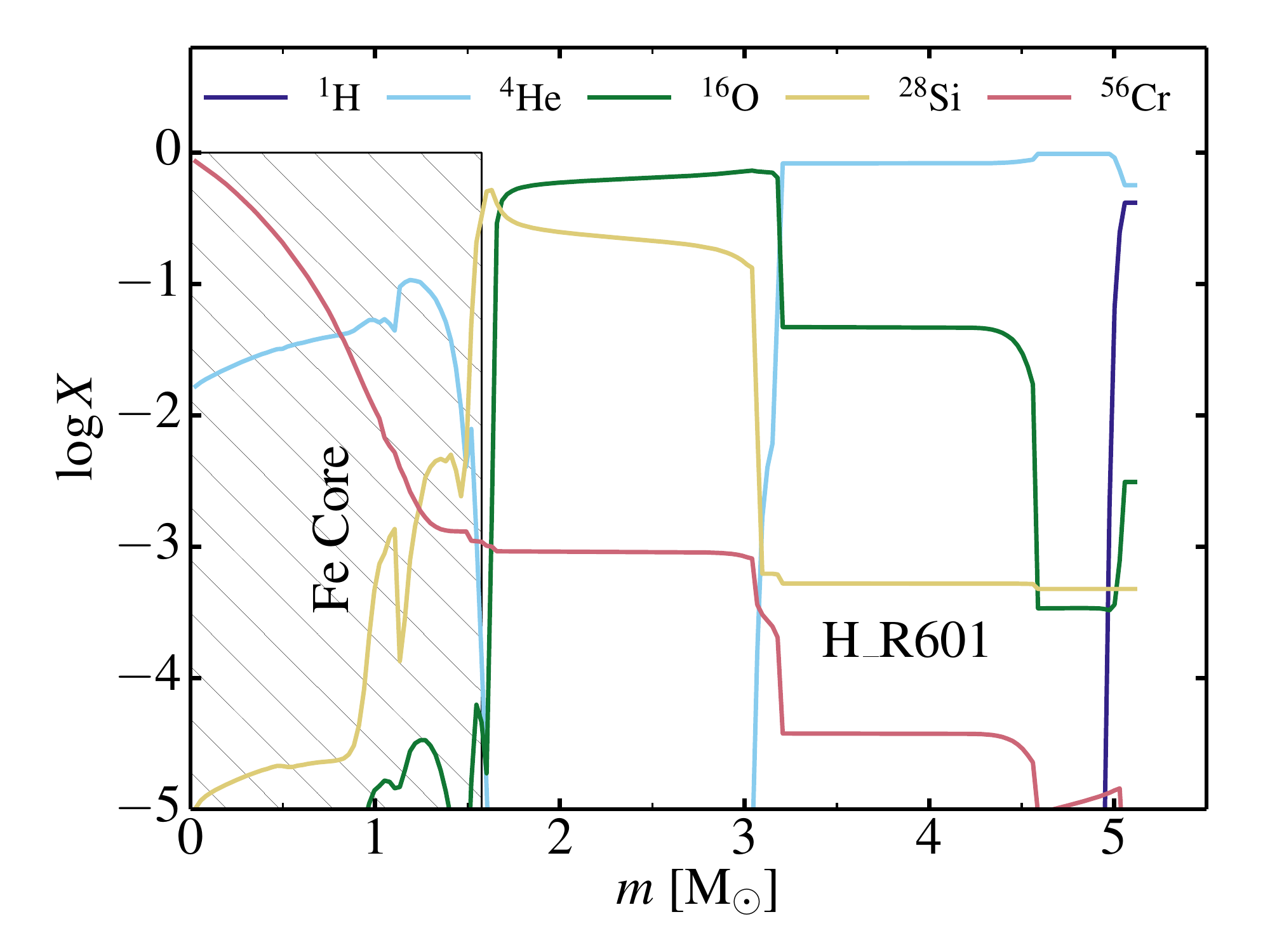}
 \includegraphics[width=0.45\hsize]{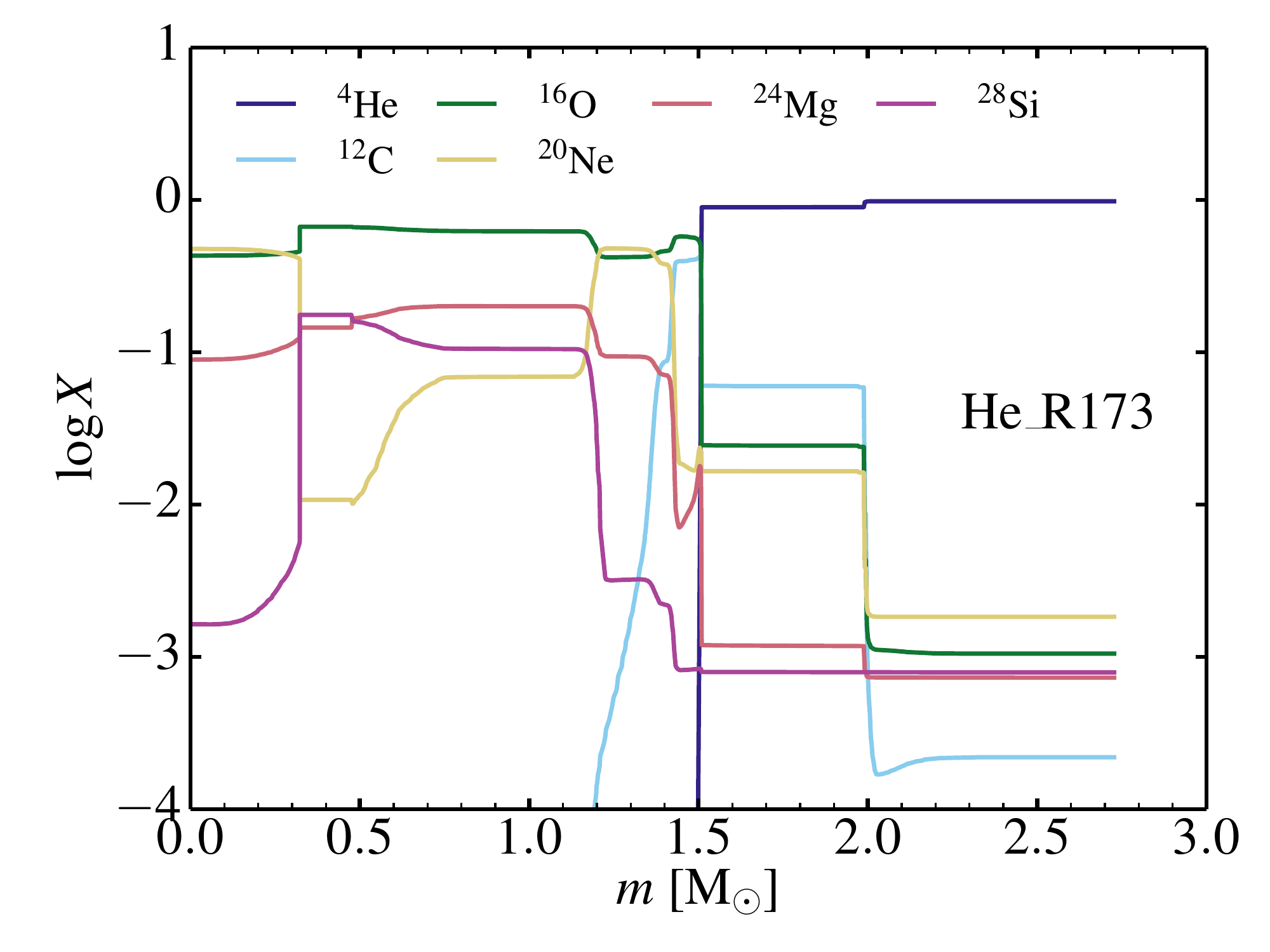}
\caption{Composition profiles for our set of models (evolved with \mesa\ or \bec)
prior to explosion. In model He\_R173,
the composition is markedly different from other models because the star evolution
is stopped at the onset of core neon burning.
\label{fig_mesa_comp_h_all}
}
\end{center}
\end{figure*}

\begin{figure*}
\begin{center}
 \includegraphics[width=0.45\hsize]{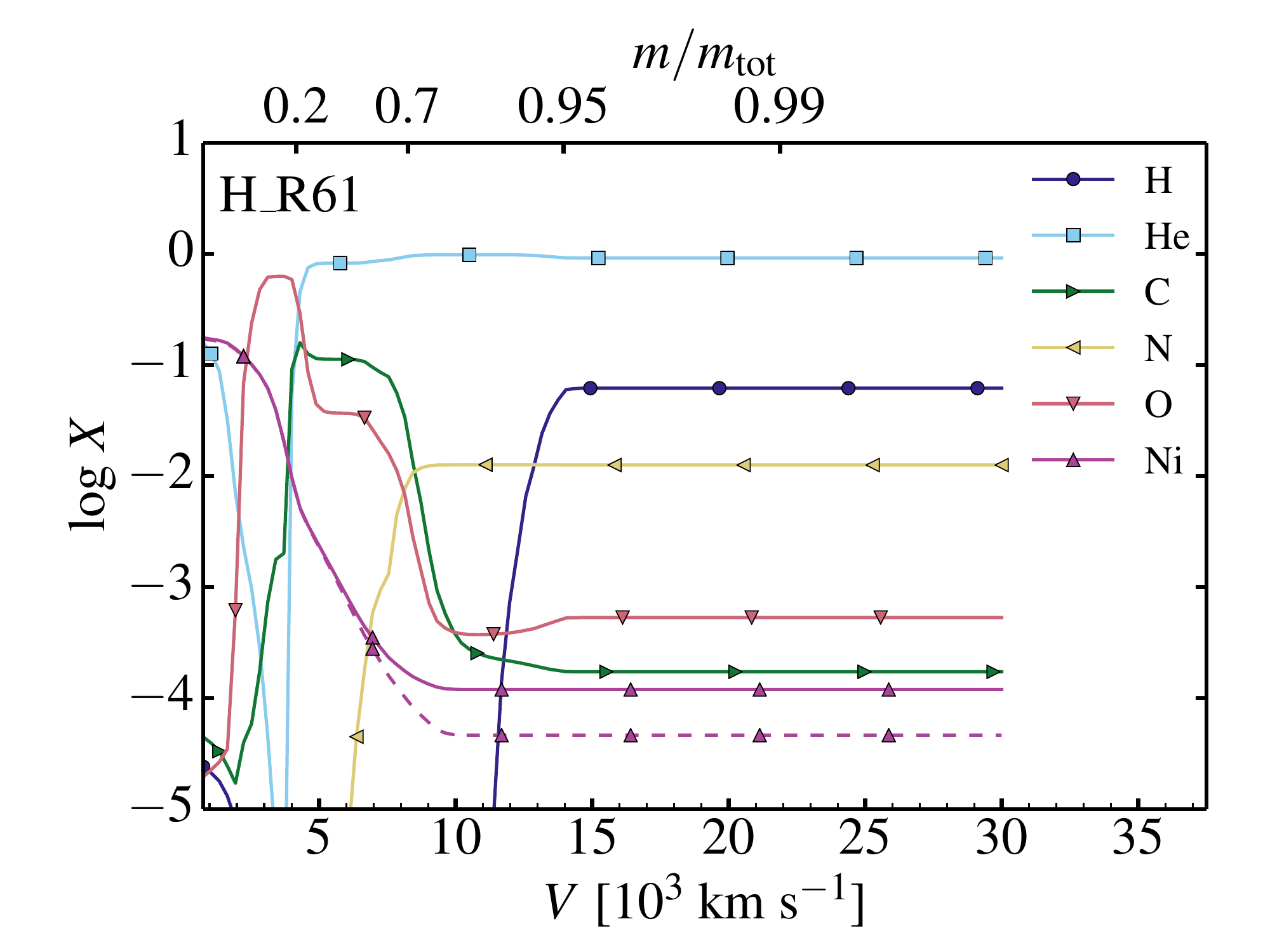}
 \includegraphics[width=0.45\hsize]{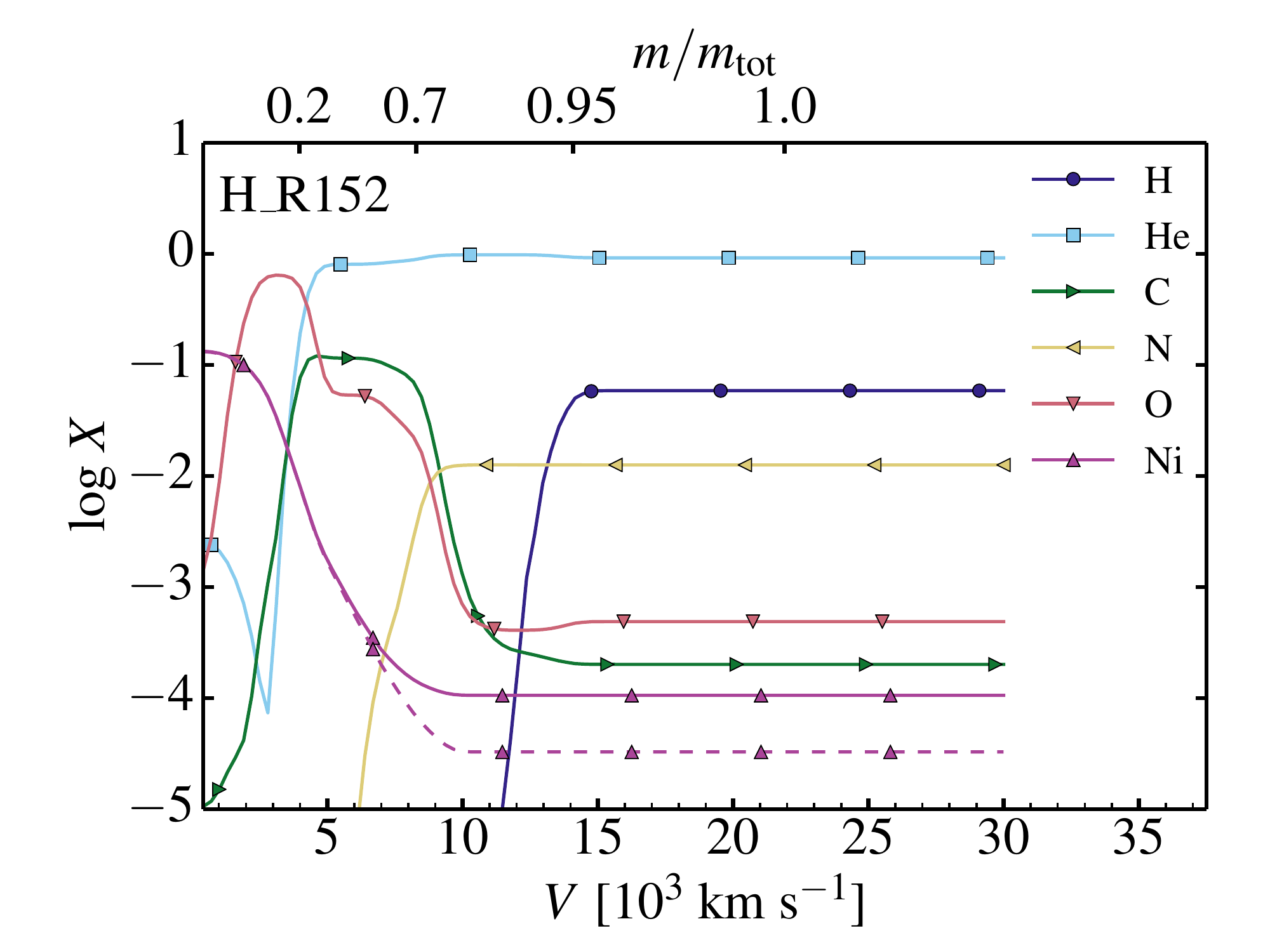}
 \includegraphics[width=0.45\hsize]{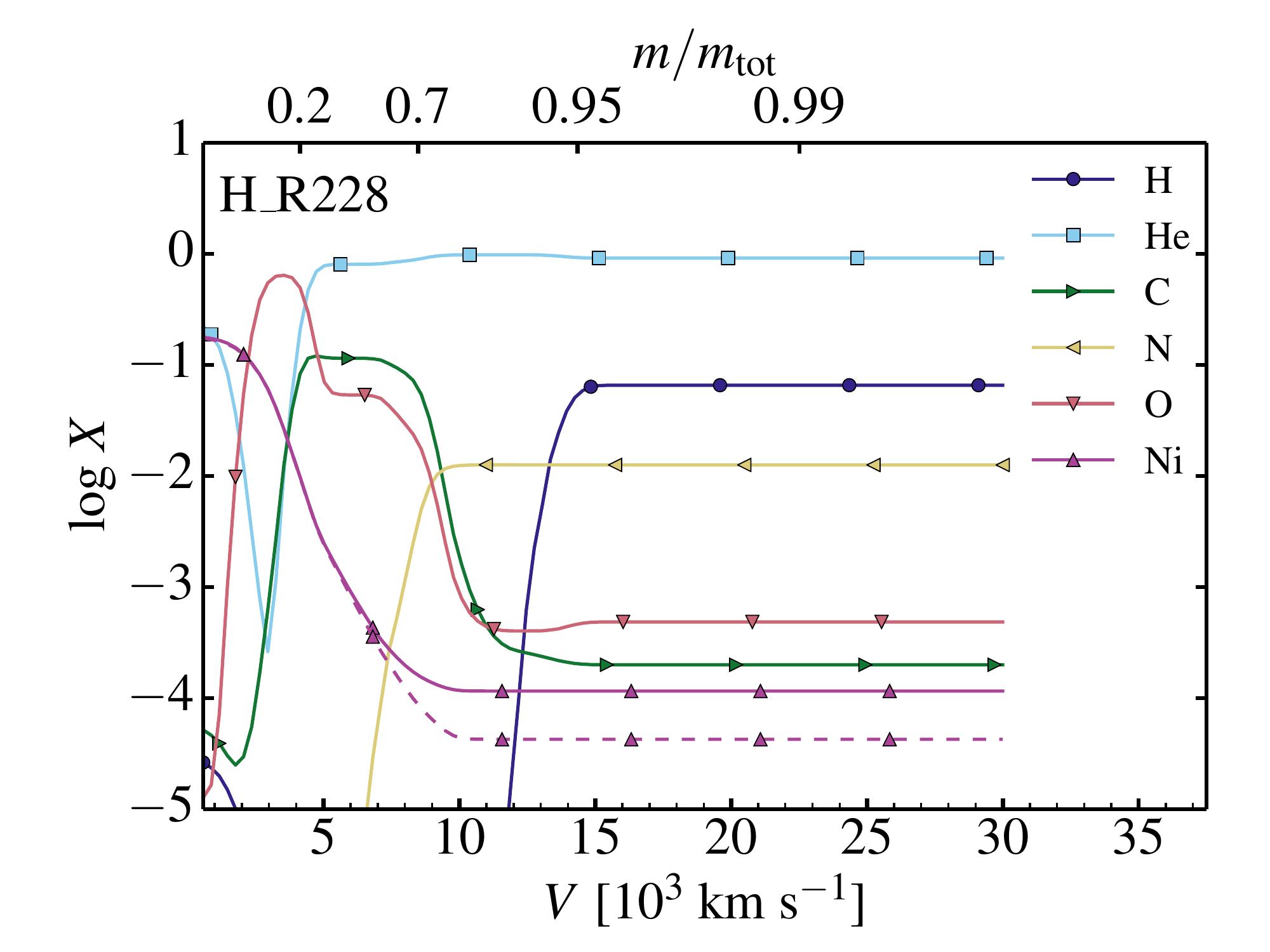}
 \includegraphics[width=0.45\hsize]{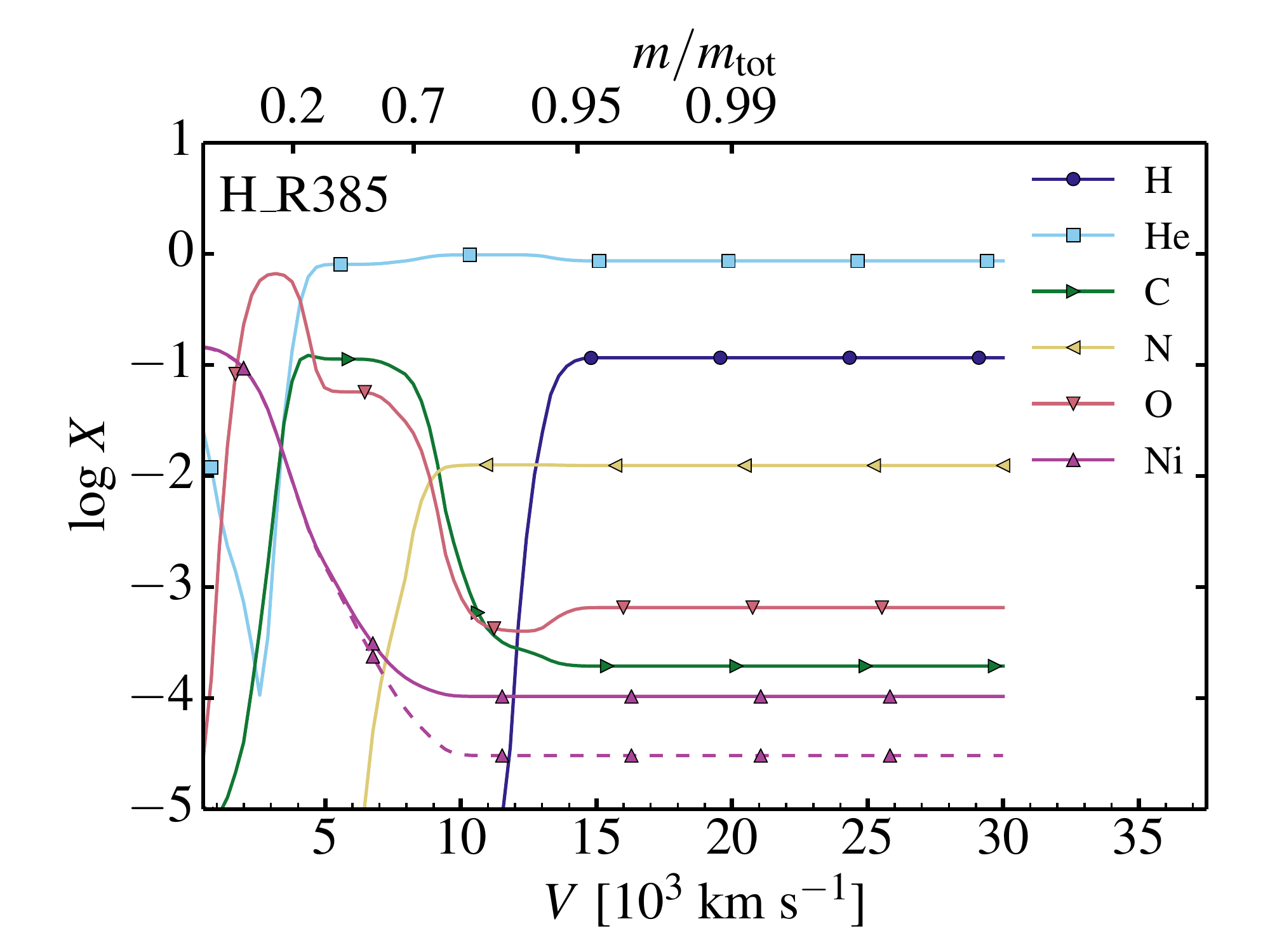}
 \includegraphics[width=0.45\hsize]{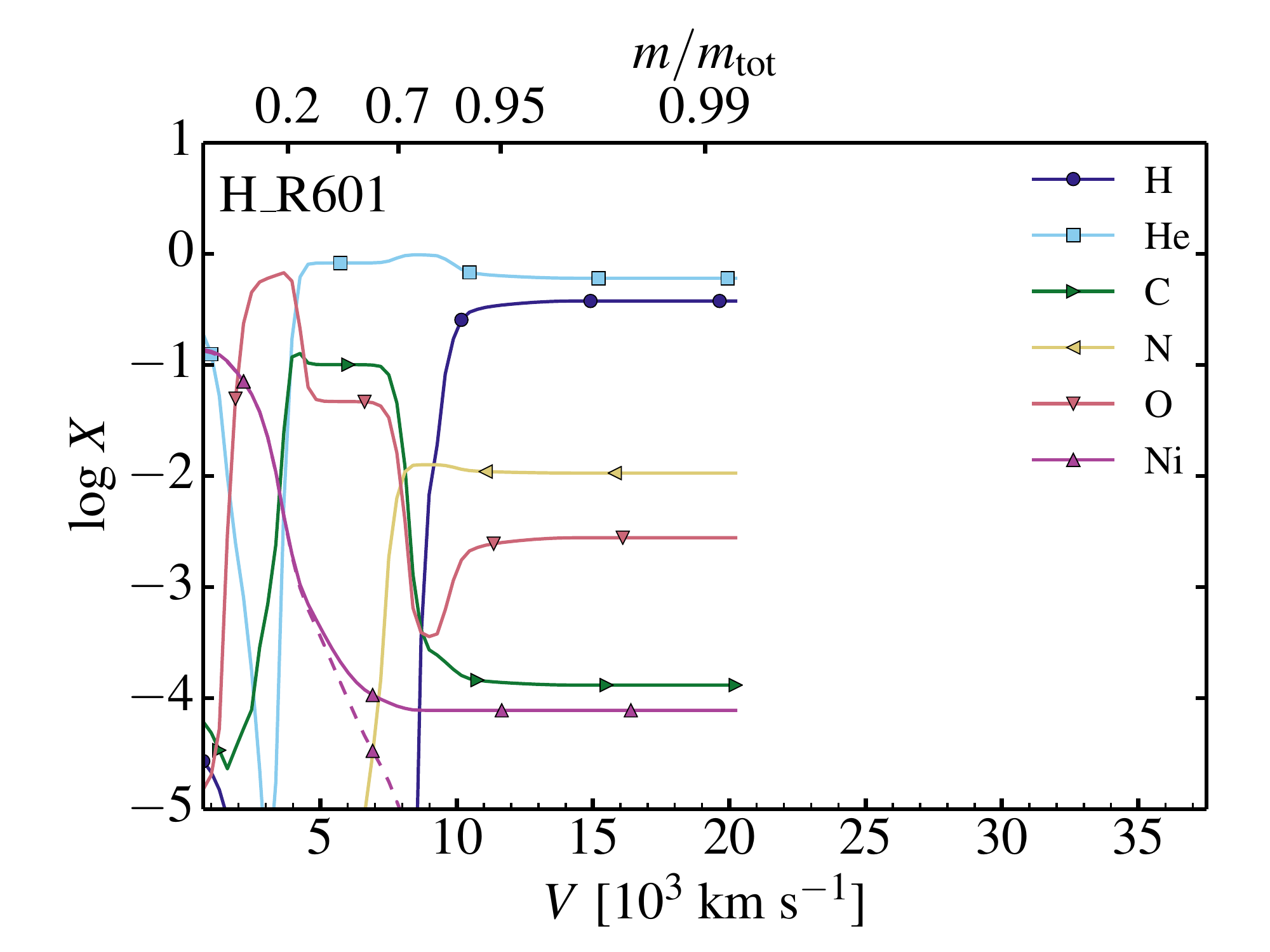}
 \includegraphics[width=0.45\hsize]{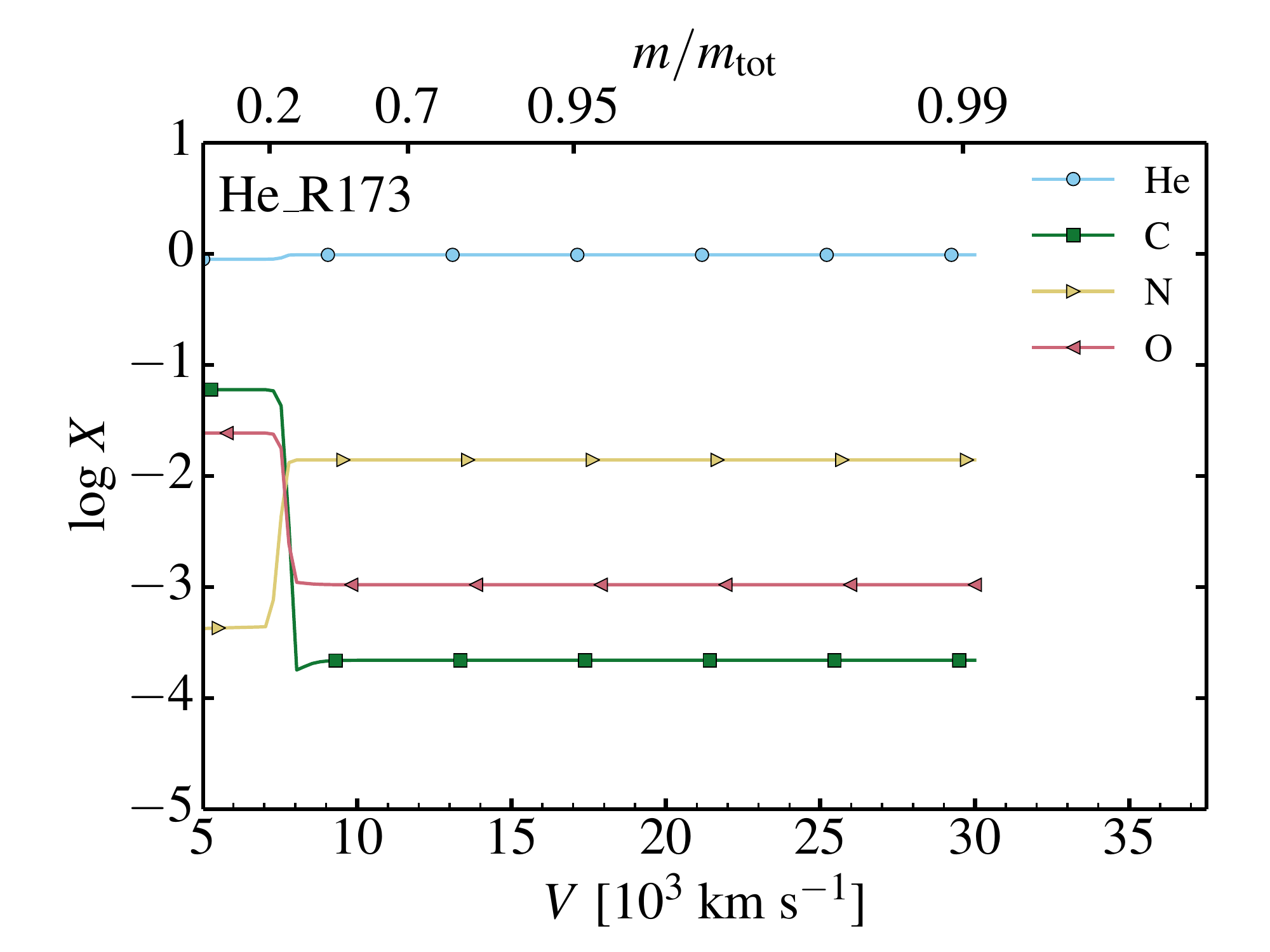}
\caption{Same as Fig.~\ref{fig_mesa_comp_h_all}, but now for the ejecta produces
with \v1d. The time is 1.2\,d after explosion and corresponds to the time of remapping
into \cmfgen.
\label{fig_v1d_comp_h_all}
}
\end{center}
\end{figure*}

In Fig.~\ref{fig_mesa_comp_h_all}, we show the composition profiles
at the time of collapse for each of the explosion models discussed in the paper.
The progenitor models are evolved from the main sequence with the codes \mesa\
or \bec\ (see Section~\ref{sect_presn}).
In Fig.~\ref{fig_v1d_comp_h_all}, we show the composition profiles
for the corresponding ejecta, computed with \v1d\ (see Section~\ref{sect_v1d}).

\section{Montage of spectra for all models}

\begin{figure*}
\begin{center}
 \includegraphics[width=0.45\hsize]{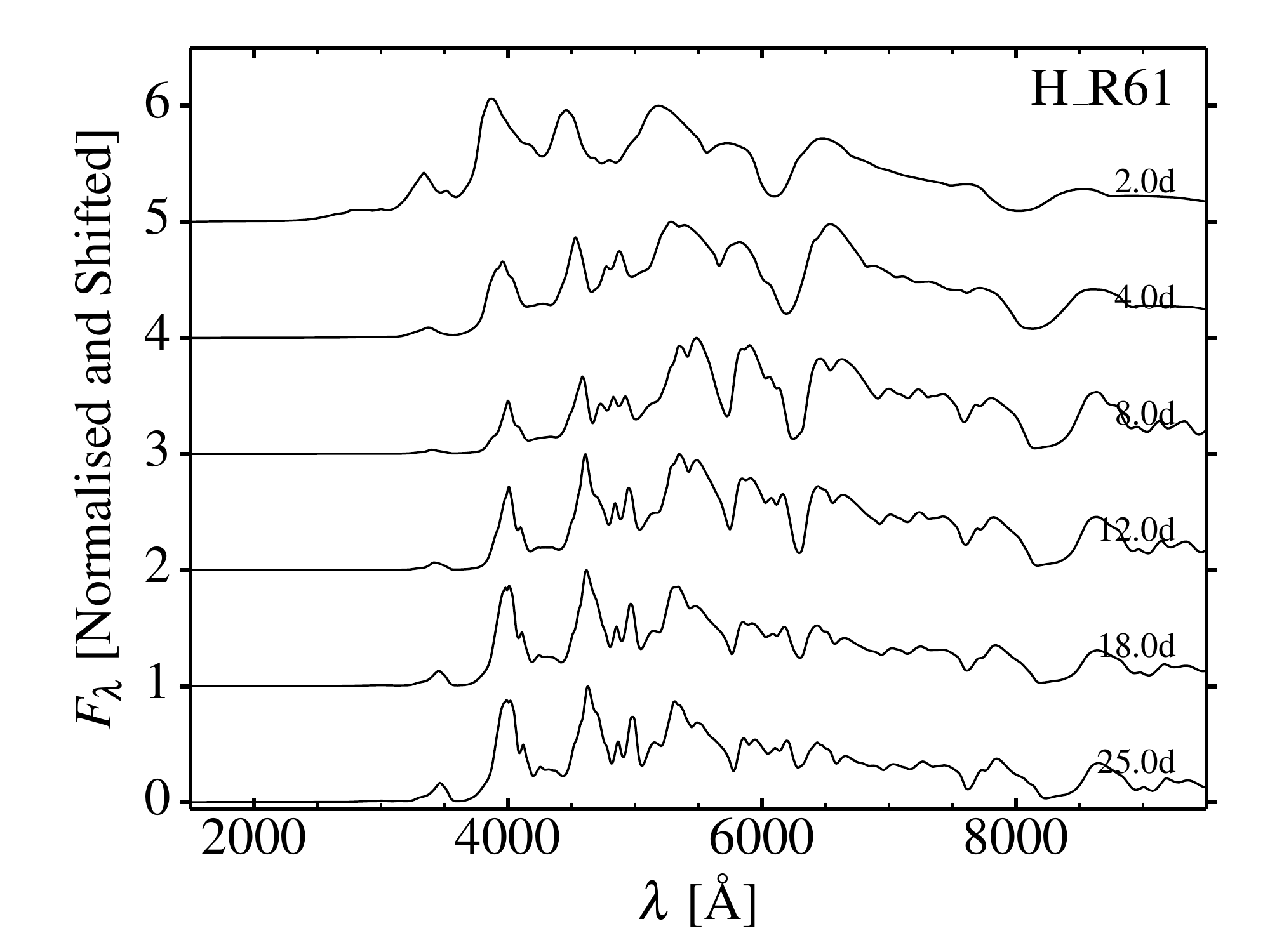}
 \includegraphics[width=0.45\hsize]{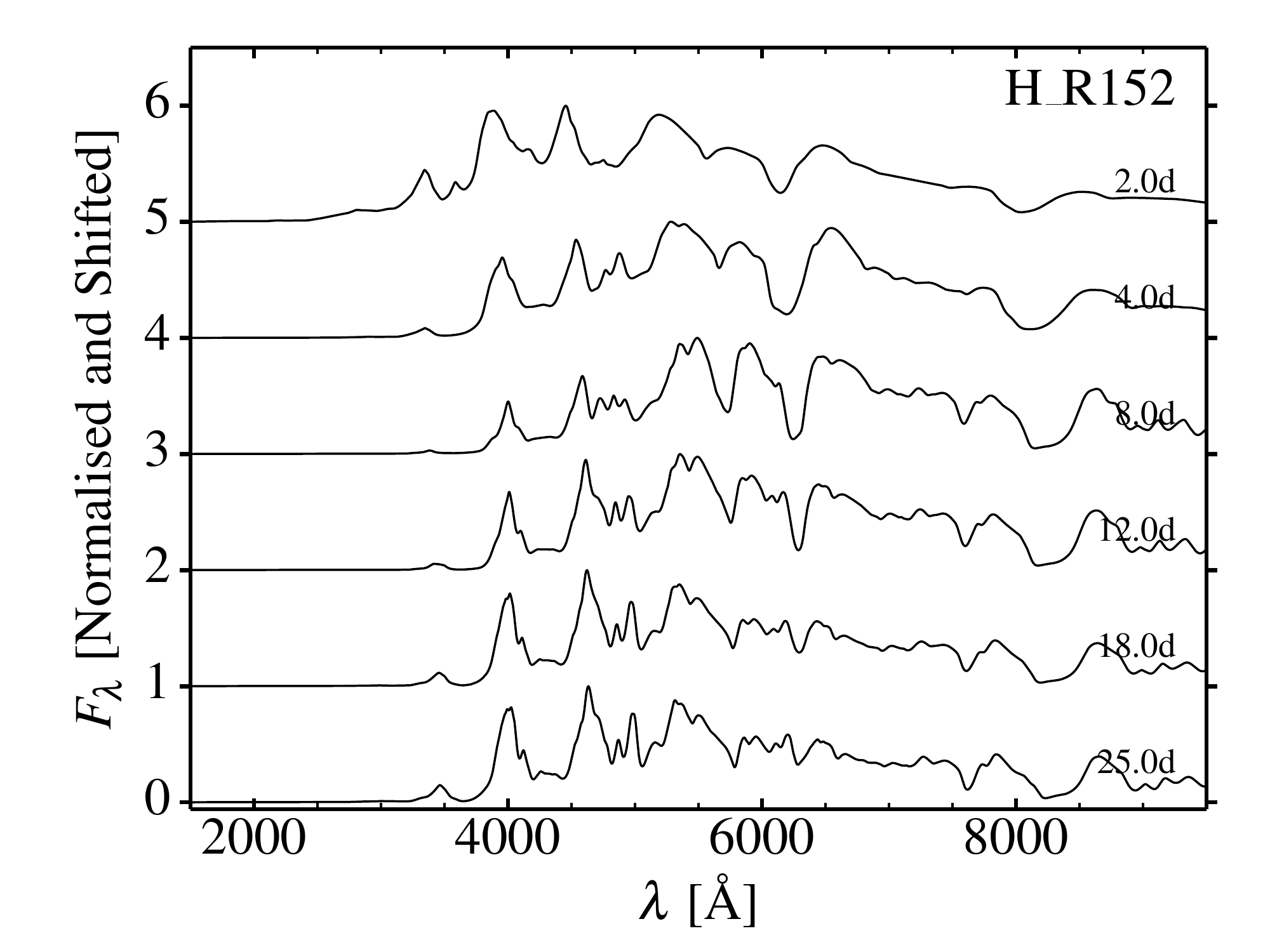}
 \includegraphics[width=0.45\hsize]{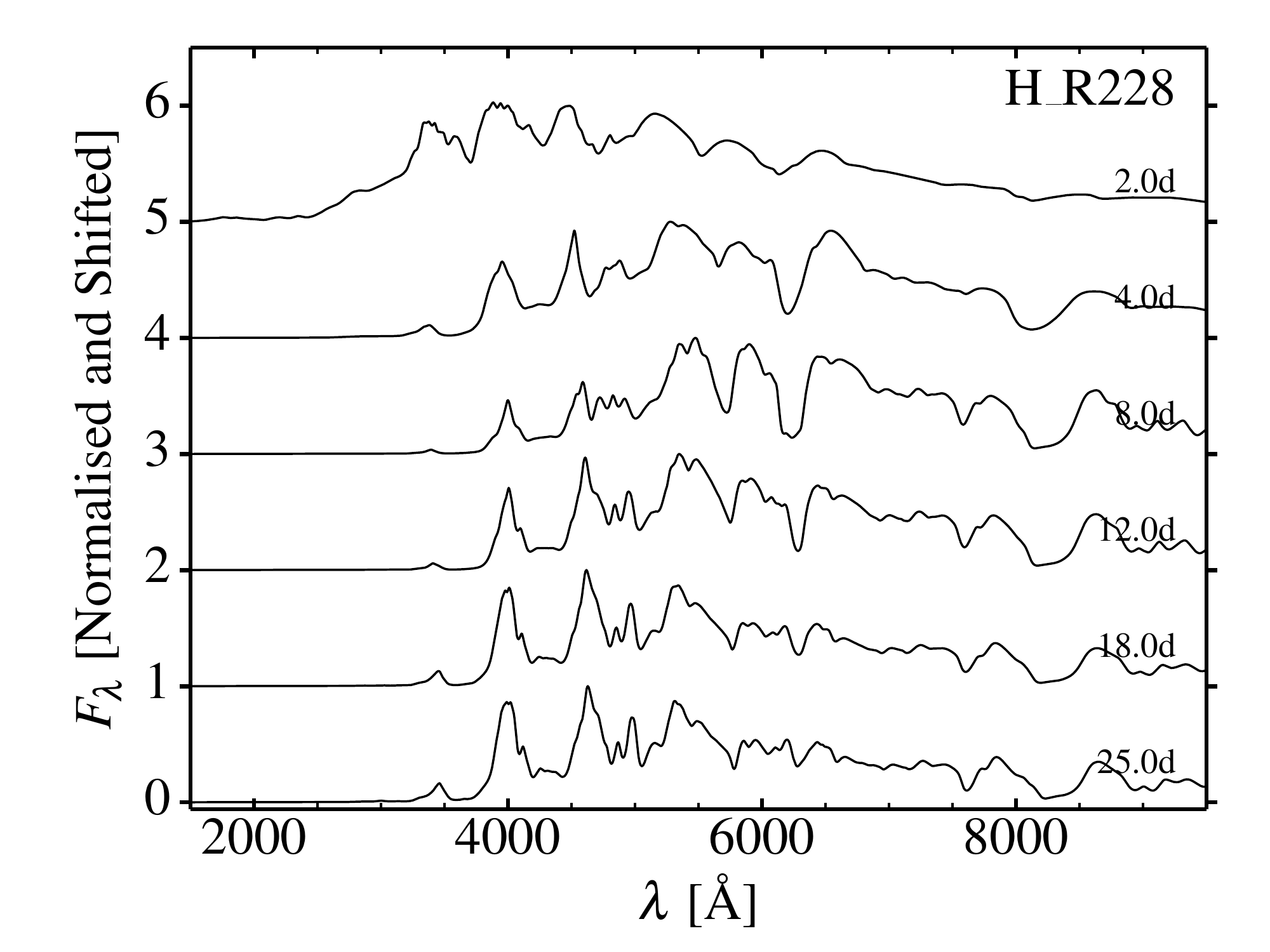}
 \includegraphics[width=0.45\hsize]{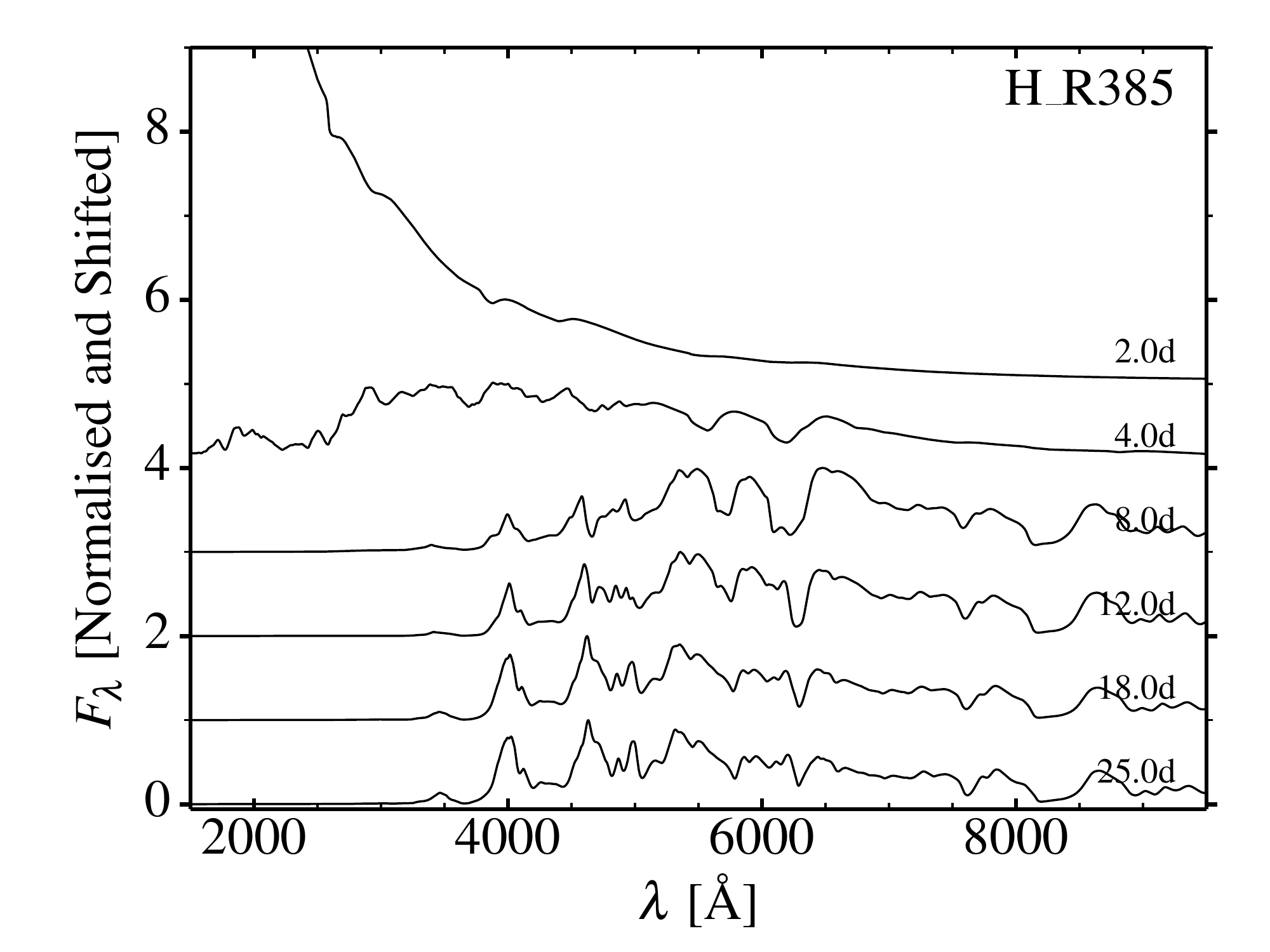}
 \includegraphics[width=0.45\hsize]{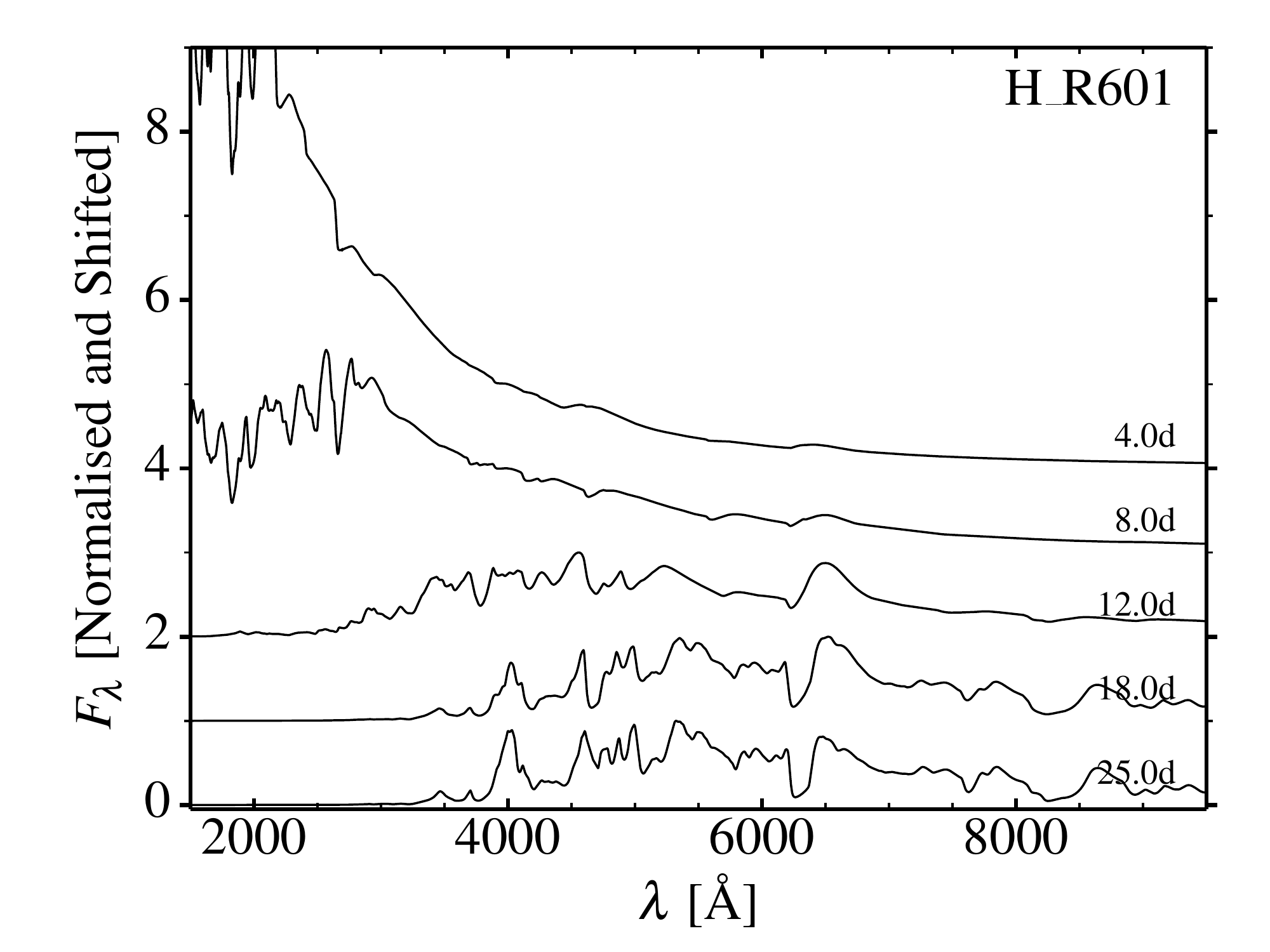}
 \includegraphics[width=0.45\hsize]{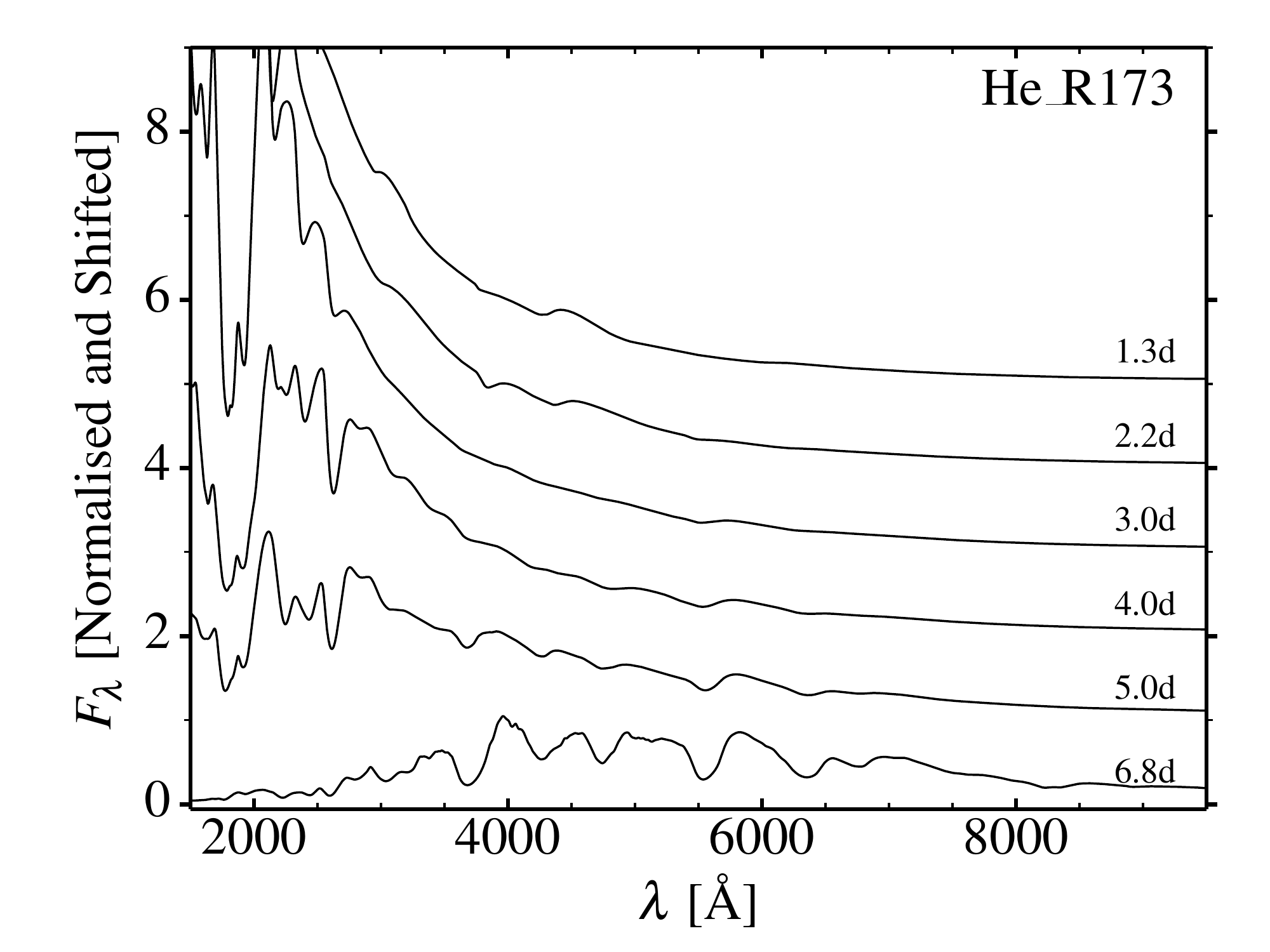}
\caption{Illustration of the spectral evolution computed with \cmfgen\ for models
H\_R61, H\_R152, H\_R228, H\_R385, and H\_R601 (H-rich models),
and for the He-giant star model He\_173 (H-deficient).
\label{fig_spec_appendix}
}
\end{center}
\end{figure*}

In Fig.~\ref{fig_spec_appendix}, we show a montage of spectra
computed with \cmfgen\ for each model, from $1-2$\,d after
explosion until the time of bolometric maximum for the H-rich
models or up to a week for the H-poor model (see Sections~\ref{sect_he_giant}
and \ref{sect_rad} for discussion).

\label{lastpage}

\end{document}